\documentclass[preprintnumbers,amsmath,amssymb,prd,notitlepage,twocolumn,nofootinbib]{revtex4-1}

\pdfoutput=1
\usepackage[english]{babel}
\usepackage[utf8]{inputenc}
\usepackage{graphicx}
\usepackage{subfigure}
\usepackage{units}
\usepackage{xfrac}
\usepackage{ulem}

\usepackage{color}

\newcommand\lya{Ly$\alpha$}
\newcommand\lyaf{LyaF}
\newcommand{\lcdm}{$\Lambda$CDM}
\newcommand{\olcdm}{$o\Lambda$CDM}
\newcommand{\wlcdm}{$w$CDM}
\newcommand{\wcdm}{$w$CDM}
\newcommand{\owcdm}{$o w $CDM}
\newcommand{\wwacdm}{$ w_0 w_a $CDM}
\newcommand{\wowacdm}{$ w_0 w_a $CDM}
\newcommand{\owwacdm}{$o w_0 w_a $CDM}
\newcommand{\owowacdm}{$o w_0 w_a $CDM}
\newcommand{\nucdm}{$\nu$CDM}
\newcommand{\nuCDM}{$\nu$CDM}

\newcommand{\apar}{\alpha_\parallel}
\newcommand{\aperp}{\alpha_\perp}
\newcommand{\hubunits}{\,{\rm km}\,{\rm s}^{-1}\,{\rm Mpc}^{-1}}
\newcommand{\hmpc}{h^{-1}\,{\rm Mpc}}
\newcommand{\kms}{\,{\rm km}\,{\rm s}^{-1}}
\newcommand{\rhocrit}{\rho_{\rm crit}}
\newcommand{\OmegaDE}{\Omega_{\rm de}}
\newcommand{\rhoDE}{\rho_{\rm de}}
\newcommand{\neff}{N_{\rm eff}}
\newcommand{\eV}{\,{\rm eV}}

\begin{document}

\title{Cosmological implications of baryon acoustic oscillation (BAO)
measurements}

\author{\'Eric Aubourg$^{1 }$}
\author{Stephen Bailey$^{2 }$}
\author{Julian~E. Bautista$^{1 }$}
\author{Florian Beutler$^{2 }$}
\author{Vaishali Bhardwaj$^{3,2 }$}
\author{Dmitry Bizyaev$^{4 }$}
\author{Michael Blanton$^{5 }$}
\author{Michael Blomqvist$^{6 }$}
\author{Adam S. Bolton$^{7 }$}
\author{Jo Bovy$^{8 }$}
\author{Howard Brewington$^{4 }$}
\author{J. Brinkmann$^{4 }$}
\author{Joel R. Brownstein$^{7 }$}
\author{Angela Burden$^{9 }$}
\author{Nicol\'as G. Busca$^{1,10,11 }$}
\author{William Carithers$^{2 }$}
\author{Chia-Hsun Chuang$^{12 }$}
\author{Johan Comparat$^{12 }$}
\author{Rupert A.C. Croft$^{13,14 }$}
\author{Antonio J. Cuesta$^{15,16 }$}
\author{Kyle S. Dawson$^{7 }$}
\author{Timoth\'ee Delubac$^{17 }$}
\author{Daniel J. Eisenstein$^{18 }$}
\author{Andreu Font-Ribera$^{2 }$}
\author{Jian Ge$^{19 }$}
\author{J.-M. Le Goff$^{20 }$}
\author{Satya Gontcho A Gontcho$^{16 }$}
\author{J. Richard Gott,~III$^{21 }$}
\author{James E. Gunn$^{21 }$}
\author{Hong Guo$^{22,7 }$}
\author{Julien Guy$^{23,2 }$}
\author{Jean-Christophe Hamilton$^{1 }$}
\author{Shirley Ho$^{13 }$}
\author{Klaus Honscheid$^{24,25 }$}
\author{Cullan Howlett$^{9 }$}
\author{David Kirkby$^{6 }$}
\author{Francisco S. Kitaura$^{26 }$}
\author{Jean-Paul Kneib$^{17,27 }$}
\author{Khee-Gan Lee$^{28 }$}
\author{Dan Long$^{4 }$}
\author{Robert H. Lupton$^{21 }$}
\author{Mariana Vargas Maga\~na$^{1 }$}
\author{Viktor Malanushenko$^{4 }$}
\author{Elena Malanushenko$^{4 }$}
\author{Marc Manera$^{9,29 }$}
\author{Claudia Maraston$^{9 }$}
\author{Daniel Margala$^{6 }$}
\author{Cameron K. McBride$^{18 }$}
\author{Jordi Miralda-Escud\'{e}$^{30,16 }$}
\author{Adam D. Myers$^{31 }$}
\author{Robert C. Nichol$^{9 }$}
\author{Pasquier Noterdaeme$^{32 }$}
\author{Sebasti\'an E. Nuza$^{26 }$}
\author{Matthew D. Olmstead$^{7 }$}
\author{Daniel Oravetz$^{4 }$}
\author{Isabelle P\^aris$^{33 }$}
\author{Nikhil Padmanabhan$^{15 }$}
\author{Nathalie Palanque-Delabrouille$^{2,20 }$}
\author{Kaike Pan$^{4 }$}
\author{Marcos Pellejero-Ibanez$^{34,35 }$}
\author{Will J. Percival$^{9 }$}
\author{Patrick Petitjean$^{32 }$}
\author{Matthew M. Pieri$^{36 }$}
\author{Francisco Prada$^{12,37,38 }$}
\author{Beth Reid$^{2,39 }$}
\author{James Rich$^{20 }$}
\author{Natalie A. Roe$^{2 }$}
\author{Ashley J. Ross$^{9,25 }$}
\author{Nicholas P. Ross$^{40 }$}
\author{Graziano Rossi$^{41,20 }$}
\author{Jose Alberto Rubi\~{n}o-Mart\'{i}n$^{34,35 }$}
\author{Ariel G. S\'anchez$^{42 }$}
\author{Lado Samushia$^{43,44 }$}
\author{Ricardo Tanaus\'u G\'enova Santos$^{34 }$}
\author{Claudia G. Sc\'occola$^{12,34,45 }$}
\author{David J. Schlegel$^{2 }$}
\author{Donald P. Schneider$^{46,47 }$}
\author{Hee-Jong Seo$^{25,48 }$}
\author{Erin Sheldon$^{49 }$}
\author{Audrey Simmons$^{4 }$}
\author{Ramin A. Skibba$^{50 }$}
\author{An\v{z}e Slosar$^{49 }$}
\author{Michael A. Strauss$^{21 }$}
\author{Daniel Thomas$^{9 }$}
\author{Jeremy L. Tinker$^{5 }$}
\author{Rita Tojeiro$^{9 }$}
\author{Jose Alberto Vazquez$^{49 }$}
\author{Matteo Viel$^{33,51 }$}
\author{David A. Wake$^{52,53 }$}
\author{Benjamin A. Weaver$^{5 }$}
\author{David H. Weinberg$^{25,54 }$}
\author{W. M. Wood-Vasey$^{55 }$}
\author{Christophe Y\`eche$^{20 }$}
\author{Idit Zehavi$^{56 }$}
\author{Gong-Bo Zhao$^{57,9 }$}

\date{\today. Authors' institutions can be found in Appendix A.}

\begin{abstract}
  We derive constraints on cosmological parameters and tests of dark
  energy models from the combination of baryon acoustic oscillation
  (BAO) measurements with cosmic microwave background (CMB) data and a
  recent reanalysis of Type Ia supernova (SN) data. In particular, we
  take advantage of high-precision BAO measurements from galaxy
  clustering and the Lyman-$\alpha$ forest (LyaF) in the SDSS-III
  Baryon Oscillation Spectroscopic Survey (BOSS). Treating the BAO
  scale as an uncalibrated standard ruler, BAO data alone yield a high
  confidence detection of dark energy; in combination with the CMB
  angular acoustic scale they further imply a nearly flat
  universe. Adding the CMB-calibrated physical scale of the sound
  horizon, the combination of BAO and SN data into an ``inverse
  distance ladder'' yields a measurement of $H_0=67.3\pm1.1\hubunits$,
  with 1.7\% precision.  This measurement assumes standard
  pre-recombination physics but is insensitive to assumptions about
  dark energy or space curvature, so agreement with CMB-based
  estimates that assume a flat \lcdm\ cosmology is an important
  corroboration of this minimal cosmological model. For constant dark
  energy ($\Lambda$), our BAO+SN+CMB combination yields matter density
  $\Omega_m=0.301 \pm 0.008$ and curvature $\Omega_k=-0.003 \pm
  0.003$.  When we allow more general forms of evolving dark energy,
  the BAO+SN+CMB parameter constraints are always consistent with flat
  \lcdm\ values at $\approx 1\sigma$. While the overall $\chi^2$ of
  model fits is satisfactory, the LyaF BAO measurements are in
  moderate ($2-2.5\sigma$) tension with model predictions.  Models
  with early dark energy that tracks the dominant energy component at
  high redshift remain consistent with our expansion history
  constraints, and they yield a higher $H_0$ and lower matter
  clustering amplitude, improving agreement with some low redshift
  observations. Expansion history alone yields an upper limit on
  the summed mass of neutrino species, $\sum m_\nu<0.56\eV$ (95\%
  confidence), improving to $\sum m_\nu < 0.25\eV$ if we include the
  lensing signal in the Planck CMB power spectrum.  In a flat \lcdm\
  model that allows extra relativistic species, our data combination
  yields $N_{\rm eff}=3.43 \pm 0.26$; while the LyaF BAO data prefer
  higher $N_{\rm eff}$ when excluding galaxy BAO, the galaxy BAO alone
  favor $N_{\rm eff} \approx 3$.  When structure growth is
  extrapolated forward from the CMB to low redshift, standard dark
  energy models constrained by our data predict a level of matter
  clustering that is high compared to most, but not all, observational
  estimates.
\end{abstract}

\maketitle
\section{Introduction}

Acoustic oscillations that propagate in the pre-recombination universe imprint a
characteristic scale in the clustering of matter, providing a cosmological
``standard ruler'' that can be measured in the power spectrum of cosmic
microwave background (CMB) anisotropies
and in maps of large-scale structure at lower
redshifts \cite{Sakharov66,Peebles70,Sunyaev70,Blake03,Seo03}.  While distance
scale measurements with Type Ia supernovae (SNIa) are calibrated against
systems in the local Hubble flow \cite{Hamuy96,Riess98,Perlmutter99}, the baryon
acoustic oscillation (BAO) scale is computed from first principles, using
physical parameters (such as the radiation, matter, and baryon densities) that
are well constrained by CMB data.  The difference between absolute and relative
measurements, the sharpening of BAO precision with increasing redshift, and the
entirely independent systematic uncertainties make BAO and SNe highly
complementary tools for measuring the cosmic expansion history and testing dark
energy models.  In spectroscopic surveys, BAO measurements in the line-of-sight
dimension allow direct determinations of the expansion rate $H(z)$, in addition
to the constraints from transverse clustering on the comoving angular diameter
distance $D_M(z) \propto \int_0^z cH^{-1}(z) dz$ in a flat spatial 
metric.\footnote{We use the notation $D_M(z)$ to
  refer to the comoving angular diameter distance, which is also referred to in
  the literature as the proper motion distance \cite{Hogg99}.  This notation
  avoids confusion with the proper angular diameter distance
  $D_A(z)=D_M(z)/(1+z)$.}

The first clear detections of low-redshift BAO
\cite{Cole05,Eisenstein05} came from galaxy clustering analyses of the
Two Degree Field Galaxy Redshift Survey (2dFGRS, \cite{Colless01}) and
the luminous red galaxy (LRG, \cite{Eisenstein01}) sample of the Sloan
Digital Sky Survey (SDSS, \cite{York00}).  Analyses of the final
2dFGRS and SDSS-II redshift surveys yielded BAO distance measurements
with aggregate precision of 2.7\% at $z \approx 0.275$
\cite{Percival10}, subsequently sharpened to 1.9\%
\cite{Padmanabhan12} by application of reconstruction methods
\cite{Eisenstein07} that suppress non-linear degradation of the BAO
feature.  The WiggleZ survey \cite{Drinkwater10} pushed BAO
measurements to higher redshifts, achieving 3.8\% aggregate precision
from galaxies in the redshift range $0.4 < z < 1.0$ \cite{Blake11}.
The Six Degree Field Galaxy Survey (6dFGS, \cite{Jones09}) took
advantage of its 17,000 deg$^2$ sky coverage to provide a BAO
measurement at low redshift, achieving 4.5\% precision at $z=0.1$
\cite{Beutler11}.  A recent reanalysis that applies reconstruction to
the main galaxy sample \cite{Strauss02} of SDSS-II obtained 3.8\%
precision at $z=0.15$ \cite{Ross:2014qpa}, in a sky area that has
minimal ($<3\%$) overlap with 6dFGS.

The Baryon Oscillation Spectroscopic Survey 
(BOSS, \cite{Dawson13})
of SDSS-III \cite{Eisenstein11} has two defining objectives:
to measure the BAO distance scale with one-percent precision from a
redshift survey of 1.5 million luminous galaxies at $z=0.2-0.7$,
and to make the first BAO measurement at $z>2$ using 3-dimensional
structure in the \lya\ forest absorption towards a dense grid of 160,000
high-redshift quasars.
This paper explores the cosmological implications of BAO measurements
from the BOSS Data Release 11 (DR11) data sample, in combination with
a variety of other cosmological data.  The measurements themselves,
including detailed discussion of statistical uncertainties and extensive
tests for systematic errors, have been presented in previous papers.
For the galaxy survey, \cite{Anderson14} report a 1.4\% measurement of
$D_M(z)$ and a 3.5\% measurement of $H(z)$ at $z=0.57$ (1$\sigma$ 
uncertainties, with a correlation coefficient of $-0.52$), and
a 2.0\% measurement of $D_V(z) \equiv [D_M^2(z) \times cz/H(z)]^{1/3}$
from lower redshift BOSS galaxies at $z=0.32$.  The $D_V(z)$ precision
at $z=0.57$ is 1.0\%.  For the \lya\ forest (often abbreviated as \lyaf\
below), we combine constraints
from the auto-correlation function, with 2.6\% precision on $H(z)$
and 5.4\% precision on $D_M(z)$ \cite{Delubac14}, and the
quasar-forest cross-correlation, with precision of 3.3\% on $H(z)$
and 3.7\% on $D_M(z)$ \cite{FontRibera14}, both at an effective redshift
$z \approx 2.34$.
While some cosmological analysis appears in these
papers, the combination of galaxy and \lya\ forest BAO measurements
and the addition of
other data allow us to constrain broader classes of cosmological
models and to search for deviations from standard assumptions.

The combination of BAO measurements with precise CMB measurements from
the Planck and WMAP satellites already yields tight
constraints on the parameters of the \lcdm\ cosmological model
(inflationary cold dark matter with a cosmological constant and
zero space curvature\footnote{Throughout the paper, the notation
\lcdm\ refers to spatially flat models; cosmological constant models
allowing non-zero curvature are denoted \olcdm.})
and on one-parameter extensions of this model that
allow, e.g., non-zero curvature, an evolving dark energy density, or a
cosmologically significant neutrino mass \cite{PlanckXVI}.
We also take advantage of another major recent advance, a comprehensive
reassessment of the SNIa distance scale by \cite{Betoule14} using
data from the 3-year Supernova Legacy Survey \cite{Conley11}
and SDSS-II Supernova Survey \cite{Frieman08,Sako14} samples and
additional data at low and high redshifts.
We will examine the consistency of the BAO and SNIa results for relative
distances and the constraints on $H_0$ that emerge from an ``inverse
distance ladder'' that combines the two data sets, in essence using
SNIa to transfer the absolute calibration of the BAO scale from the 
intermediate redshifts where it is precisely measured down to $z=0$.
Our primary focus will be on the cosmological parameter
constraints and model tests that come from combining the BAO
and SNIa data with Planck CMB data.\footnote{We use the Planck 2013
data, which were publicly available at the time of our analysis
and paper submission.  Because best-fit parameter values from 
the Planck 2015 data are similar to those from the Planck 2013 data
\cite{Planck2015XIII}, we expect that using the 2015 data would make
little difference to our results, though with some modest improvements
in parameter uncertainties.  We will present analyses that use the
Planck 2015 data in concert with BOSS DR12 BAO measurements in future work.}
When fitting models to these
data, we will also examine their predictions for observable
measures of structure growth and compare the results to inferences from
weak lensing, clusters, redshift-space distortions, 
and the 1-d \lya\ power spectrum.
The interplay of BAO, CMB, and SNIa constraints, and the more general
interplay between measurements of expansion history and structure
growth, are reviewed at length by \cite{Weinberg13}, along with detailed
introductions to the methods themselves.
In particular, Section 4 of \cite{Weinberg13} provides a thorough description
of the BAO method and its motivation.

Section \ref{sec:data} describes the basic methodology of our analysis,
including the relevant underlying equations, and reviews the BAO,
CMB, and SN measurements that we adopt for our constraints, concluding
with variants of ``BAO Hubble diagrams'' that illustrate our qualitative
results.  Section~\ref{sec:bao-as} presents the constraints obtainable by
assuming that the BAO scale is a standard ruler independent of redshift
without computing its physical scale; in particular, we demonstrate that galaxy
and \lyaf\ BAO alone yield a convincing detection of dark energy and that
addition of the angular scale of the CMB acoustic peaks requires a nearly
flat universe if dark energy is a cosmological constant.  
Section~\ref{sec:invdistladder} presents our inverse distance ladder
determination of $H_0$, which assumes standard recombination physics
but does not assume a specific dark energy model or a flat universe.
Section~\ref{sec:deconstraints} describes our constraints on the parameters
of standard dark energy models, while Section~\ref{sec:alternatives} considers
models that allow early dark energy, decaying dark matter, cosmologically
significant neutrino mass, or extra relativistic species.  We compare the
predictions of our BAO+SN+CMB constrained models to observational estimates
of matter clustering in Section~\ref{sec:growth} and summarize our overall
conclusions in Section~\ref{sec:conclusions}.

\section{Methodology, Models and Data Sets}
\label{sec:data}

\subsection{Methodology}
\label{sec:methodology}

A homogenenous and isotropic cosmological model is 
specified by the curvature parameter $k$
entering the Friedman-Robertson-Walker metric
\begin{equation}
  ds^2 = -d t^2 +  a^2(t) \left[ \frac{dr^2}{1-kr^2} +r ^2 d\Omega^2\right]~,
\end{equation}
which governs conversion between radial and transverse distances,
and by the evolution of $a(t) = (1+z)^{-1}$.
In General Relativity (GR), this evolution is governed by 
the Friedmann equation \cite{Friedman22}, which can be written in
the form
\begin{equation}
  \frac{H^2(a)}{H^2_0} = \frac{\rho(a)}{\rho_0} + \Omega_k a^{-2}~,
\end{equation}
where $H \equiv \dot{a}/a$ is the Hubble parameter, $\rho(a)$ is
the total energy density (radiation + matter + dark energy), and
the subscript 0 denotes the present day ($a=1$).
We define the density parameter of component $x$ by the ratio
\begin{equation}
  \Omega_x = {\rho_x \over \rhocrit} = {8\pi G \over 3H^2} \rho_x ~
\end{equation}
and the curvature parameter
\begin{equation}
  \Omega_k = 1 - \sum \Omega_x ~,
\end{equation}
where the sum is over all matter and energy components and
$\Omega_k = 0$ for a flat ($k=0$) universe.
Density parameters and $\rhocrit$ always refer to values at $z=0$
unless a dependence on $a$ or $z$ is written explicitly, 
e.g., $\Omega_x(z)$.  We will frequently refer to the Hubble
constant $H_0$ through the dimensionless ratio
$h \equiv H_0/100\hubunits$.
The dimensionless quantity $\omega_x \equiv \Omega_x h^2$ is
proportional to the physical density of component $x$ at the
present day.  

Given the curvature parameter and $H(z)$ from the Friedmann equation,
the comoving angular diameter distance can be computed as
\begin{equation}
\begin{split}
  D_M(z) &= {c \over H_0} S_k \left( {D_C(z) \over c/H_0} \right) \\
  &\approx
   D_C(z)\left[1+{1\over 6}\Omega_k\left({D_C(z) \over c/H_0}\right)^2\right],
\end{split}
\label{eqn:dm}
\end{equation}
where the line-of-sight comoving distance is
\begin{equation}
  D_C(z) = {c \over H_0}\int_0^z dz' {H_0 \over H(z')}
\label{eqn:dcomove}
\end{equation}
where
\begin{equation}
S_k (x) =
\begin{cases}
 \sin(\sqrt{-\Omega_k} x)/\sqrt{-\Omega_k} & \Omega_k<0, \\
 \sinh(\sqrt{\Omega_k}x)/\sqrt{\Omega_k} & \Omega_k>0, \\
 x & \Omega_k=0. \\
\end{cases}
\end{equation}
Positive $k$ corresponds to negative $\Omega_k$.
We do not use the small $\Omega_k$ approximation of
equation~(\ref{eqn:dm}) in our calculations, but we provide it here to
illustrate that for small non-zero curvature the change in distance is
linear in $\Omega_k$ and quadratic in $D_C(z)$.

Curvature affects $D_M(z)$ both through its influence on
$H(z)$ and through the geometrical factor in equation~(\ref{eqn:dm}).
The luminosity distance (relevant to supernovae) is $D_L = D_M(1+z)$.

The energy components considered in our models are pressureless
(cold) dark matter, baryons, radiation, neutrinos, and dark energy.
The densities of CDM and baryons scale as $a^{-3}$; we refer
to the density parameter of these two components together as
$\Omega_{cb}$.  The energy density of neutrinos with non-zero
mass scales like radiation at early times and like matter at late times, with
\begin{equation}
\begin{split}
  \frac{\rho_{\nu+r} (a)}{\rho_{\rm crit}} =& 
  \frac{8 \pi^3 k_B^4 G}{45 \hbar^3 c^5 H_0^2} \times \\
  & 
  \left[T_{\rm CMB}(a)^4 + T_\nu(a)^4 \sum_i I(m_i c^2 / k_B T_\nu(a)) \right]
  ~,
\end{split}
\end{equation}
where both CMB temperature $T_{\rm CMB}$ and neutrino temperature
scale inversely with scale factor, and the neutrino temperature is
given by $T_\nu = T_{\rm CMB}
\left(\frac{4}{11}\right)^{\nicefrac{1}{3}} g_c$, where
$g_c=\left(\nicefrac{3.046}{3}\right)^{\nicefrac{1}{4}}$ accounts for
small amount of heating of neutrinos due to electron-positron
annihilation.  The sum in the above expression is over 
neutrino species with masses $m_i$.  The integral $I$ is given by
\begin{equation}
  I(r) = \frac{15}{\pi^4} \int_0^\infty \frac{\sqrt{x^2+r^2}}{e^x+1} x^2 dx
\end{equation}
and must be evaluated numerically. For massless neutrinos
$I(0)=\nicefrac{7}{8}$, while in the limit of very massive neutrinos
$I(r)\sim 45 \zeta(3) (2\pi^4)^{-1} r$ (for $r\gg 1$; here $\zeta(3)$
is the Riemann function), i.e., scaling proportionally with $a$ so
that neutrinos behave like pressureless matter.  When we refer to the
$z=0$ matter density parameter $\Omega_m$, we include the
contributions of radiation (which is small compared to the
uncertainties in $\Omega_m$) and neutrinos (which are non-relativistic
at $z=0$), so that $\Omega_m + \OmegaDE + \Omega_k \equiv
1$. Following the Planck Collaboration \cite{PlanckXVI}, we adopt
$\sum m_\nu = 0.06\,$ eV with one massive and two massless neutrino
species in all models except the one referred to as
\nucdm, where it is a free parameter. The
default implies $\omega_{\nu} = 6.57\times10^{-4}$ including massless
species and $\omega_{\nu}=6.45\times10^{-4}$ excluding them. The
effect of finite neutrino temperature at $z=0$ is a very small
$10^{-4}$ relative effect.  The adopted values are close to the
minimum value allowed by neutrino oscillation experiments.

We consider a variety of models for the evolution of the energy
density or equation-of-state parameter $w = p_{\rm de}/\rho_{\rm de}$.
Table~\ref{tab:models} summarizes the primary models discussed in the
paper, though we consider some additional special cases in
Section~\ref{sec:alternatives}.   \lcdm\ represents a flat universe
with a cosmological constant ($w=-1$).  \olcdm\ extends this model to
allow non-zero $\Omega_k$.  \wcdm\ adopts a flat universe and constant
$w$, and \owcdm\ generalizes to non-zero $\Omega_k$.  \wowacdm\ and
\owowacdm\ allow $w(a)$ to evolve linearly with $a(t)$, $w(a) = w_0 +
w_a(1-a)$.  PolyCDM adopts a quadratic polynomial form for $\rho_{\rm
  de}(a)$ and allows non-zero space curvature, to provide a highly
flexible description of the effects of dark energy at low
redshift. Finally, Slow Roll Dark Energy is an example of a
one-parameter evolving-$w$ model, based on a quadratic dark energy
potential.

We focus in this paper on parameter constraints and model tests
from measurements of cosmic distances and expansion rates,
which we refer to collectively as ``expansion history'' or
``geometric'' constraints.
We briefly consider comparisons to measurements of low-redshift
matter clustering in Section~\ref{sec:growth}.
In this framework, the crucial roles of CMB anisotropy measurements
are to constrain the parameters (mainly $\omega_m$ and $\omega_b$)
that determine the BAO scale and to determine the angular diameter
distance to the redshift of recombination.  
For most of our analyses, this approach allows us to use a highly compressed
summary of CMB constraints, discussed in Section~\ref{sec:cmb} below,
and to compute parameter constraints with a simple and fast
Markov Chain Monte Carlo (MCMC) code that computes expansion rates
and distances from the Friedmann equation.
The code is publicly available with data used in this paper at
\texttt{https://github.com/slosar/april}.

\begin{table*}
  \centering
  \begin{tabular}{c|l|c|c}
    Name & Friedmann equation ($\nicefrac{H^2}{H^2_0}$) & Curvature  &
    Section(s)\\
    \hline
    \lcdm  & $\Omega_{cb} a^{-3} + \Omega_\Lambda
    +{\rho_{\nu+r}(z)}/{\rho_{\rm crit}}$ & no & \ref{sec:bao-as}-\ref{sec:deconstraints}\\
    \olcdm & $\Omega_{cb} a^{-3} + \Omega_\Lambda
    +{\rho_{\nu+r}(z)}/{\rho_{\rm crit}} + \Omega_k a^{-2}$ & yes & \ref{sec:bao-as}-\ref{sec:deconstraints}\\
    \wcdm & $\Omega_{cb} a^{-3} + \OmegaDE a^{-3(1+w)}
    +{\rho_{\nu+r}(z)}/{\rho_{\rm crit}}  $  & no & \ref{sec:deconstraints} \\
    \owcdm & $\Omega_{cb} a^{-3} + \OmegaDE a^{-3(1+w)}
    +{\rho_{\nu+r}(z)}/{\rho_{\rm crit}} + \Omega_k a^{-2} $  & yes & \ref{sec:deconstraints}\\
    \wowacdm & $\Omega_{cb} a^{-3} + \OmegaDE
    a^{-3(1+w_0+w_a)}\exp[{-3w_a(1-a)}] +{\rho_{\nu+r}(z)}/{\rho_{\rm
        crit}}  $  & no & \ref{sec:deconstraints}\\
    Slow Roll Dark Energy &  $\Omega_{cb} a^{-3}  +{\rho_{\nu+r}(z)}/{\rho_{\rm crit}}  + 
    \Omega_{DE}\left[ {a^{-3}}/(\Omega_m a^{-3}+\Omega_{DE})
    \right]^{\delta w_0/\Omega_{DE}}$ & no & \ref{sec:deconstraints}\\
    \hline
    \owowacdm & $\Omega_{cb} a^{-3} + \OmegaDE
    a^{-3(1+w_0+w_a)}\exp[{-3w_a(1-a)}] +{\rho_{\nu+r}(z)}/{\rho_{\rm
        crit}} + \Omega_k a^{-2} $  & yes & \ref{sec:invdistladder}-\ref{sec:deconstraints}\\
    PolyCDM & $\Omega_{cb} a^{-3} + (\Omega_1 + \Omega_k) a^{-2} +
    \Omega_2 a^{-1}  + (1-\Omega_{cb}-\Omega_k-\Omega_1-\Omega_2)  $
    & yes\footnote{with Gaussian prior $\Omega_k=0\pm0.1$}& \ref{sec:invdistladder}\\
    \hline
    Early Dark Energy  &  See relevant section. &no& \ref{sec:ede}\\
    Decaying Dark Matter & See relevant section.  & no & \ref{sec:ddm}\\
    \nuCDM & free neutrino mass
    ($\Sigma m_{\nu} < 1\,{\rm eV}$)  & no & \ref{sec:mnu}\\
    $\Delta N_{\rm eff}~\Lambda$CDM & non-standard
    radiation component ( $2<N_{\rm eff}<5$) & no & \ref{sec:neff}\\
    Tuned Oscillation &  See relevant section.  & no & \ref{sec:tun} \\
 
  \end{tabular}
  \caption{Models considered in the paper and section in the paper
    where they are discussed. The top section is the minimal
    cosmological model (with and without curvature) and various
    extensions in the dark energy sector. The middle group are two
    models used to mimic non-parametric methods (i.e.,
    flexible models where the only de-facto assumption is smoothness of
    the expansion history). The bottom group are various extension of
    the minimal model to which we are sensitive only in conjuction
    with the CMB data. Throughout,   $\Omega_{cb}$ is the $z=0$
    density parameter of baryons + CDM and $\rho_{\nu+r}(z)$ is the energy
    density of radiation + massive neutrinos.  All models except
    \nucdm\ and $\Delta N_{\rm eff}\Lambda$CDM 
    adopt $\sum m_\nu = 0.06\,$eV and the 
    standard radiation content $\neff=3.046$.}
  \label{tab:models}
\end{table*}

\subsection{BAO data}
\label{sec:bao}

The BAO data in this work are summarized in Table \ref{tab:data}
and more extensively discussed below.

\begin{table*}
  \centering
  \setlength{\tabcolsep}{1em}
  \begin{tabular}{c|c|cccc} 
   Name & Redshift &  $D_V/r_d$         &   $D_M/r_d$ &   $D_H/r_d$ &  $r_{\rm off}$ \\
   \hline
   6dFGS     			& 0.106   &  $3.047 \pm 0.137$ &       --    &  --            		&   --      \\ 
  
   MGS				& 0.15 	  &  $4.480 \pm 0.168$ &	-- 	& --			&   --	   \\	
   \hline
   BOSS \texttt{LOWZ} Sample        & 0.32    &  $8.467 \pm 0.167$  &       --     & -- 		&   --     \\
   BOSS \texttt{CMASS}  Sample      & 0.57    &    --	            &  $14.945 \pm 0.210$ & $20.75 \pm 0.73 $ & $-0.52$ \\
   \hline
   \lyaf\ auto-correlation     & 2.34 &    --               &  $37.675 \pm 2.171$ & $9.18 \pm 0.28 $ & $-0.43$ \\
   \lyaf-QSO cross correlation     & 2.36 & --             &  $36.288 \pm 1.344$ & $9.00 \pm 0.30 $   & $-0.39$ \\
   Combined \lyaf          & 2.34    &    --               &  $36.489 \pm 1.152$ & $9.145 \pm 0.204 $ & $-0.48$\\ 
\hline
  \end{tabular} 

  \caption{
    BAO constraints used in this work.  These values are taken from
    \cite{Beutler11} (6dFGS), \cite{Ross:2014qpa} (MGS), \cite{Anderson14} (BOSS galaxies),
    \cite{Delubac14} (BOSS \lyaf\ auto-correlation), and
    \cite{FontRibera14} (BOSS \lyaf\ cross-correlation).
    For our likelihood calculations, we adopt Gaussian approximations for
    6dFGS and LOWZ (with 6dFGS truncated at $\Delta \chi^2=4$), while
    for others we use the full $\chi^2$ look-up tables.
    The \lyaf\ auto-correlation and cross-correlation
    results are used directly; the combined \lyaf\
    numbers are provided here for convenience.
  }
  \label{tab:data}
\end{table*}

The robustness of BAO measurements arises from the fact that a
sharp feature in the correlation function (or an oscillatory feature
in the power spectrum) cannot be readily mimicked by systematics, 
whether observational or astrophysical, as these should be
agnostic about the BAO scale and hence smooth over the relevant part
of the correlation function (or power spectrum). 
In most current analyses, the BAO scale is determined by adopting
a fiducial cosmological model that translates angular and redshift separations
to comoving distances but allowing the location of the BAO feature
itself to shift relative to the fiducial model expectation.
One then determines the likelihood of obtaining the observed
two-point correlation function or power spectrum as a function 
of the BAO offsets, while marginalizing over nuisance parameters.
These nuisance parameters characterize ``broad-band''
physical or observational effects that smoothly change the
shape or amplitude of the underlying correlation function or power spectrum,
such as scale-dependent bias of galaxies or the \lyaf, or distortions
caused by continuum fitting or by variations in star-galaxy separation.
In an isotropic fit, the measurement is encoded in the $\alpha$
parameter, the ratio of the measured BAO scale to that predicted by
the fiducial model.
In an anisotropic analysis, one separately constrains $\alpha_\perp$
and $\alpha_\parallel$, the ratios perpendicular and parallel
to the line of sight. In real surveys the errors on $\alpha_\perp$ and
$\alpha_\parallel$ are significantly correlated for a given redshift
slice, but they are typically uncorrelated across different redshift slices.
While the values of $\alpha$ are referred to a specified fiducial model,
the corresponding physical BAO scales are insensitive to the choice of
fiducial model within a reasonable range.

The BAO scale is set by the radius of the sound horizon at the
drag epoch $z_d$ when photons and baryons decouple,
\begin{equation}
  r_d = \int_{z_d}^\infty {c_s(z) \over H(z)} dz~,
  \label{eqn:rdrag}
\end{equation}
where the sound speed in the photon-baryon fluid is
$c_s(z) = 3^{-1/2} c \left[1+\frac{3}{4}\rho_b(z)/\rho_\gamma(z)\right]^{-1/2}$.
A precise prediction of the BAO signal requires a
full Boltzmann code computation, but for reasonable variations
about a fiducial model the {\it ratio} of BAO scales is 
given accurately by the ratio of $r_d$ values computed from the
integral~(\ref{eqn:rdrag}).  Thus, a measurement of $\alpha_\perp$
from clustering at redshift $z$ constrains the ratio of the
comoving angular diameter distance to the sound horizon:
\begin{equation}
  D_M(z)/r_d = \alpha_\perp D_{M,{\rm fid}}(z)/r_{d,{\rm fid}}~.
\end{equation}
A measurement of $\alpha_\parallel$ constrains the Hubble parameter
$H(z)$, which we convert to an analogous quantity:
\begin{equation}
  D_H(z) = c/H(z),
\end{equation}
with
\begin{equation}
  D_H(z)/r_d = \alpha_\parallel D_{H,{\rm fid}}(z)/r_{d,{\rm fid}}~.
\end{equation}
An isotropic BAO analysis measures some effective combination of these
two distances.  If redshift-space distortions are weak, which is a
good approximation for luminous galaxy surveys after reconstruction
but not for the \lyaf, then the constrained quantity is the volume
averaged distance
\begin{equation}
  D_V(z) = \left[z D_H(z) D_M^2(z) \right]^{1/3},
\end{equation}
with
\begin{equation}
  D_V(z)/r_d = \alpha D_{V,{\rm fid}}(z)/r_{d,{\rm fid}}.
\end{equation}

There are different conventions in use for defining $r_d$, which
differ at the 1-2\% level, but ratios of $r_d$ for different cosmologies
are independent of the convention provided one is consistent throughout.
In this work we adopt the \texttt{CAMB} convention for $r_d$, i.e., 
the value that is reported by the linear perturbations code
\texttt{CAMB}\cite{Lewis00}. In practice, we use the
numerically calibrated approximation
\begin{equation}
  r_d \approx {55.154 \exp\left[-72.3(\omega_\nu+0.0006)^2\right]
    \over 
    \omega_{cb}^{0.25351} \, \omega_b^{0.12807} }
    ~{\rm Mpc}~.
   \label{eqn:rd}
\end{equation}
This approximation is accurate to 0.021\% 
for a standard radiation background with 
$\neff=3.046$, $\sum m_\nu < 0.6\eV$, and
values of $\omega_b$ and $\omega_{cb}$ within $3\sigma$ of values
derived by Planck. 
It supersedes a somewhat less
accurate (but still sufficiently accurate)
approximation from \cite{Anderson14} (their eq.~55).
Note that $\omega_\nu = 0.0107(\sum m_\nu / 1.0\,{\rm eV})$, 
and a 0.5 (1.0) eV neutrino mass changes $r_d$ by $-0.26\%$
($-0.92\%$) for fixed $\omega_{cb}$.
For neutrino masses in the range allowed by current cosmological
constraints, the CMB constrains $\omega_{cb}$ rather
than $\omega_{cb}+\omega_\nu$ because neutrinos remain relativistic
at recombination, even though they are non-relativistic at $z=0$.
For the case of extra relativistic species, a useful fitting formula is
\begin{equation}
  r_d \approx {56.067\, \exp\left[-49.7 (\omega_\nu+0.002)^2\right]
    \over
    \omega_{cb}^{0.2436} \,
    \omega_b^{0.128876} \, [1+(\neff-3.046)/30.60] } ~{\rm Mpc}~,
  \label{eqn:rdneff}
\end{equation}
which is accurate to 0.119\% if we restrict
to neutrino masses in the range $0<\sum m_\nu<0.6\eV$ and $3<\neff<5$. 
Increasing $\neff$ by unity decreases $r_d$ by about 3.2\%.

For \lcdm\ models (with $\sum m_\nu = 0.06\eV$, $\neff=3.046$) constrained by
Planck, $r_d = 147.49 \pm 0.59$ Mpc.  This 0.4\% uncertainty is only slightly
larger for \olcdm, \owcdm, or even \owowacdm\ (see Table
\ref{tab:models} for cosmological model definitions), because the relevant quantities
$\omega_{cb}$ and $\omega_b$ are constrained by the relative heights of the
acoustic peaks, not by their angular locations.  The inference of matter energy
densities from peak heights thus depends on correct understanding of physics in
the pre-recombination epoch, where curvature and dark energy are negligible in
any of these models. 

BAO measurements constrain cosmological parameters through their
influence on $r_d$, their influence on $D_H(z)$ via the 
Friedmann equation, and their influence on 
$D_M(z)$ via equation~(\ref{eqn:dm}).
For standard models, the 0.4\% error on $r_d$ from Planck
is small compared to current BAO measurement errors, so the
constraints come mainly through $D_H$ and $D_M$.
From the Friedmann equation, we see that $D_H(z)$ is directly
sensitive to the total energy density at redshift $z$, while
$D_M(z)$ constrains an effective average of the energy density
and is also sensitive to curvature.

Measurements in Table~\ref{tab:data} are reported in terms of
$D_V/r_d$, $D_M/r_d$, and $D_H/r_d$, using the $r_d$ convention
of equation~(\ref{eqn:rdrag}).  Expressed in these terms,
the results are independent of the fiducial cosmologies assumed
in the individual analysis papers.  
Note that some of the referenced papers quote values
of $D_A/r_d$ rather than $D_M/r_d$, differing by a factor $(1+z)$.
An anisotropic analysis yields 
anti-correlated errors on $D_M$ and $D_H$, and the correlation
coefficients are reported in Table~\ref{tab:data}.
Each sample spans a range of redshift, and the quoted effective 
redshift is usually weighted by statistical contribution to the BAO 
measurement.  Because redshift-space positions are scaled to comoving 
coordinates based on a fiducial cosmological model, and BAO measurements are
obtained as ratios relative to that fiducial model,
the values of the effective redshift in Table~\ref{tab:data}
can be treated as exact, e.g., one should compare the
BOSS CMASS numbers to model predictions computed at 
$z=0.5700.$

\subsubsection{Galaxy BAO Measurements}
\label{sec:galaxybao}

The most precise BAO measurements to date come from analyses
of the BOSS DR11 galaxy sample by \cite{Anderson14}.
BOSS uses the same telescope \cite{Gunn06} as the original SDSS,
with spectrographs \cite{Smee13} that were substantially
upgraded to improve throughput and increase multiplexing
(from 640 fibers per plate to 1000).
Redshift completeness for the primary BOSS sample
is nearly 99\%, with typical redshift uncertainty of
a few tens of $\kms$ \cite{Bolton12}.
The DR11 sample has a footprint of 8377 deg$^2$, compared
to 10,500 deg$^2$ expected for the final BOSS galaxy
sample to appear in DR12.

BOSS targets two distinct samples of luminous galaxies
selected by different flux and color cuts \cite{Dawson13}:
CMASS, designed to approximate a constant threshold
in galaxy stellar mass in the range $0.43 < z < 0.7$,
and LOWZ, which provides roughly three times the
density of the SDSS-II LRG sample in the range $0.15 < z < 0.43$.
Analysis of both samples incorporates reconstruction
\cite{Eisenstein07,Padmanabhan12} to sharpen the BAO
peak by partly reversing non-linear effects, thus improving
measurement precision.  For CMASS, we use results of the
anisotropic analysis by \cite{Anderson14}, which yields
$D_M/r_d = 14.945 \pm 0.210$ (1.4\% precision) and
$D_H/r_d = 20.75 \pm 0.73$ (3.5\% precision) with 
anti-correlated errors ($r=-0.52$).

The LOWZ sample does not have sufficient statistical power
for a robust anisotropic analysis, so we use the measurement
of $D_V = 8.467 \pm 0.167$ at $z=0.320$
(discussed in detail by \cite{Tojeiro14}).
We have not included results from the SDSS-II LRG or WiggleZ
surveys cited in the introduction because these partly overlap
the BOSS volume and are not statistically independent.
We do include results of a new analysis \cite{Ross:2014qpa} that uses
reconstruction to achieve a 3.8\% $D_V/r_d$ measurement from
the SDSS main galaxy sample (MGS) at effective redshift $z = 0.150$,
which should be nearly independent of the BOSS LOWZ measurement.
We also include the 6dFGS measurement of $D_V/r_d = 3.047 \pm 0.137$
(4.5\% precision) at $z = 0.106$.  Because the 6dFGS BAO 
detection is of moderate statistical significance and we do
not have a full $\chi^2$ surface for it, we truncate
its $\chi^2$ contributions at $\Delta\chi^2 = 4$ to guard
against non-Gaussian tails of the error distribution.
In practice the 6dFGS measurement carries little
statistical weight in our constraints.
These galaxy BAO measurements are listed in the first
four lines of Table~\ref{tab:data}.

\subsubsection{BOSS \lya\ forest BAO Measurements}
\label{sec:lyabao}

The BAO scale was first measured at higher redshift ($z\sim2.4$) from
the auto-correlation of the \lya\ forest fluctuations in the spectra
of high-redshift quasars from BOSS DR9
(\cite{Busca13},\cite{Slosar13},\cite{Kirkby13}) following the
pioneering work of measuring 3D fluctuations in the forest
\cite{2011JCAP...09..001S}. 
Here we use the
results from \cite{Delubac14}, who present an improved measurement
using roughly twice as many quasar spectra from BOSS DR11.  
The DR11 quasar catalog will be made publicly available simultaneously
with the DR12 catalog in 2015.  The catalog construction is similar
to that of the DR10 quasar catalog described by 
\cite{2014A&A...563A..54P}.  The BOSS quasar selection criteria 
are described by \cite{2012ApJS..199....3R} and the
background methodology papers 
\cite{2012ApJ...749...41B,2011ApJ...743..125K,Richards09,2010A&A...523A..14Y}.

The measurement of \lyaf\ 
BAO peak positions is marginalized over parameters
describing broad-band distortion of the correlation function using the
methodology of \cite{Kirkby13}.
Because of the low effective bias factor of the \lyaf, redshift-space
distortion strongly enhances the BAO peak in the line-of-sight direction.
The measurement of $D_H/r_d$ is therefore more precise (3.1\%) than
that of $D_M/r_d$ (5.8\%), as seen in line 5 of Table~\ref{tab:data}.
The errors of these two measurements are anti-correlated, with 
$r_{\rm off} = -0.43$, and the optimally measured combination
$D_H^{0.7}D_M^{0.3}$ is determined with a precision of $\sim 2\%$.
While the overall signal-to-noise ratio of the BAO measurement is high,
the detection significance for transverse separations ($\mu < 0.5$)
is only moderate, as one can see in Figure~3 of \cite{Delubac14}.

At the same redshift, BAO have also been measured in the cross-correlation of
the \lya\ forest with the density of quasars in BOSS DR11
\cite{FontRibera14}.
While the number of quasar-pixel pairs
is much lower than the number of pixel-pixel pairs in the auto-correlation
function, the clustering signal is much stronger because of the high
bias factor of quasars.  For the same reason, redshift-space distortion
is much weaker in the cross-correlation, and in this case the measurement
precision is comparable for $D_M/r_d$ (3.7\%) and for $D_H/r_d$ (3.3\%),
(line 6 of Table~\ref{tab:data}).
The higher precision of the transverse measurement makes the
cross-correlation measurement an especially valuable complement to the 
auto-correlation measurement.

Even though these results are derived from the same volume, we can consider
them as independent because their uncertainties are not dominated by
cosmic variance.  They are dominated instead by the combination of
noise in the spectra and sparse sampling of the structure in the
survey volume, both of which affect the auto-correlation and
cross-correlation almost independently.  A number of tests using mock
catalogs and several analysis procedures are presented in
\cite{Delubac14}, finding good agreement between error estimates from
the likelihood function and from the variance in mock catalogs.  This
independence allows us to add the $\chi^2$ surfaces from both
publications, which are publicly available at
\url{http://darkmatter.ps.uci.edu/baofit/}.  While we use these two
$\chi^2$ surfaces separately, the last line of Table~\ref{tab:data}
lists the $D_M/r_d$ and $D_H/r_d$ constraints from the combined
measurement, with respective precision of 3.2\% and 2.2\% and a
correlation coefficient $r_{\rm off}=-0.48$.

We caution that BAO measurement from \lyaf\ data is a relatively
new field, pioneered entirely by BOSS, in contrast to the now mature 
subject of galaxy BAO measurement, which has been studied both
observationally and theoretically by many groups.
\cite{Delubac14} present numerous tests using mock catalogs and
different analysis procedures, finding good agreement between
error estimates from the likelihood surface and from mock catalog
variance and identifying no systematic effects that are 
comparable to the statistical errors.
However, the analysis uses only 100 mock catalogs, limiting the
external tests of the tails of the error distribution.
The systematics and error estimation of
the cross-correlation measurement have also been less thoroughly
examined than those of the auto-correlation measurement,
though continuing investigations within the BOSS collaboration
find good agreement with the errors and covariances reported in the
publications above.
On the theoretical side, \cite{Pontzen14} and \cite{Gontcho14}
have examined the potential impact of UV background fluctuations
on \lyaf\ BAO measurement, finding effects that are much
smaller than the current statistical errors.

We anticipate significant improvements in the \lyaf\ analyses
of the Data Release 12 sample, thanks to the larger data set
and ongoing work on broadband distortion modeling, 
larger mock catalog samples, and spectro-photometric calibration.
For the current paper, we adopt the BAO 
likelihood surfaces as reported in \cite{FontRibera14} and
\cite{Delubac14}.  

\subsection{Cosmic Microwave Background Data}
\label{sec:cmb}

In this paper we focus on constraints on the expansion history of the
homogeneous cosmological model. For this purpose, we compress the
Cosmic Microwave Background (CMB) measurements to the variables
governing this expansion history. This approach greatly simplifies the
required computations, allowing us to fit complex models that have a
simple solution to the Friedmann equation without the need to
numerically solve for the evolution of perturbations.  
It is also physically illuminating, making clear what relevant
quantities the CMB determines and distinguishing expansion history
constraints from those that depend on the evolution of clustering.
For some models or special cases we use more complete CMB results obtained
by running the industry-standard \texttt{cosmomc}
\cite{2002PhRvD..66j3511L} or by relying on the publicly available
Planck MCMC 
chains.\footnote{\texttt{http://wiki.cosmos.esa.int/planckpla/index.php/ Cosmological\_Parameters}}

The CMB plays two distinct but important roles in our analysis.
First, we treat the CMB as a ``BAO experiment'' at redshift
$z_\star=1090$, measuring the angular scale of the sound horizon
at very high redshift.
Here we
ignore the small dependence of the last-scattering
redshift $z_\star$ on cosmological parameters and the fact that the
relevant scale for the CMB is $r_\star$ rather than the drag redshift
$r_d$ that sets the BAO scale in low-redshift structure.  We have
checked that both approximations are valid to around $0.1\sigma$ 
for the case of BAO and Planck data and the \lcdm\ model. 
In its second important role,
the CMB calibrates the absolute length of the BAO ruler
through its determination of $\omega_b$ and $\omega_{cb}$.

Inspired by the existence of well-known degeneracies in CMB data
\cite{1999MNRAS.304...75E, 2002PhRvD..66f3007K, Wang13}, we compress
the CMB measurements into three variables: $\omega_b$, $\omega_{cb}$ and
$D_M(1090)/r_d$.  The mean vector and the $3\times3$ covariance
matrix are used to describe the CMB constraints by a simple Gaussian
likelihood.  In order to calibrate these variables, we rely on the
publicly available Planck chains. In particular, we use the
\texttt{base\_Alens} chains with the \texttt{planck\_lowl\_lowLike}
dataset corresponding to the Planck dataset with low-$\ell$ WMAP
polarization (referred to in this paper as Planck+WP).
We find that the data vector
\begin{equation}
  \mathbf{v} = \left(
    \begin{matrix}
      \omega_b \\
      \omega_{cb} \\
      D_M(1090)/r_d\\
    \end{matrix} 
 \right)
\label{eqn:cmbvector}
\end{equation}
can be described by a Gaussian
likelihood with mean 
\begin{equation}
  \mu_{\mathbf{v}} = \left(
    \begin{matrix}
      0.02245\\
      0.1386\\
      94.33\\
    \end{matrix}
    \right)
\label{eqn:cmbmean}
\end{equation}
and covariance
\begin{equation}
  C_{\mathbf{v}} = \left(
    \begin{matrix}
1.286 \times 10^{-7} &  -6.033 \times 10^{-7} &  1.443 \times 10^{-5} \\
-6.033 \times 10^{-7} &  7.542 \times 10^{-6} &  -3.605 \times 10^{-5} \\
1.443 \times 10^{-5} &  -3.605 \times 10^{-5} &  0.004264 \\
    \end{matrix}
\right).
\label{eqn:cmbcovar}
\end{equation}
The fractional diagonal errors on $\omega_b$, $\omega_{cb}$, and
$D_M(1090)/r_d$ are 1.5\%, 1.9\%, and 0.06\%, respectively.  
We similarly compress the WMAP 9-yr data into 
\begin{equation}
  \mu_{\mathbf{v}} = \left(
    \begin{matrix}
      0.02259\\
      0.1354\\
      94.51\\
    \end{matrix}
    \right)
\label{eqn:cmbmeanW9}
\end{equation}
and covariance
\begin{equation}
  C_{\mathbf{v}} = \left(
    \begin{matrix}
2.864 \times 10^{-7} &  -4.809 \times 10^{-7} &  -1.111 \times 10^{-5} \\
-4.809 \times 10^{-7} &  1.908 \times 10^{-5} &  -7.495 \times 10^{-6} \\
-1.111 \times 10^{-5} &  -7.495 \times 10^{-6} &  0.02542 \\
    \end{matrix}
\right).
\label{eqn:cmbcovarW9}
\end{equation}
For
reference in interpreting the cosmological constraints from CMB+BAO
data, especially in Section~\ref{sec:alternatives} below, note
that the contributions to $D_M(1090)$ accumulate over a wide range of
redshift,
with 14\%, 25\%, 38\%, 47\%, 69\%, 88\%, and 99\% of the integral 
in equation~(\ref{eqn:dcomove}) 
coming from redshifts $z < 0.5$, 1.0, 2.0, 3.0, 10, 50, and 640,
respectively.

The \texttt{base\_Alens} model corresponds to the basic flat
$\Lambda$CDM cosmology with explicit marginalisation over the
foreground lensing potential. Our decision to use the flat model
was intentional, since we found that in curved models
there is significant non-Gaussian correlation of $\omega_b$ and
$\omega_m$ with curvature.  Because our BAO data inevitably collapse
more complex models to nearly flat ones, use of the flat data is more
appropriate.  We have tested the data compression in a couple of
simple cases by comparing results of BAO+CMB data to \texttt{cosmomc}
chains and found less than $0.5\sigma$ differences in
best-fit parameter values between using compressed and full
chains. The residual differences 
are driven by the fact that our compressed likelihood
attempts to extract purely geometric information from the CMB data
(for example, values of $\omega_b$ and $\omega_m$ are different at
roughly the same level between chains that marginalise over lensing
potential and those that do not).  For BAO-only data combinations
the results are completely consistent.

Throughout the paper, we refer to the constraints represented
by equations (\ref{eqn:cmbvector})-(\ref{eqn:cmbcovar}) simply as
``Planck'' (although they also include information from
WMAP polarization measurements).
In Section~\ref{sec:bao-as} we treat the CMB as a BAO experiment measuring
$D_M(1090)/r_d$, but we eliminate its calibration of the absolute BAO scale
by artificially blowing up
the errors on $\omega_b$ and $\omega_{cb}$; we denote this case as
``$+$Planck $D_M$''. 
Conversely, in Section~\ref{sec:invdistladder} we use the CMB 
information on $\omega_b$ and $\omega_{cb}$ to set the
size of our standard ruler $r_d$ but omit the $D_M(1090)/r_d$
information by artificially inflating its errors;
we denote this case as ``$+r_d$''.
When we use a full Planck chain instead of the compressed information,
we adopt the notation ``Planck (full)'' and specify what additional
parameters (such as $A_{\rm lens}$, $\neff$, or tensor-to-scalar ratio $r$)
are being varied in the chain.

If one assumes a flat universe, a cosmological constant ($w=-1$),
standard relativistic background ($\neff = 3.046$), and minimal
neutrino mass ($\sum m_\nu = 0.06\eV$), then the CMB data
summarized by equations~(\ref{eqn:cmbvector})-(\ref{eqn:cmbcovar})
also provide a precise constraint on the Hubble parameter $h$, 
and thus on $\Omega_{cb}$, $\Omega_b$, and $\Omega_\Lambda$.  
At various points in the paper we refer to a ``fiducial'' Planck
\lcdm\ model for which we adopt $\omega_b h^2=0.022032$,
$\Omega_m=0.3183$, and $h=0.6704$, which are the best fit parameters for
``Planck+WP'' combination as cited in the Table 2 of \cite{PlanckXVI}.
The CMB constraints on $h$ and $\Omega_m$ become much weaker
if one allows $w \neq -1$ or $\Omega_k \neq 0$, 
so for more general models BAO data or other constraints are
needed to restore high precision on cosmological parameters.

\subsection{Supernova Data}
\label{sec:supernovae}

A comprehensive set of relative luminosity distances of 740 SNIa was
presented in \cite{Betoule14}, based on a joint calibration and
training set of the SDSS-II Supernova Survey \cite{Sako14} and the
Supernova Legacy Survey (SNLS) 3-year data set \cite{Conley11}. The
374 supernovae from SDSS-II and 239 from SNLS were combined with 118
nearby supernovae from
\cite{Contreras10,Hicken09,Jha06,Altavilla04,Riess99,Hamuy96} and nine
high-redshift supernovae discovered and studied by HST \cite{Riess07}.
We use this set, dubbed Joint Light-curve Analysis (JLA), rather than
the Union 2.1 compilation of \cite{Suzuki2012} because of the
demonstrated improvement in calibration and corresponding reduction in
systematic uncertainties presented in \cite{Betoule14}.

While \cite{Betoule14} also provide a full {\tt cosmomc} 
module and a covariance
matrix in relevant parameters, we here instead use their 
compressed representation of
relative distance constraints due to conceptual simplicity and
a drastic increase in computational
speed when combining with other cosmology probes.
The compressed information consists of a piece-wise linear function fit
over 30 bins (leading to 31 nodes) spaced evenly in $\log{z}$ (to 
minimum $z \sim 0.01$) with a $31 \times 31$
covariance matrix that {\it includes all of the systematics} from the
original analysis.
SNIa constrain {\it relative} distances, so the remaining
marginalization required to use this compressed
respresentation in a comological analysis is over the fiducial absolute 
magnitude of a SNIa, $M_B$.
In Section~\ref{sec:invdistladder} we also utilize a similar
compression of the Union 2.1 SN data set, which we have
constructed in analogous fashion.

\subsection{Visualizing the BAO Constraints}
\label{sec:visualization}

\begin{figure*}
  \centering
    \includegraphics[width=\linewidth]{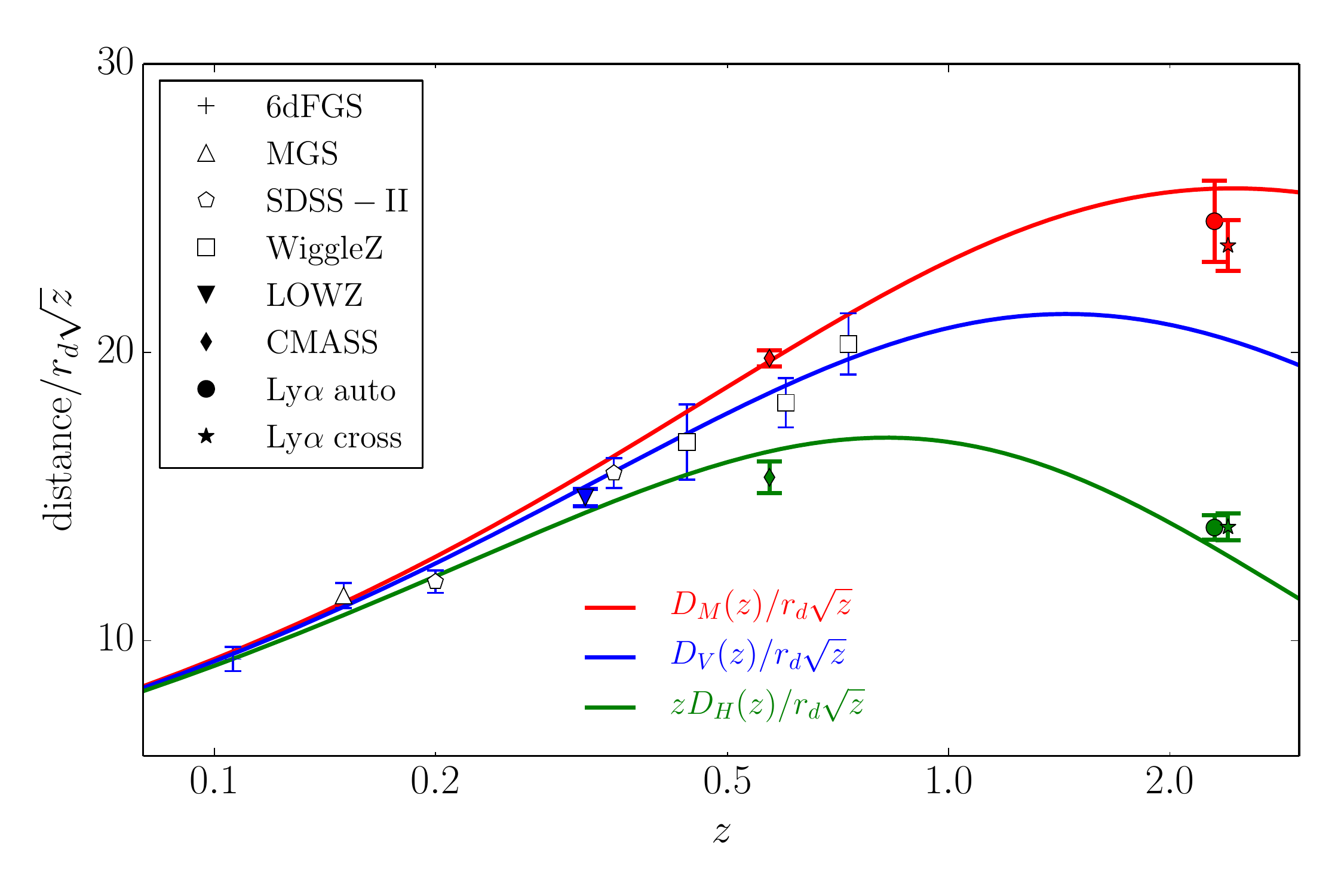}

    \caption{The BAO ``Hubble diagram'' from a world collection of
      detections.  Blue, red, and green points show BAO measurements
      of $D_V/r_d$, $D_M/r_d$, and $z D_H/r_d$, respectively, from the
      sources indicated in the legend.  These can be compared to the
      correspondingly colored lines, which represents predictions of
      the fiducial Planck \lcdm\ model (with $\Omega_m=0.3183$,
      $h=0.6704$, see Section~\ref{sec:cmb}).  The scaling by
      $\sqrt{z}$ is arbitrary, chosen to compress the dynamic range
      sufficiently to make error bars visible on the plot.  Filled
      points represent BOSS data, which yield the most precise BAO
      measurements at $z < 0.7$ and the only measurements at $z>2$.
      For visual clarity, the Ly$\alpha$ cross-correlation points have
      been shifted slightly in redshift; auto-correlation points are
      plotted at the correct effective redshift.}
  \label{fig:hubblediagram}
\end{figure*}

Figure~\ref{fig:hubblediagram} shows the ``Hubble diagram''
(distance vs.\ redshift) from a variety of recent BAO measurements
of $D_V/r_d$, $D_M/r_d$, or $z D_H/r_d$; these three
quantities converge at low redshift.
In addition to the data listed in Table~\ref{tab:data}, we show
measurements from the DR7 data set of SDSS-II by \cite{Percival10}
and from the WiggleZ survey by \cite{Blake11}, which are not
included in our cosmological analysis because they are not
independent of the (more precise) BOSS measurements in 
similar redshift ranges.
Curves represent the predictions of the fiducial Planck \lcdm\ model,
whose parameters are determined independently of the BAO measurements
but depend on the assumptions of a flat universe and a cosmological constant.
Overall, there is impressively good agreement between the
CMB-constrained \lcdm\ model and the BAO measurements, especially
as no parameters have been adjusted in light of the BAO data.
However, there is noticeable tension between the Planck \lcdm\ model
and the \lyaf\ BAO measurements.

\begin{figure}
\begin{center}
\includegraphics[width=1.0\linewidth]{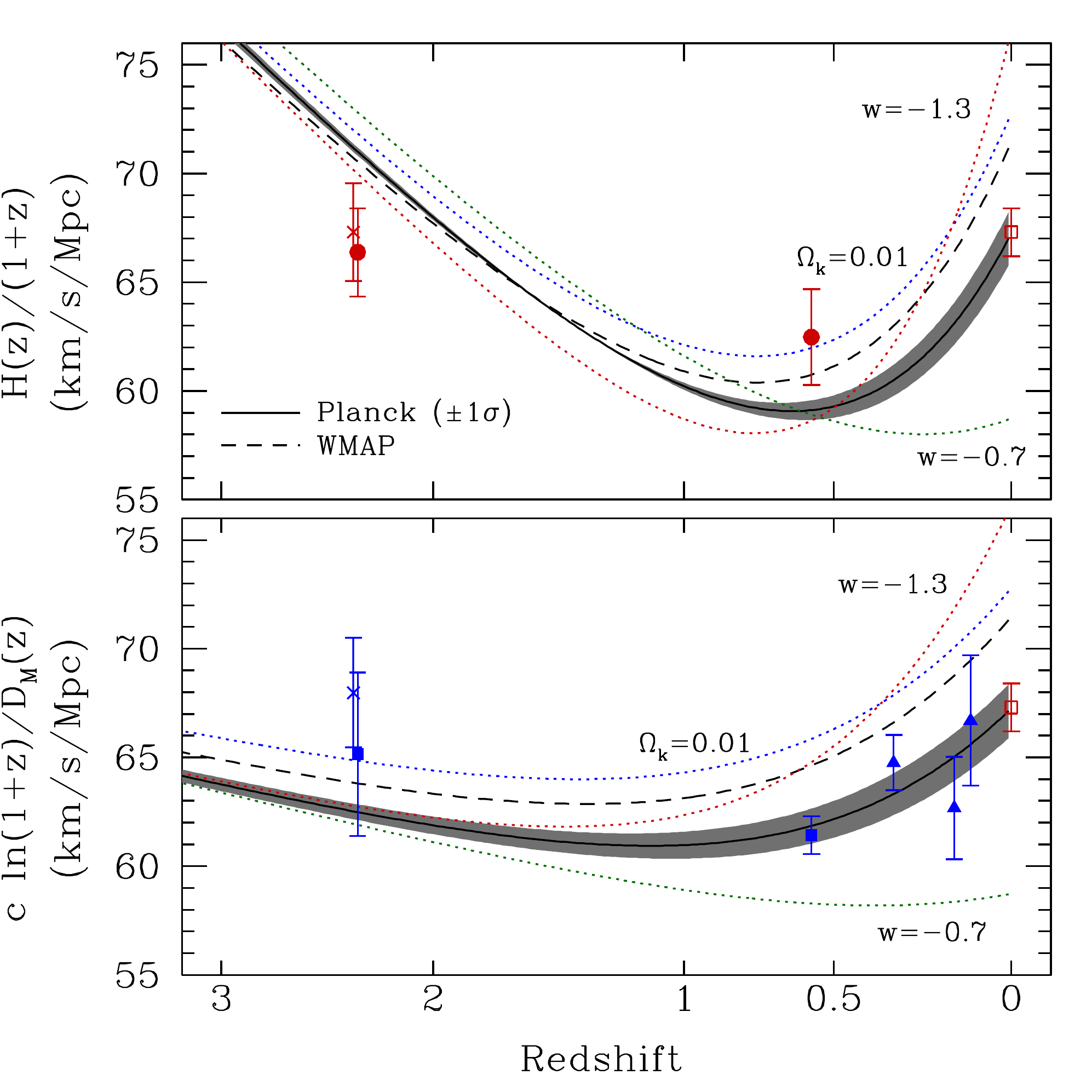}
\end{center}
\caption{\label{fig:Ha}%
BAO measurements and model predictions of $H(z)$ and $D_M(z)$
as a function of redshift, with physically informative scalings.
The top panel shows $H(z)/(1+z)$, the proper velocity between
two objects 1 comoving Mpc apart.
The bottom panel shows $c\ln(1+z)/D_M(z)$, a scaling
that matches a constant line $H(z)=(1+z) H_0$ in the top panel
to the same constant line in the bottom panel for a flat universe.
Filled circles and squares show the BOSS CMASS and \lyaf\ measurements
of $H(z)$ and $D_M(z)$, respectively; we show the \lyaf-quasar
cross-correlation as crosses to distinguish from the \lyaf\ auto-correlation
measurments.
Filled triangles in
the bottom panel show the BOSS LOWZ and MGS measurements of $D_V(z)$
converted to $D_M(z)$.  Open squares show the value of
$H_0 = 67.3\pm 1.1\hubunits$ determined from the combination 
of BAO and SNIa data described in Section~\ref{sec:invdistladder}.
The grey swath in both panels is the prediction from the Planck 
\lcdm\ cosmology including $1\sigma$ parameter errors;
in the top panel, one can easily see the model transition from
deceleration to acceleration at $z \approx 0.6$.  
The dashed line shows the \lcdm\ prediction using the
best-fit WMAP parameters, which has lower $\Omega_m h^2$.
Dotted curves show models that match the best-fit Planck values of
$\omega_{cb}$, $\omega_b$, and $D_M(1090)/r_d$ but
have $\Omega_k=0.01$ (blue), $w=-0.7$ (green), or $w=-1.3$ (red).
The $x$-axis is set to $\sqrt{1+z}$ both for display purposes
and so that a pure matter universe ($\Omega_m=1$)
appears as a decreasing straight line on the top panel.  
}
\end{figure}

Figure~\ref{fig:Ha} displays a subset of these BAO measurements with
scalings that elucidate their physical content.
In the upper panel, we plot $H(z)/(1+z)$, which is
the proper velocity between two objects with a constant comoving
separation of 1 Mpc.
This quantity is declining in a decelerating universe and
increasing in an accelerating universe.
We set the $x$-axis to be $\sqrt{1+z}$, which makes
$H(z)/(1+z)$ a straight line of slope $H_0$ in an
Einstein-de Sitter ($\Omega_m=1$) model.
For the transverse BAO measurements in the lower panel, we plot
$c \ln(1+z)/D_M(z)$,
chosen so that a constant (horizontal) line in the $H(z)/(1+z)$ 
plot would produce
the same constant line in this panel, assuming a flat Universe.
This quantity would decrease monotonically in
a non-accelerating flat cosmology.
The quantities in both the upper and lower panels approach $H_0$
as $z$ approaches zero, independent of other cosmological parameters.
We convert the BOSS LOWZ and MGS measurements of $D_V(z)$ to $D_M(z)$
in the lower panel assuming the fiducial Planck \lcdm\ parameters;
this is a robust approximation because all acceptable cosmologies
produce similar scaling at these low redshifts.
Note that the $H(z)$ and $D_M(z)$ measurements from a given data set
(i.e., at a particular redshift) are covariant, in the sense that
the points on these panels are anti-correlated (see Table~\ref{tab:data}).  
For example,
if $H(z)$ at $z=2.34$ were scattered upward by a
statistical fluctuation, then the $z=2.34$ point in the lower panel 
would be scattered downward.

As discussed below in Section~\ref{sec:invdistladder}, the galaxy BAO and
JLA supernova data can be combined to yield an ``inverse distance ladder''
measurement of $H_0$, which utilizes the CMB measurements of
$\omega_{cb}$ and $\omega_b$ but no other CMB information.
This value of $H_0$ is robust to a wide range of assumptions about
dark energy evolution and space curvature, although it does assume
a standard radiation background for the calculation of $r_d$.
We plot the resulting determination of $H_0 = 67.3 \pm 1.1 \hubunits$
as the open square in both panels.

The grey swath in both panels of Figure~\ref{fig:Ha}
represents the 1$\sigma$ region for the
fiducial Planck \lcdm\ model, with the top panel clearly 
showing the transition from deceleration to acceleration at $z \approx 0.6$.
Formally, we are scaling both panels by $(r_d/r_{d,{\rm fid}})$, so that the
comparison of the BAO data points to the CMB prediction is invariant
to changes in the sound horizon.  
The galaxy BAO measurements of $D_M(z)$ from BOSS and MGS
are in excellent agreement with the predictions of this model
(as are the other measurements shown previously in 
Fig.~\ref{fig:hubblediagram}), 
and the combination of BAO and SNe yields an $H_0$ value in excellent
agreement with this model's prediction.  
The expansion rate
$H(0.57)$ from CMASS is high compared to the model prediction, at
moderate significance.  
Compared to Planck, the best-fit value of $\Omega_m h^2$ from
the 9-year WMAP analysis \cite{WMAP9} is lower, 0.143 vs. 0.137,
implying lower $\Omega_m$ and slightly higher $h$ for a \lcdm\ model.
The model using these best-fit parameters, shown by the dashed lines,
agrees better with the CMASS $H(z)$ measurement but
is in tension with the distance data, especially the CMASS value of
$D_M(0.57)$.

The \lya\ forest measurements are much more
difficult to reconcile with the \lcdm\ model: compared to the
Planck curve, the \lyaf\ BAO $H(z)$ is low and $[D_M(z)]^{-1}$ is high.  
It is important to keep the error anti-correlation
in mind when assessing significance ---
if $H(z)$ fluctuates up then $1/D_M(z)$ will fluctuate down,
which tends to reduce the tension relative to the CMB.
However, our subseqent analyses
(and those already reported by \cite{Delubac14}) will show that
the discrepancy is significant at the $2-2.5\sigma$ level.
The dotted curves show predictions of cosmological models with
$\Omega_k=0.01$ or $1+w = \pm 0.3$.
While changing curvature or the dark energy equation-of-state
can improve agreement with some of the data points,
it worsens agreement with other data points, 
and on the whole (as demonstrated quantitatively in 
Section~\ref{sec:deconstraints}) such variations do not
noticeably improve the fit to the combined CMB, BAO, and SN data.

Not plotted in Figure~\ref{fig:Ha}
is the value of $D_M(1090)$ that comes from the angular
acoustic scale in the CMB.  Connecting the acoustic scale measured in 
CMB anisotropy to that measured in large-scale structure does require model
assumptions about structure formation at the recombination epoch.
However, it would be difficult to move the relative calibration significantly
without making substantial changes to the CMB damping tail,
which is already well constrained by observations.
Using the ratio of $D_M(1090)/r_d$ in equation~(\ref{eqn:cmbmean})
and $r_d = 147.49$ Mpc, we find $c\ln(1+z)/D_M(z) = 151\hubunits$ at $z=1090$
with percent level accuracy,
a factor of two larger than any of the low-redshift values in
Figure~\ref{fig:Ha}.
On their own, the BAO data in Figure~\ref{fig:Ha} clearly
favor a universe that transitions from deceleration at $z>1$
to acceleration at low redshifts, and this evidence becomes
overwhelming if one imagines the corresponding CMB measurements
off the far left of the plot.
We quantify these points in the following section.

It is tempting to consider a flat cosmology with a constant $H/(1+z)$ 
as an alternative model of these data \cite{Melia12}.  
Note that although this form of $H(z)$ occurs in coasting (empty)
cosmologies in general relativity, those models have open curvature
and hence a sharply different $D_M(z)$.
But even for the flat model, the data are not consistent with a 
constant $H(z)/(1+z)$, first because
the increase in $c\ln(1+z)/D_M(z)$ from $z=0.57$ to $z=0.0$ is statistically significant,
and second because of the factor of two change of this quantity relative 
to that inferred from the CMB angular
acoustic scale.  The change from $z=0.57$ to $z=0$ is more significant
than the plot indicates because the data points are correlated;
this occurs because the $H_0$ value results from normalizing the SNe
distances with the BAO measurements.
We measure the ratio of the values, $H_0 D_M(0.57)/c\ln(1.57)$, to 
be $1.080\pm 0.014$ from the combination of BAO and SNe datasets, 
a 5.5$\sigma$ rejection of a constant hypothesis and an indication 
of the strength of the SNe data in detecting the low-redshift 
accelerating expansion.

\section{BAO as an uncalibrated ruler}
\label{sec:bao-as}

\subsection{Convincing Detection of dark energy from BAO data alone}

For quantitative contraints, we start by considering BAO data alone
with the simple assumption that the BAO scale is a standard comoving
ruler, whose length is independent of redshift and orientation but is
{\it not} necessarily the value computed using CMB parameter
constraints. A similar analysis has been presented in
\cite{2013MNRAS.436.1674A}.  In this case, a simple dimensional
analysis shows that in addition to fractional densities in cosmic
components, one can constrain the dimensionless quantity $P=c / (H_0
r_d)$.

\begin{figure}
  \centering
    \includegraphics[width=\linewidth]{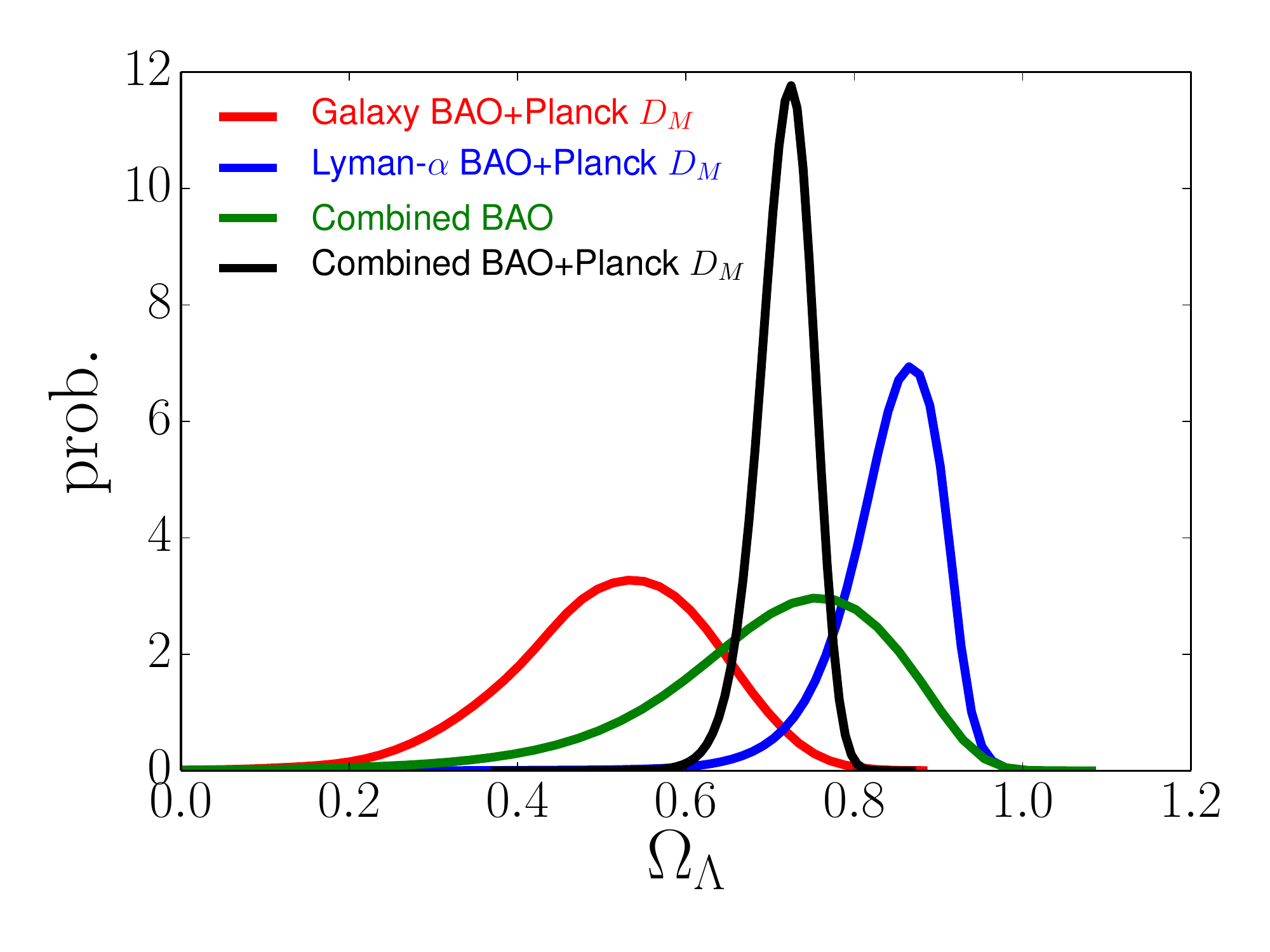} 

    \includegraphics[width=\linewidth]{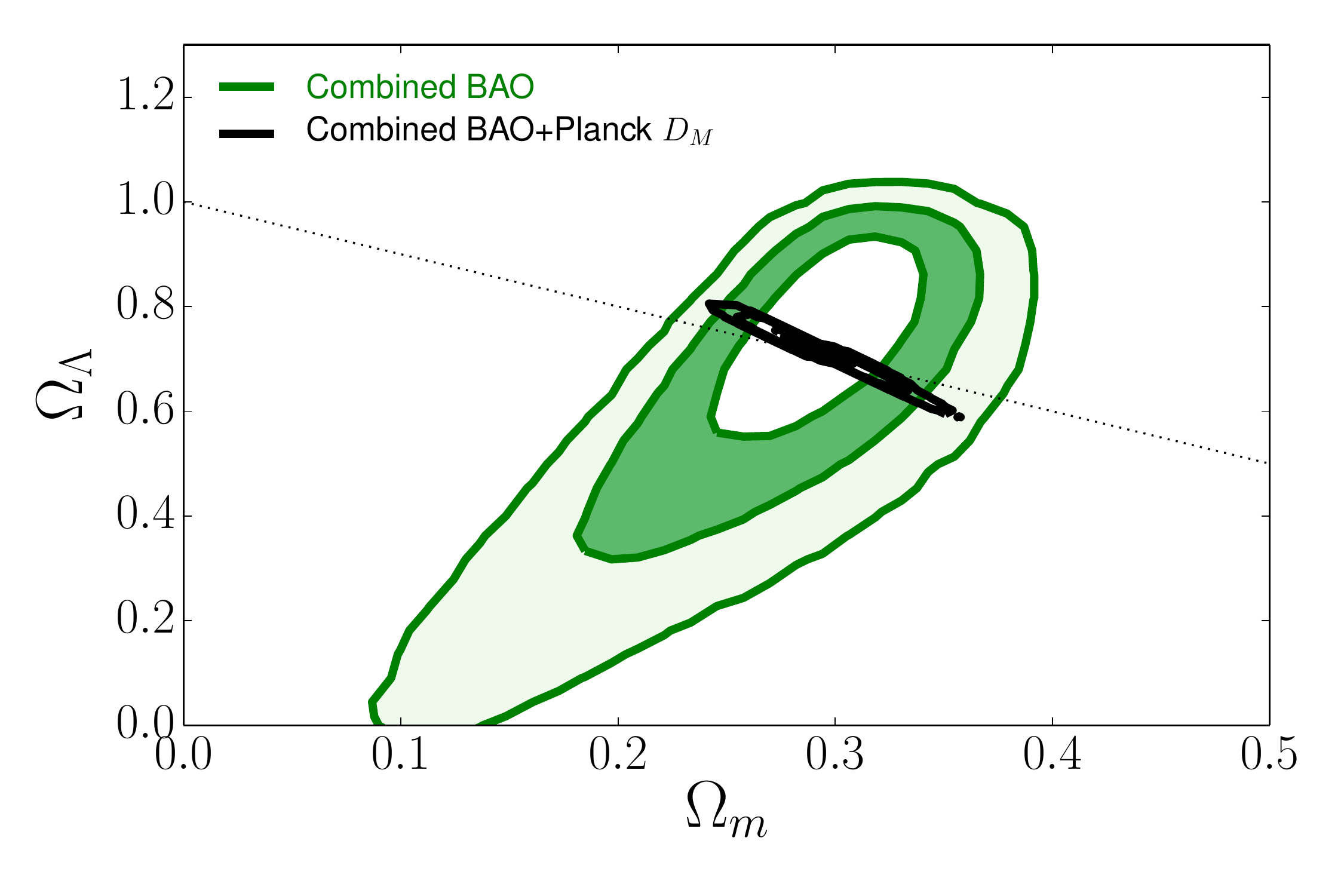} 

    \includegraphics[width=\linewidth]{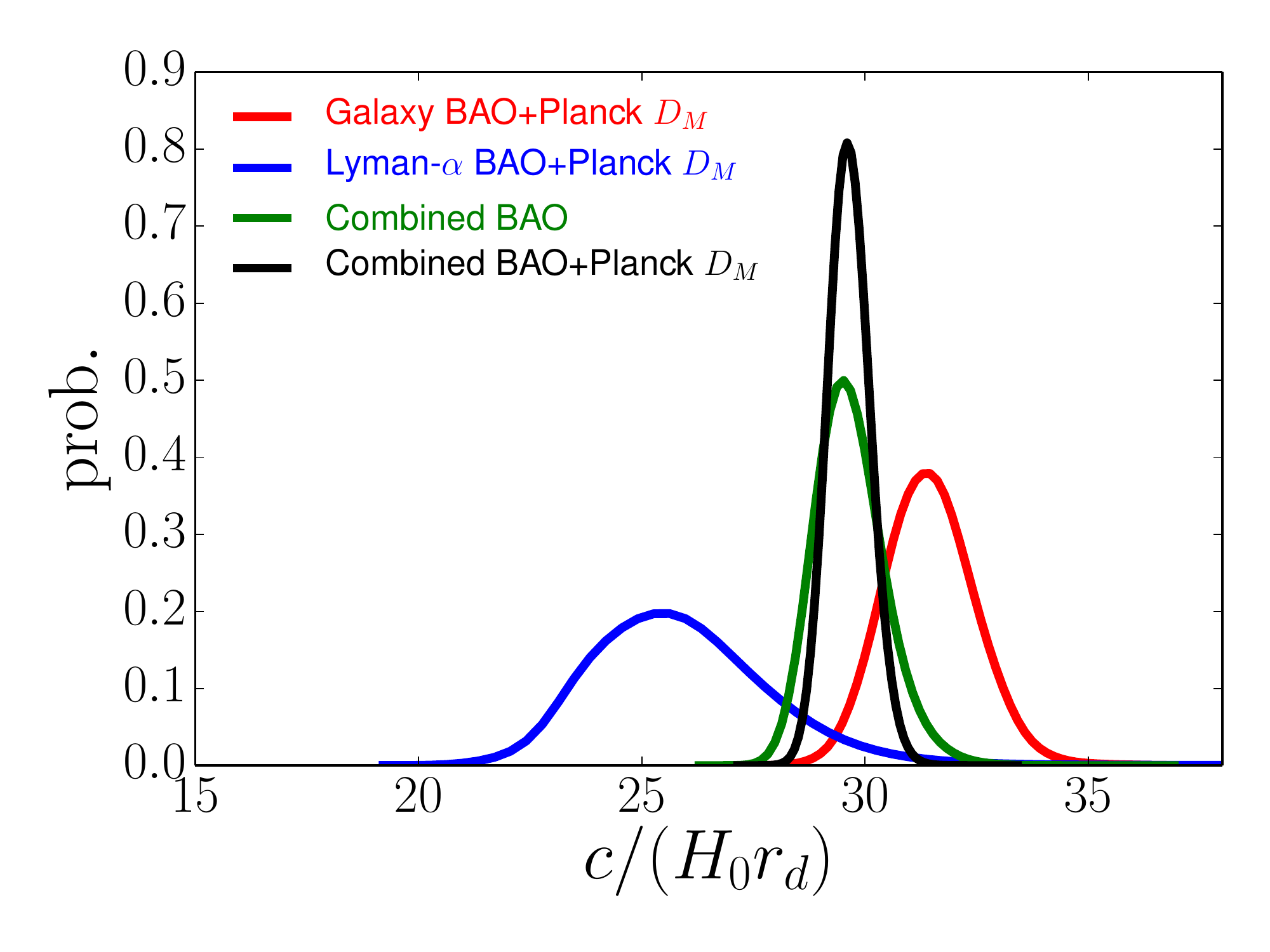} \\

    \caption{Constraints from BAO on the parameters of \olcdm\ models,
      treating the BAO scale as a redshift-independent standard
      ruler of unknown length.  Green curves/contours in each
      panel show the combined constraints from galaxy and \lyaf\ BAO,
      with no CMB information.  Black curves/contours include
      the measurement of $D_M(1090)/r_d$ from the CMB acoustic
      scale, again with no assumption about the value of $r_d$
      except that it is the same scale as the lower redshift measurements.
      This combination of BAO measurements yields precise 
      constraints on $\Omega_\Lambda$ (top panel) and the dimensionless
      quantity $c/(H_0 r_d)$ (bottom panel), and it requires a low 
      density ($\Omega_m \approx 0.29$), nearly flat universe
      (middle panel).  Blue and red curves in the top and
      bottom panels show the result of combining the CMB BAO 
      measurement with either the galaxy or \lyaf\ BAO
      measurement separately.  The dotted line in the middle panel
      marks $\Omega_m+\Omega_\Lambda=1$.
      }
  \label{fig:bat1}
\end{figure}

Figure \ref{fig:bat1} presents constraints on relevant quantities
in \olcdm\ models, which assume that dark energy is a 
cosmological constant but allow $\Omega_\Lambda=0$ and arbitrary $\Omega_k$.
The combination of galaxy and \lyaf\ BAO measurements
yields a marginalised constraint of 
$\Omega_\Lambda= 0.73^{+0.25}_{-0.68}$ at 99.7\% confidence,
implying a $>3\sigma$ detection of dark energy
from BAO alone {\it without} CMB data.

These constraints become much tighter if we assume that the CMB is measuring the
{\it same} acoustic scale, functioning as an additional BAO experiment at a much
higher redshift.  As discussed in Section~\ref{sec:cmb}, we implement this case
by retaining the high-precision CMB measurement of $D_M(1090)/r_d$ but
drastically inflating (by a factor of 100) the CMB errors on $\omega_{cb}$ and
$\omega_b$, so that the value of $r_d$ itself remains effectively unknown.  
Combining the CMB measurement with galaxy or \lyaf\ BAO alone
yields a strong detection of non-zero $\Omega_\Lambda$, but with different
central values reflecting the tensions already discussed in
Section~\ref{sec:visualization} and examined further in
Sections~\ref{sec:deconstraints}--\ref{sec:alternatives}.
Combining all three measurements yields a marginalised
$\Omega_\Lambda=0.72^{+0.030}_{-0.034}$ (at 68\% confidence, with reasonably
Gaussian errors), implying $>20\sigma$ preference for a low-density universe
dominated by dark energy.  
The dimensionless quantity $P=c/(H_0 r_d) =
29.63^{+0.48}_{-0.45}$ is determined with 1.6\% precision.  Most importantly,
this data combination also requires a nearly flat universe,
with a total density $\Omega_m+\Omega_\Lambda = 1.011^{+0.014}_{-0.016}$ 
determined to 1.5\% and consistent with the critical density.
Thus, with the minimal assumption that the BAO scale is a standard ruler,
these data provide strong support for the standard cosmological
model.

\subsection{External calibration of $r_d$}
\label{sec:rd}

\begin{figure}
  \centering
  \includegraphics[width=\linewidth]{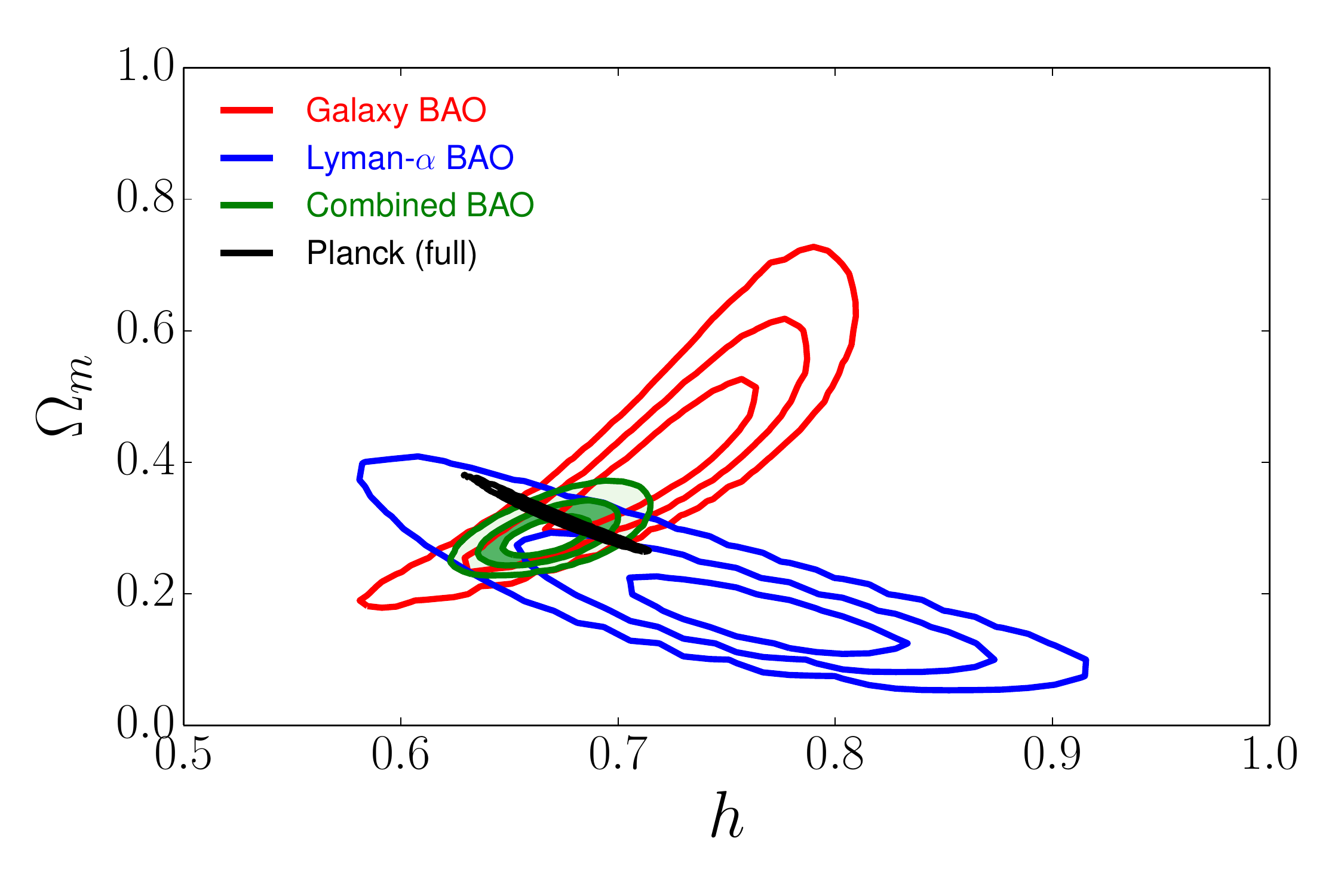}
  \caption{Constraints on $\Omega_m$ and $h$ in a flat \lcdm\ model
  from galaxy BAO (red), \lyaf\ BAO (blue), and the combination
  of the two (green), using a BBN prior on $\omega_b$ and standard
  physics to compute the sound horizon $r_d$ but incorporating
  no CMB information.  Contours are plotted at 68\%, 95\%,
  and 99.7\% confidence (the {\it interior} white region of
  the green ``donut'' is 68\%).
  Black contours show the entirely independent constraints on
  $\Omega_m$ and $h$ in \lcdm\ from full Planck CMB chains.
  }
  \label{fig:lcdm}
\end{figure}

We proceed further by computing the sound horizon scale $r_d$ from the
standard physics of the pre-recombination universe but adopting
empirical constraints external to the CMB.  In particular, we adopt a
prior on the baryon density of $\omega_b=0.02202\pm0.00046$ determined
from big bang nucleosynthesis (BBN) and the observed primordial 
deuterium abundance \cite{Cooke14}, and we assume a standard relativistic
background ($\neff = 3.046$, $\omega_\nu \approx 0$).  For any values
of $\Omega_m$ and the Hubble parameter $h$ that arise in our MCMC
chain, we can then compute the value of $r_d$ from
equation~(\ref{eqn:rd}).  Compared to the previous section, the
addition of the physical scale allows us to convert the measured value
of $c/(H_0 r_d)$ into a measurement of the dimensional parameter
$H_0$.  In practice, we derive constraints in a separate MCMC run
where, instead of a flat prior on $P$, we have a flat prior on $h$ and
the above prior on $\omega_b$. We also fix the curvature parameter 
$\Omega_k$ to zero.
Results are presented in Figure~\ref{fig:lcdm}.  The red
(galaxy BAO) and blue (\lyaf\ BAO) contours in this figure use no CMB
information at all, but they {\it do} assume a spatially flat universe
in contrast to Figure~\ref{fig:bat1}.

The point of this exercise is the following. The homogeneous part of the minimal
\lcdm\ model has just two adjustable parameters, $\Omega_m$ and $h$, which
matches the two degrees of freedom offered by a measurement of anisotropic BAO
at a single redshift.
(The weak BBN prior is required to fix the magnitude of $r_d$, but it does
not affect the expansion history.) 
One can therefore get meaningful constraints from
either galaxy BAO or \lyaf\ BAO alone, though this is no longer
true if one allows non-zero curvature and therefore introduces a 
third parameter.
There is substantial $\Omega_m-h$
degeneracy for either measurement individually, but both are generally
compatible with standard values of these parameters.  The tension of the \lyaf\
BAO with the Planck \lcdm\ model manifests itself here as a best fit at
relatively low matter density and high Hubble parameter.  Combining the galaxy
and \lyaf\ measurements produces a precise measurement of both
$\Omega_m$ and the Hubble parameter {\it coming from BAO alone}, 
independent of CMB data.  In combination, 
we find $h=0.67\pm0.013$ and $\Omega_m=0.29\pm0.02$ (68\% confidence).
The small black ellipse in Figure~\ref{fig:lcdm} shows the Planck constraints
for \lcdm, computed from full Planck chains, which are in excellent agreement
with the region allowed by the joint BAO measurements.

\section{BAO, SNIa, and the Inverse Distance Ladder}
\label{sec:invdistladder}

The traditional route to measuring the Hubble constant $H_0$
is built on a distance ladder anchored in the nearby Universe:
stellar distances to galaxies
within $\sim 20\,$Mpc are used to calibrate secondary indicators,
and these in turn are used to measure distances to galaxies
``in the Hubble flow,'' i.e., far enough away that peculiar
velocities are a sub-dominant source of uncertainty when
inferring $H_0=v/d$ \cite{Freedman10}.
The most powerful implementations of this program in recent
years have used Cepheid variables --- calibrated by direct parallax,
by distance estimates to the LMC, or by the maser distance to
NGC 4258 --- to determine distances to host galaxies of
SNIa, which are the most precise of the
available secondary distance indicators 
\cite{Riess11,Freedman12,Humphreys13}.

Because the BAO scale can be computed in absolute units from basic
underlying physics, the combination of BAO with SNIa allows a
measurement of $H_0$ via an ``inverse distance ladder,'' anchored at
intermediate redshift.  The BOSS BAO data provide absolute values of
$D_V$ at $z=0.32$ and $D_M$ at $z=0.57$ with precision of 2.0\% and
1.4\%, respectively.  The JLA SNIa sample provides a high-precision
relative distance scale, which transfers the BAO measurement down to
low redshift, where $H_0$ is simply the slope of the distance-redshift
relation.  Equivalently, this procedure calibrates the absolute
magnitude scale of SNIa using BAO distances instead of the 
Cepheid distance scale.
Although the extrapolation from the BAO redshifts to
low redshifts depends on the dark energy model, the SNIa relative
distance scale is precisely measured over a well sampled redshift
interval which includes the BAO redshifts, so this extrapolation
introduces practically no uncertainty even when the dark energy model
is extremely flexible.  CMB data enter the inverse distance ladder by
constraining the values of $\omega_m$ and $\omega_b$ and thus allowing
computation of the sound horizon scale $r_d$.

\begin{figure}
  \centering
  \includegraphics[width=\linewidth]{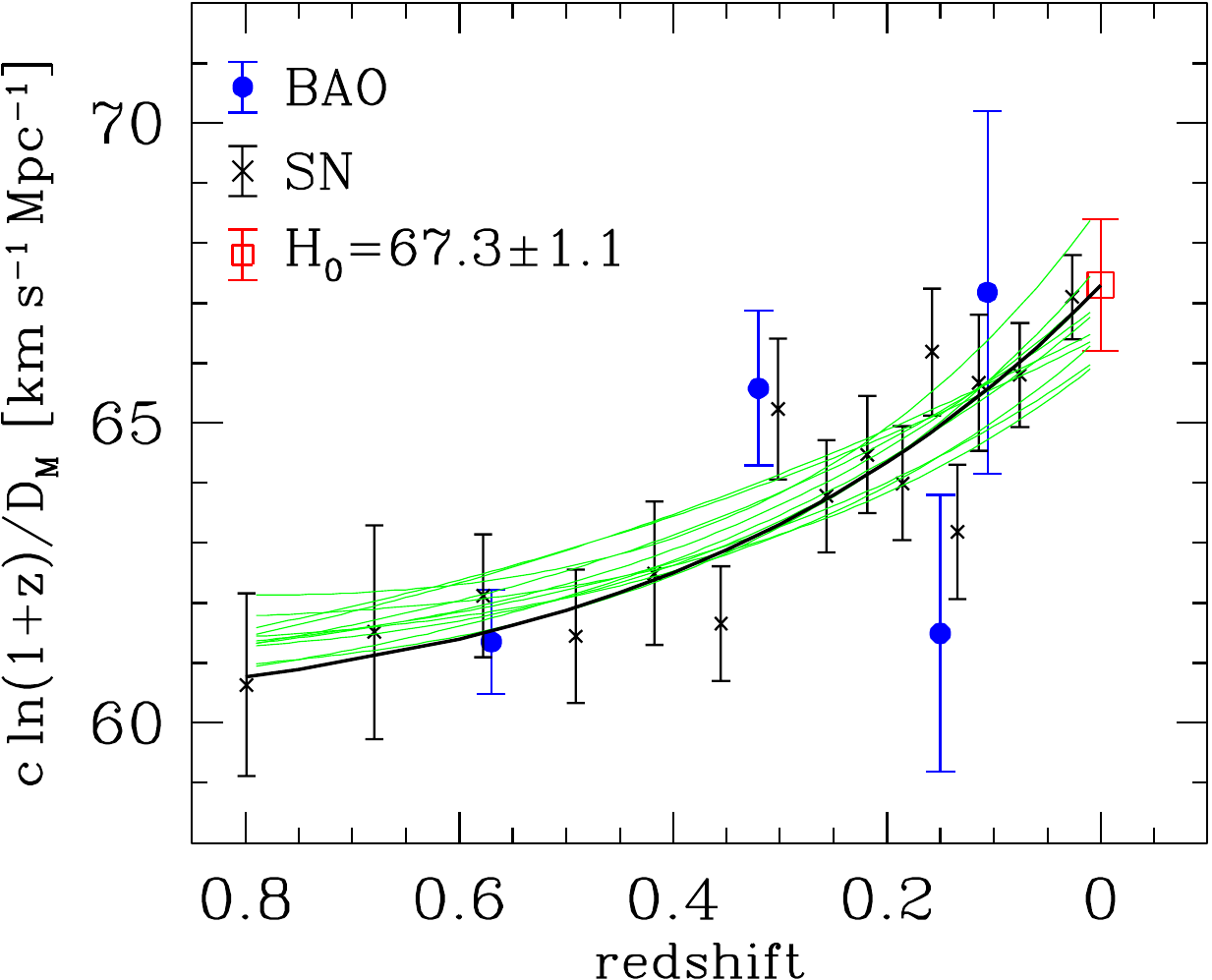}
  \caption{ Determination of $H_0$ by the ``inverse distance ladder''
    combining BAO absolute distance measurements and SNIa relative
    distance measurements, with CMB data used to calibrate the sound
    horizon scale $r_d$.  The quantity $c\ln(1+z)/D_M(z)$
    converges to $H_0$ at $z=0$.  Filled circles show the four BAO
    measurements, normalized with $r_d = 147.49\,$Mpc; for the three
    lower redshift points, $D_V$ has been converted to $D_M$ assuming
    \lcdm.  Crosses show the SNIa measurements, with error bars
    representing diagonal elements of the covariance matrix.  Because
    the absolute luminosity of SNIa is not known {\it a priori}, the
    SNIa points are free to shift vertically by a constant factor,
    which is chosen here to produce the best joint fit with the BAO data.
    The red square and error bar shows the value $H_0 = (67.3\pm
    1.1)\hubunits$ determined by the full inverse distance ladder
    procedure described in the text.  The black curve shows the
    prediction for a \lcdm\ model with $\Omega_m=0.3$ and the best-fit
    $H_0$, and green curves show ten PolyCDM models randomly selected
    from our MCMC chain that have $\Delta\chi^2 < 4$ relative to the
    best-fit PolyCDM model.  This $H_0$ determination assumes standard
    pre-recombination physics to evaluate $r_d$.  For non-standard
    energy backgrounds (e.g., extra relativistic species or early
    dark energy) the more general result is described by
    equation~(\ref{eqn:h0model}).
    }
  \label{fig:invdistladder}
\end{figure}

Figure~\ref{fig:invdistladder} provides a conceptual illustration
of this approach, zeroing in on the $z < 1$ portion of the 
Hubble diagram.  Filled points show $c\ln(1+z)/D_M(z)$ from the
CMASS, LOWZ, MGS, and 6dFGS BAO measurements, where for illustrative
purposes only we have converted the latter three measurements
from $D_V(z)$ to $D_M(z)$ using Planck \lcdm\ parameters.
The error bars on these points include the 0.4\% uncertainty
in $r_d$ arising from the uncertainties in the Planck determination
of $\omega_m$ and $\omega_b$, but this is a small 
contribution to the error budget.  Crosses show the binned
SNIa distance measurements, with the best absolute magnitude
calibration from the joint BAO+SNIa fit.  We caution that 
systematic effects introduce error correlations across redshift
bins in the SNIa data, which are accounted for in our full analysis.
To allow flexibility in the dark energy model, we adopt the
PolyCDM parameterization described in Section~\ref{sec:methodology},
imposing a loose Gaussian prior $\Omega_k = 0 \pm 0.1$ to suppress
high curvature models that are clearly inconsistent with the CMB.
Thin green curves in 
Figure~\ref{fig:invdistladder} show $c\ln(1+z)/D_M(z)$ for
ten PolyCDM models that have $\Delta \chi^2 < 4$ relative 
to the best-fit model, 
selected from the MCMC chains described below.
The intercept of these curves at $z=0$ is the value of $H_0$.
While low-redshift BAO measurements like those of 6dFGS and MGS incur
minimal uncertainty from the extrapolation to $z=0$,
the statistical error is necessarily large because of the limited
volume at low $z$.  It is evident from Figure~\ref{fig:invdistladder}
that using SNIa to transfer intermediate redshift BAO measurements
to the local Universe yields a much more precise determination of $H_0$ 
than using only low-redshift BAO measurements, even allowing
for great flexibility in the dark energy model.  

To compute our $H_0$ constraints, we adopt the $D_M(z)$ and $H(z)$
constraints from CMASS BAO (including covariance), the $D_V(z)$
constraints from LOWZ, MGS, and 6dFGS BAO, the compressed JLA SNIa
data set with its full $31\times 31$ covariance matrix, and an $r_d$
constraint from Planck (see Section \ref{sec:cmb}).
Marginalizing over the PolyCDM parameters yields
$H_0 = 67.3 \pm 1.1\hubunits$, a 1.7\% measurement.
Even if we include the CMB angular diameter distance at its full precision,
our central value and error bar on $H_0$ change negligibly
because the flexibility of the PolyCDM model effectively
decouples low- and high-redshift information.

As a by-product of our $H_0$ measurement, we determine the
absolute luminosity of a fiducial SNIa
to be $M_B=-19.14\pm0.042$~mag.
Here 
we define a fiducial SNIa as having SALT2 (as retrained in \cite{Betoule14}) 
light-curve width and color parameters $x_1=0$ and $C=0$ and having exploded in 
a galaxy with a stellar mass $<10^{10} M_\odot.$

Our best-fit $H_0$ and its $1\sigma$ uncertainty are shown by the open
square and error bar in Figures~\ref{fig:Ha}
and~\ref{fig:invdistladder}.  To characterize the sources of error, we
have repeated our analyses after multiplying either the CMB, SN, or BAO
covariance matrix by a factor of ten (and thus reducing errors by
$\sqrt{10}$).  Reducing the CMB errors, so that they yield an
essentially perfect determination of $r_d$, makes almost no difference
to our $H_0$ error, because the 0.4\% uncertainty in $r_d$ is already
small.  Reducing either the SNIa or BAO errors shrinks the $H_0$
error by approximately a factor of two, indicating that the BAO
measurement uncertainties and the SNIa measurement uncertainties make
comparable contributions to our error budget; the errors add (roughly)
linearly rather than in quadrature because both measurements constrain
the redshift evolution in our joint fit.  If we replace PolyCDM with
\owwacdm\ in our analysis, substituting a different but still highly
flexible dark energy model, the derived value of $H_0$ drops by less
than $0.2\sigma$ and the error bar is essentially unchanged.  If we
instead fix the dark energy model to \lcdm, the central value and
error bar are again nearly unchanged, because with the dense sampling
provided by SNe the extrapolation from the BAO redshifts down to $z=0$
is also only a small source of uncertainty.  To test sensitivity to
the SN data set, we constructed a compressed description of the Union
2.1 compilation \cite{Suzuki2012} analogous to that of the JLA
compilation; substituting Union 2.1 for JLA makes negligible
difference to our best-fit $H_0$ while increasing the error bar by
about 30\% (see Table~\ref{tab:danger}).  Finally, if we substitute
the WMAP9 constraints on $\omega_m$ and $\omega_b$ for the Planck
constraints, the central $H_0$ decreases by 0.5\% (to $66.9\hubunits$)
and the error bar grows by 8\% (to $1.2\hubunits$).

To summarize, this 1.7\% determination of $H_0$ is robust to details of our
analysis, with the error dominated by the BAO and SNIa measurement
uncertainties.  The key assumptions behind this method are (a)
standard matter and radiation content, with three species of light
neutrinos, and (b) no unrecognized systematics at the level of our
statistical errors in the CMB determinations of $\omega_m$ and
$\omega_b$, in the BAO measurements, or in the SNIa measurements used
to tie them to $z=0$.  Note that the SNIa covariance matrix already
incorporates the detailed systematic error budget of \cite{Betoule14}.
The measurement systematics are arguably smaller than those that
affect the traditional distance ladder.  Thus, with the caveat that it
assumes a standard matter and radiation content, this measurement of
$H_0$ is more precise and probably more robust than current distance-ladder
measurements. 

Non-standard radiation backgrounds remain a topic
of intense cosmological investigation, and a convincing mismatch
between $H_0$ determinations from the forward and inverse distance
ladders could be a distinctive signature of non-standard
physics that alters $r_d$.  We can express our constraint in
a more model-independent form as
\begin{equation}
\label{eqn:h0model}
H_0 = (67.3 \pm 1.1) \times (147.49\,{\rm Mpc} / r_d) \hubunits.
\end{equation}
Raising $\neff$ from 3.046 to 4.0
would increase our central value of $H_0$ to $69.5\hubunits$
(eq.~\ref{eqn:rdneff}, but see further discussion in Section~\ref{sec:neff}).

\begin{figure}
  \centering
  \includegraphics[width=\linewidth]{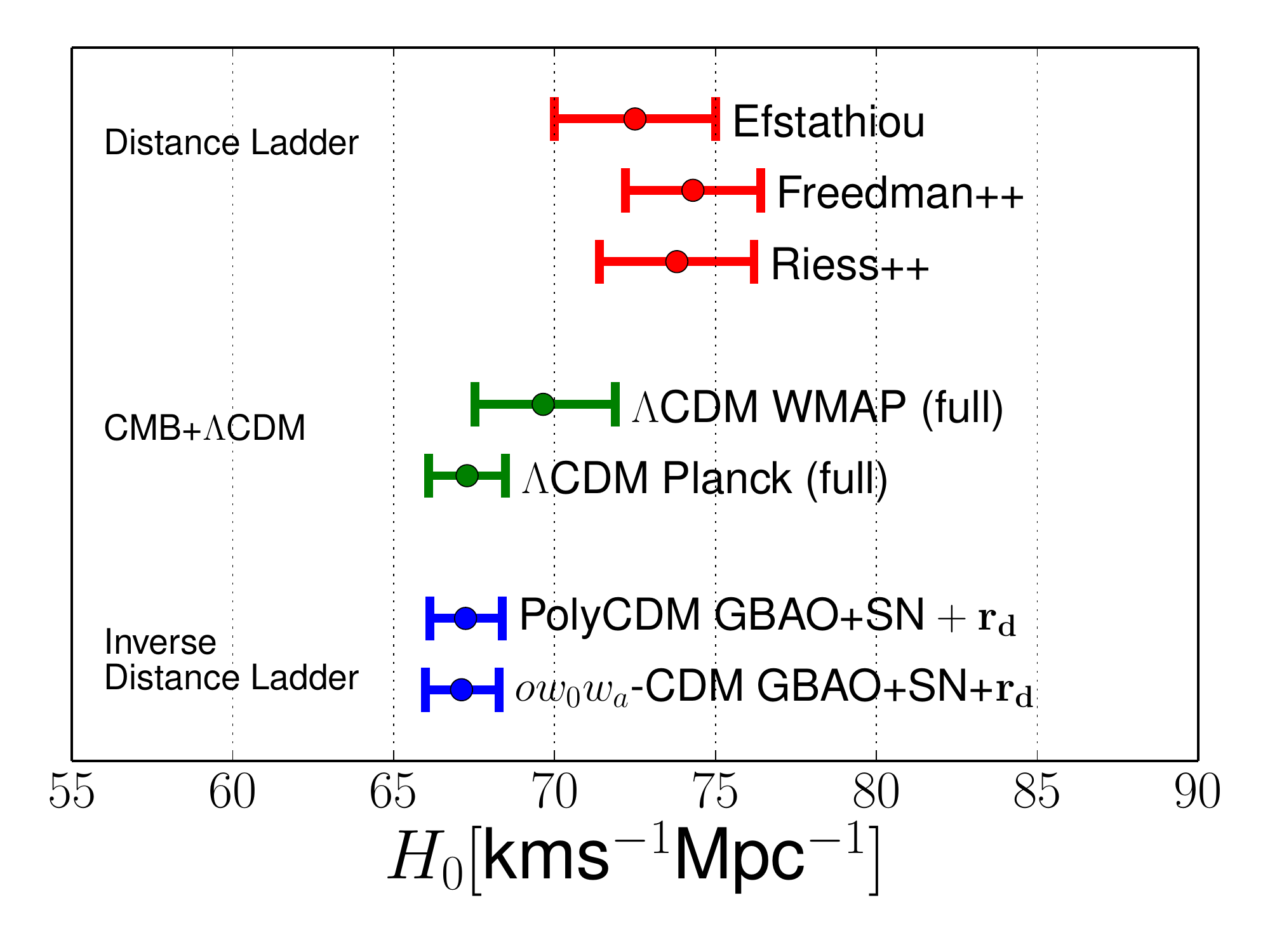}
  \caption{Constraints on the Hubble constant $H_0$ from
  this paper's inverse distance ladder analysis (blue, at bottom),
  from three direct distance ladder estimates (red, at top),
  and from Planck or WMAP CMB data assuming \lcdm\ 
  (green, middle).  All error bars are $1\sigma$.
  The inverse distance ladder estimates assume
  $r_d = 147.49 \pm 0.59$ Mpc, based on Planck constraints
  for a standard radiation background, while the green points
  make the much stronger assumptions of a flat universe with
  a cosmological constant.
   }
  \label{fig:ladder}
\end{figure}

Figure~\ref{fig:ladder} compares our $H_0$ determination to several
other values from the literature.  The lower two points show our
results using either the PolyCDM model or the \owwacdm\ model.
The top three points show recent distance-ladder determinations
from Riess et al.\ \cite{Riess11}, Freedman et al.\ \cite{Freedman12},
and a reanalysis of the Riess et al.\ data set by \cite{Efstathiou14}.
There is mild ($\approx 2\sigma$) tension between these determinations
and our value.
The central two points show the values of $H_0$ inferred
from Planck or WMAP CMB data {\it assuming a flat} \lcdm\ {\it model},
with values and uncertainties taken from the MCMC chains
provided by the Planck collaboration.
These inferences of $H_0$ are much more model dependent
than our inverse distance ladder measurement; with the 
\owwacdm\ or PolyCDM dark energy models the errors on $H_0$ from CMB
data alone increase by more than order of magnitude because of 
the CMB geometric degeneracy. Consistency of these $H_0$ values is therefore a
consistency test for the \lcdm\ model, which it passes here with
flying colors.

\begin{table}
  \centering
  \begin{tabular}{l|c|c}
    Combination & Model & $H_0$ \\
    \hline
 Galaxy BAO + SN + $r_d$ & PolyCDM & $ 67.3\pm 1.1$ \\
 Galaxy BAO + SN + $r_d$ &\owwacdm & $ 67.1\pm 1.1$ \\
 \smallskip
 Galaxy BAO + Union SN + $r_d$ & PolyCDM & $   67.3\pm 1.5$ \\
 Galaxy BAO + Union SN + $r_d$ & \owwacdm & $  67.2\pm 1.5$ \\
  \end{tabular}

  \caption{Constraints on $H_0$ (in $\hubunits$) from the
  inverse distance ladder, assuming $r_d =147.49 \pm 0.59$ Mpc
  as inferred from Planck with a standard radiation background.
  The bottom two lines substitute the Union 2.1 SN data set
  for the JLA data set.  Errorbars are $1\sigma$.}
  \label{tab:danger}
\end{table}

Our results can be compared to those of several other recent
analyses.  \cite{Cheng14} determine $H_0$ from a collection of
BAO data sets using the Planck-calibrated value of $r_d$.
They do not incorporate SNIa, but they assume a flat \lcdm\ model,
which allows them to obtain a tight constraint
$H_0 = 68.11 \pm 0.86\hubunits$.
\cite{Cuesta14} carry out a more directly comparable
inverse distance ladder measurement with essentially the
same data sets but cosmological models that are 1-parameter
extensions of \lcdm, finding $H_0 =68.0 \pm 1.2\hubunits$
for either \olcdm\ or \wcdm.
\cite{2014arXiv1409.6217H} carry out a rather different analysis that
uses age measurements for early-type galaxies to provide an
absolute timescale.  In combination with BAO and SNIa, they then
constrain the acoustic oscillation scale $r_d=101.9\pm 1.9\hmpc$
independent of CMB data or early universe physics.
Their result, which assumes an \olcdm\ cosmology, 
can be cast in a form similar to ours,
$H_0 = (69.9\pm1.3)\times (147.49\,{\rm Mpc} / r_d) \hubunits$;
the agreement implies that their stellar evolution 
age scale is consistent with the scale implied by early-universe
BAO physics.  As an $H_0$ determination, our analysis makes 
much more general assumptions about dark energy than these
other analyses, but it yields a consistent result.
It is also notable that our value of $H_0$ agrees with
the value of $67 \pm 2 \hubunits$ inferred from a 
median-statistics analysis of {\it direct} distance
ladder estimates {\it circa} 2001 (\cite{Gott01}, see
\cite{Chen11} for a 2011 update).

From Figure~\ref{fig:invdistladder}, it is visually evident
that the relative distance scales implied by our BAO and
SN are in fairly good agreement.  We have converted 
SN luminosity distances to comoving angular diameter distances
with $D_M(z) = D_L(z)/(1+z)$, a relation that holds in any
metric theory of gravity (see section 4.2 of \cite{Lampeitl10}
and references therein).
As a quantitative consistency test,
we refit the PolyCDM model with an additional free parameter
that artificially modifies the luminosity distance by
$D_L(z)\rightarrow D_L (1+z)^\beta$, finding $\beta=0.13\pm0.063$.
This result is consistent with the expected $\beta=0$ at
$2\sigma$, but there is a mild tension because the SN data
are in good agreement with \lcdm\ predictions while the
ratio of $D_M(z)$ between the CMASS and LOWZ samples is
somewhat higher than expected in \lcdm.

\section{Constraints on Dark Energy Models}
\label{sec:deconstraints}

We now turn to constraints on dark energy and space curvature
from the combination of BAO, CMB, and SNIa data.
In this section, we consider models with standard matter
and radiation content, including three neutrino species
with the minimal allowed mass $\sum m_\nu = 0.06\eV$
(although the cosmological differences between $0.06\eV$ and $0\eV$
are negligible relative to current measurement errors).
In Section~\ref{sec:alternatives}, we will consider models that
allow dynamically significant neutrino mass, extra relativistic
species, dark matter
that decays into radiation, or ``early'' dark energy
that is dynamically non-negligible even at high redshifts.

\begin{figure*}
  \centering
    \includegraphics[width=\linewidth]{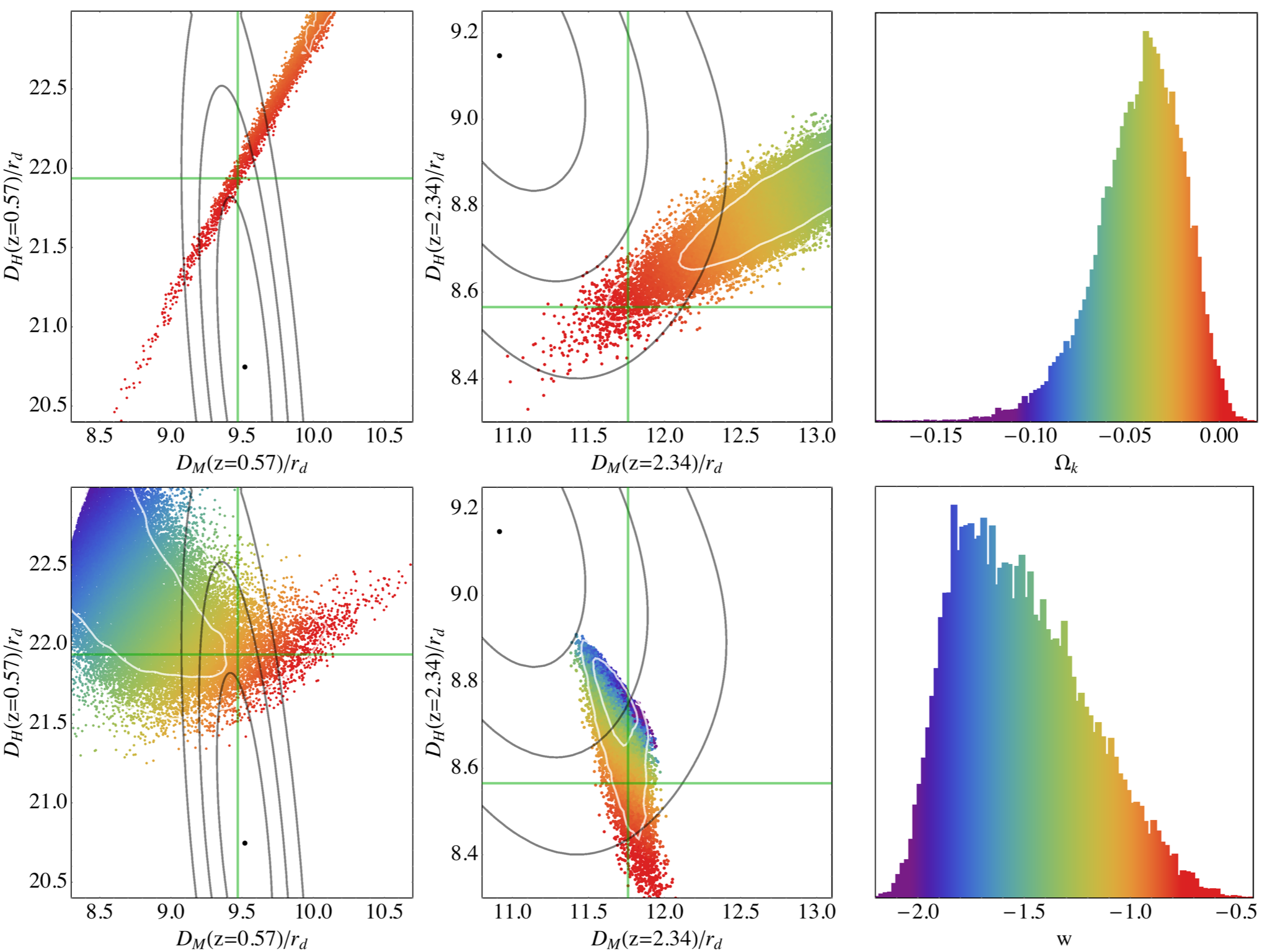}

    \caption{BAO constraints in the $D_M-D_H$ planes at $z=0.57$
	  (left) and $z=2.34$ (middle) compared to predictions of
	  \olcdm\ (top row) or \wcdm\ (bottom row) models constrained
	  by CMB data.  Black curves show 68\%, 95\%, and 99.7\% likelihood
	  contours from the CMASS and \lyaf\ BAO measurements,
	  relative to the best-fit values (black dots).
	  Colored points represent individual models from
	  Planck+WP+ACT/SPT MCMC chains, which are color-coded
	  by the value of $\Omega_k$ (top row) or $w$ (bottom row)
	  as illustrated in the right panels.
	  Green cross-hairs mark the predictions of the flat
	  \lcdm\ model that best fits the CMB data.
	  White curves show 68\% and 95\% likelihood contours
	  for the CMB data alone.
      }
  \label{fig:chain_vs_contour1}
\end{figure*}

To set the scene, Figure~\ref{fig:chain_vs_contour1} compares
the predictions of models constrained by CMB data
to the BOSS BAO constraints on $D_M$ and $D_H$ at $z=0.57$
and $z=2.34$, from CMASS galaxies and the \lyaf, respectively.
Black dots mark best-fit values of $(D_M,D_H)$, and contours are
shown at $\Delta \chi^2 = 2.30,$ 6.18, and 11.83 (coverage
fractions of 68\%, 95\%, and 99.7\% for a 2-d Gaussian).
The top row shows results for \olcdm\ models, which assume
a constant dark energy density but allow non-zero space curvature.
Here we have taken models from the Planck Collaboration MCMC
chains, based on the combination of Planck, WMAP polarization,
and ACT/SPT data.  The upper right panel shows the one-dimensional
PDF for the curvature parameter $\Omega_k$ based on the CMB
data alone.  Each point in the left and middle panels
represents a model from the chains, color-coded by the value
of $\Omega_k$ on the scale in the right panel.
The green cross-hairs mark the predicted $(D_M,D_H)$ from
the flat \lcdm\ model that best fits the CMB data alone.
This model lies just outside the 68\% contour 
for CMASS, but it is discrepant at $>95\%$ with the
\lyaf\ measurements, as remarked already by \cite{Delubac14}.
When the flatness assumption is dropped, both
the galaxy and \lyaf\ 
BAO data strongly prefer $\Omega_k$ close to zero, firmly
ruling out the slightly closed ($\Omega_k \sim -0.05$) 
models that are allowed by the CMB alone.  

The bottom row shows results for \wcdm\ models, which assume
a flat universe but allow a constant equation-of-state parameter
$w \equiv p/\rho \neq -1$ for dark energy.
The CMB data alone are consistent with a wide range of $w$ values,
and they are generally better fit with $w < -1$.  
However, the combination with CMASS BAO data sharply
limits the acceptable range of $w$, favoring values 
close to $-1.0$ (a cosmological constant).
The fit to the \lyaf\ BAO results could be significantly
improved by going to $w \leq -1.3$, but this change would be
inconsistent with the CMASS measurements.
This example illustrates a general theme of our results:
parameter changes that improve agreement with the \lyaf\ 
BAO measurements usually run afoul of the galaxy BAO measurements.

\begin{figure*}
  \centering
  \begin{tabular}{cc}
    \includegraphics[width=0.5\linewidth]{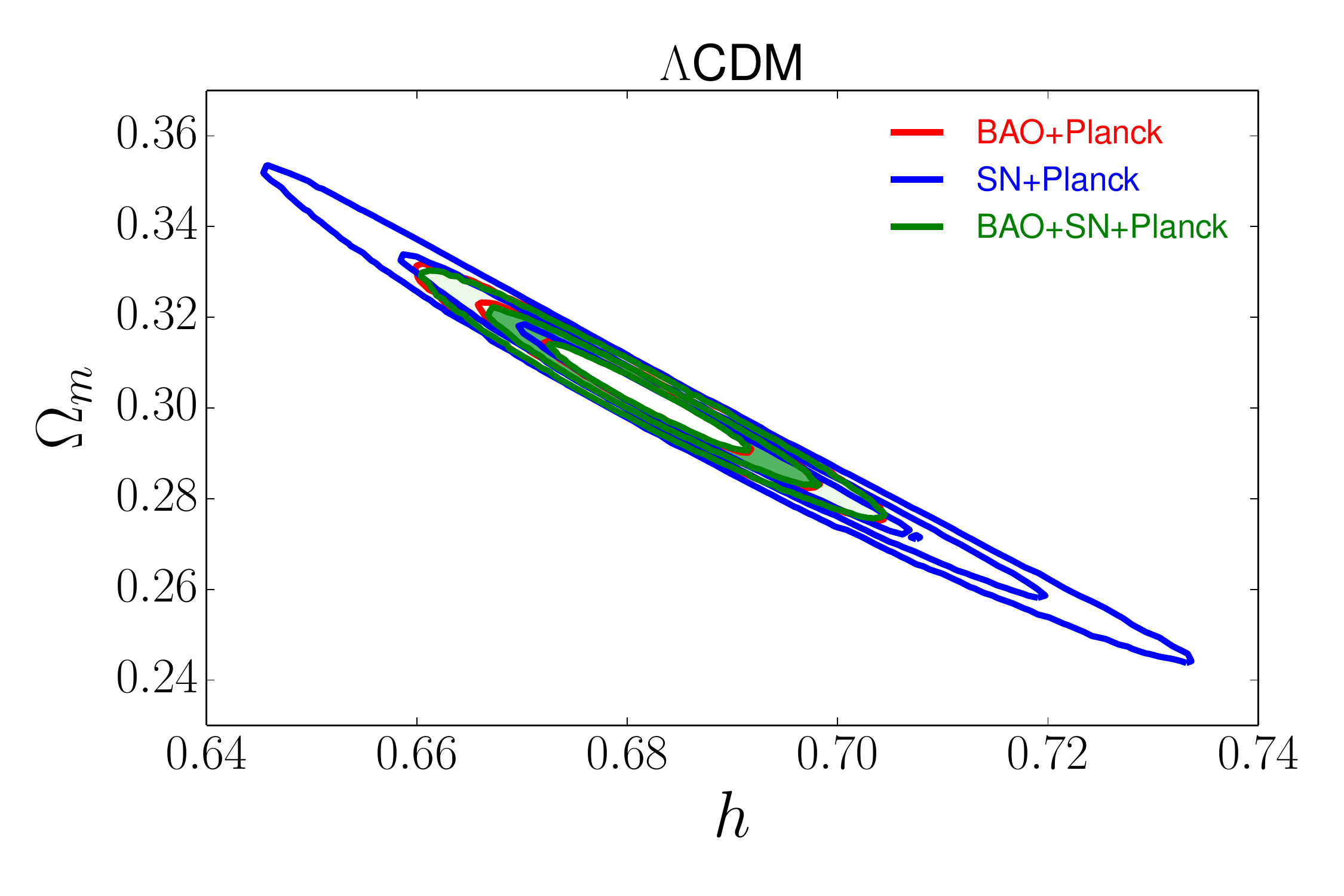} &
    \includegraphics[width=0.5\linewidth]{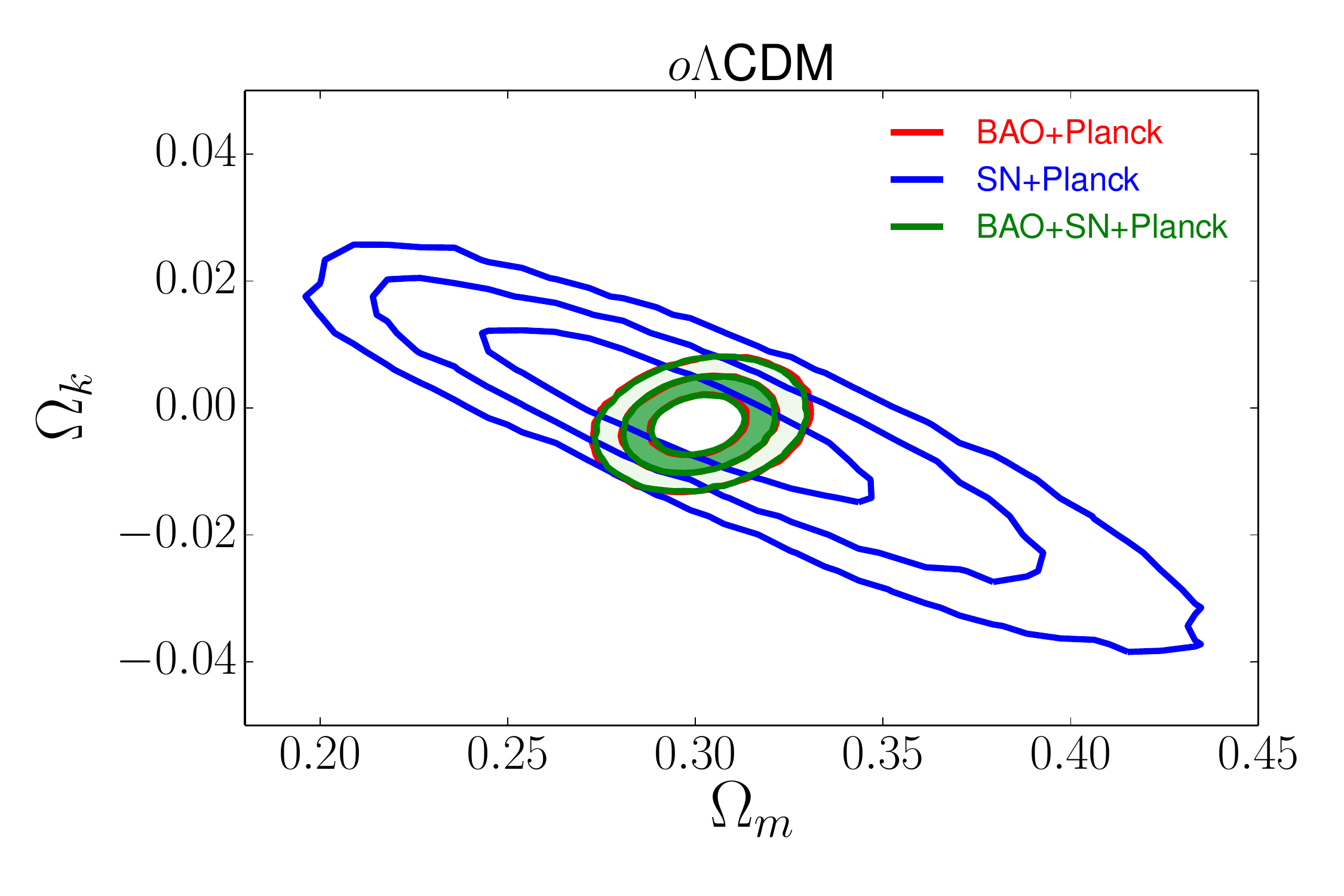} \\
    \includegraphics[width=0.5\linewidth]{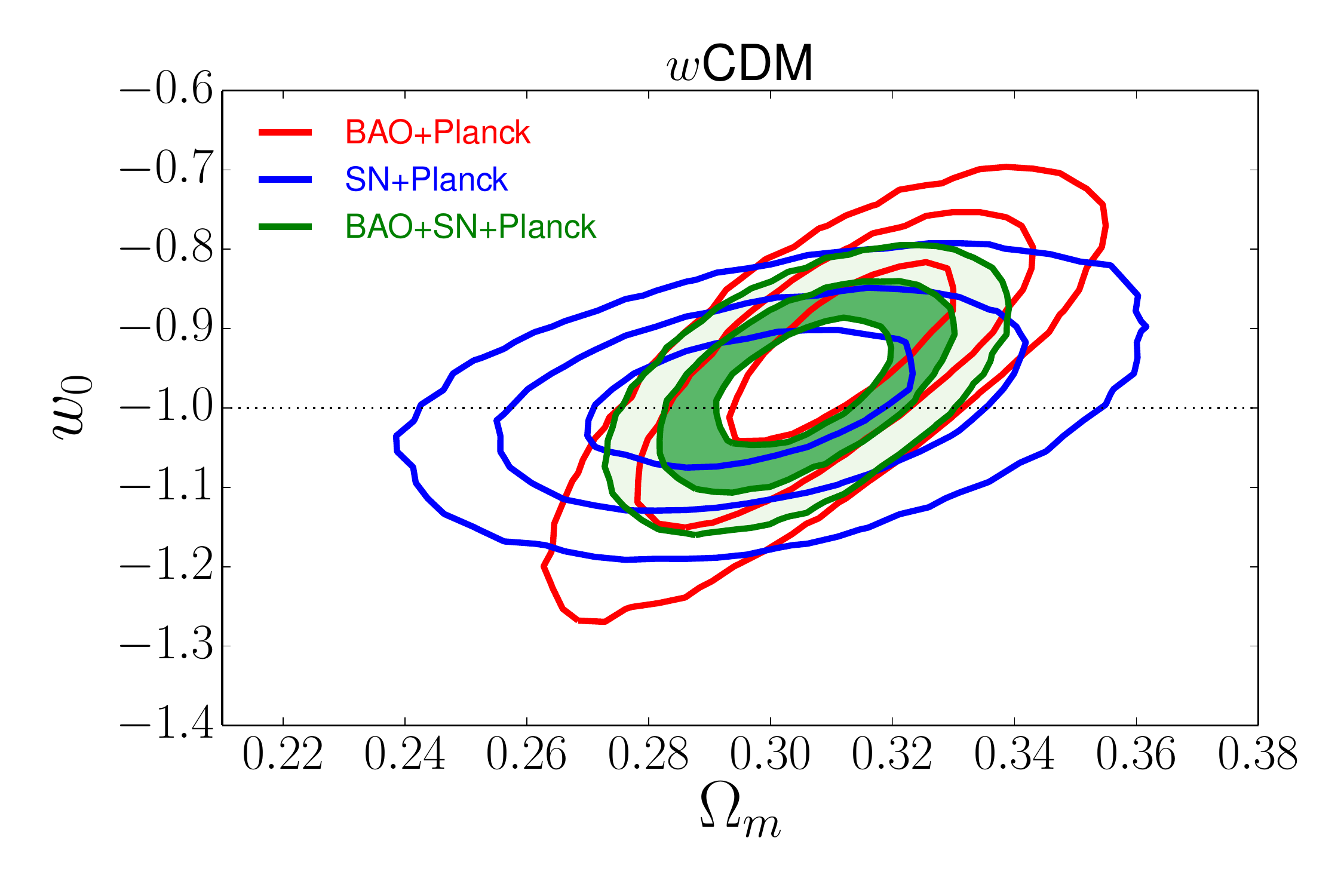} &
    \includegraphics[width=0.5\linewidth]{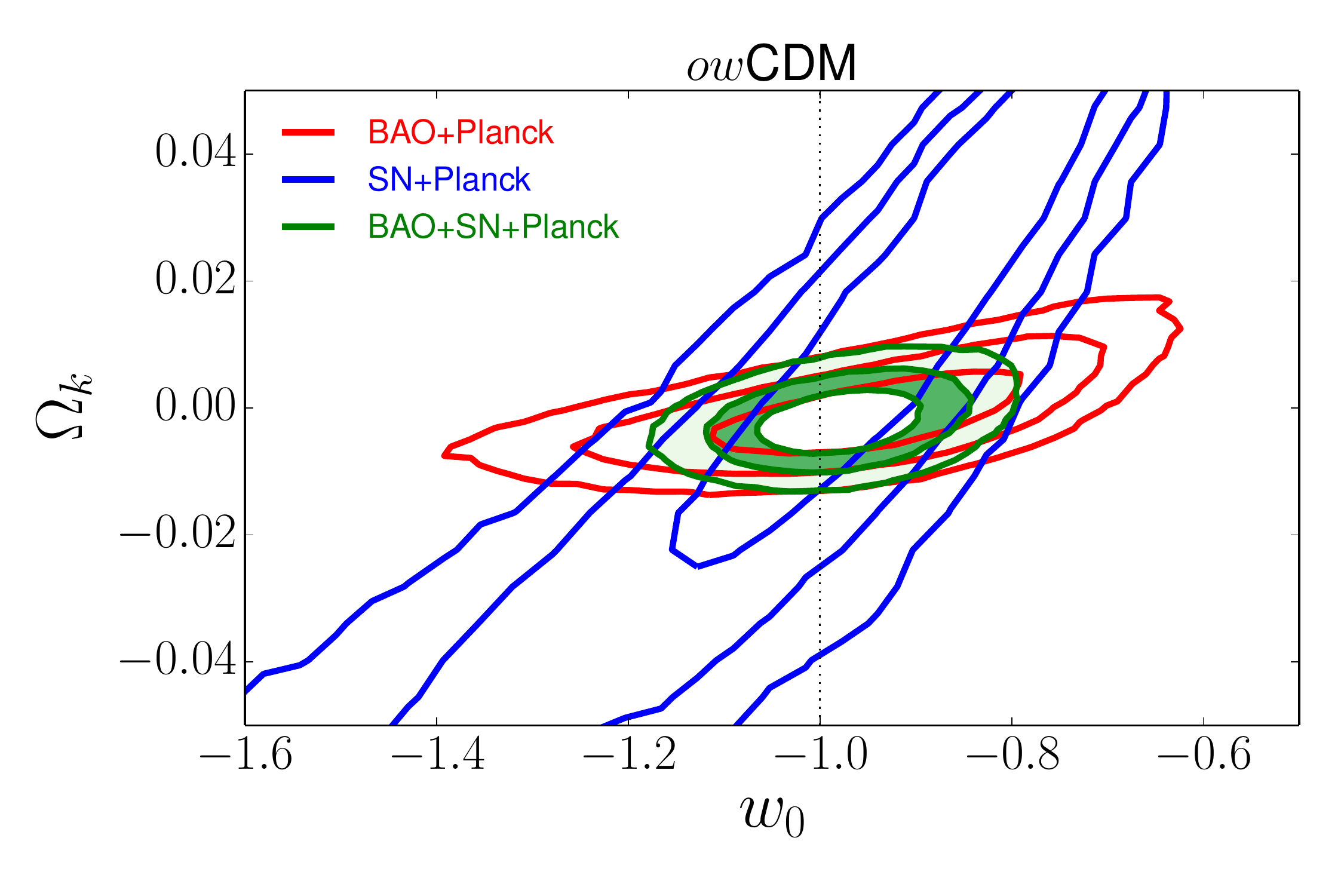} \\
    \includegraphics[width=0.5\linewidth]{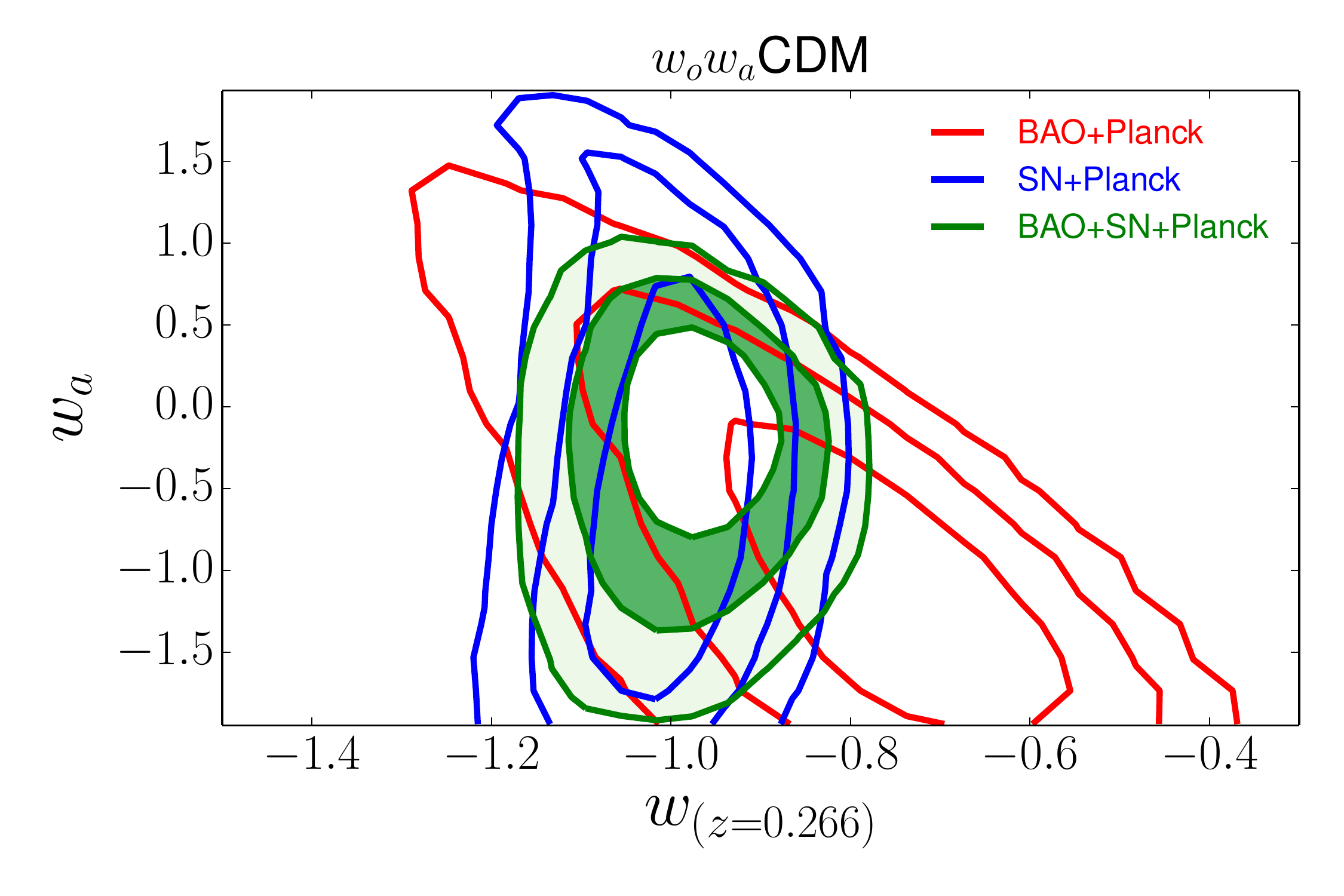}  &
    \includegraphics[width=0.5\linewidth]{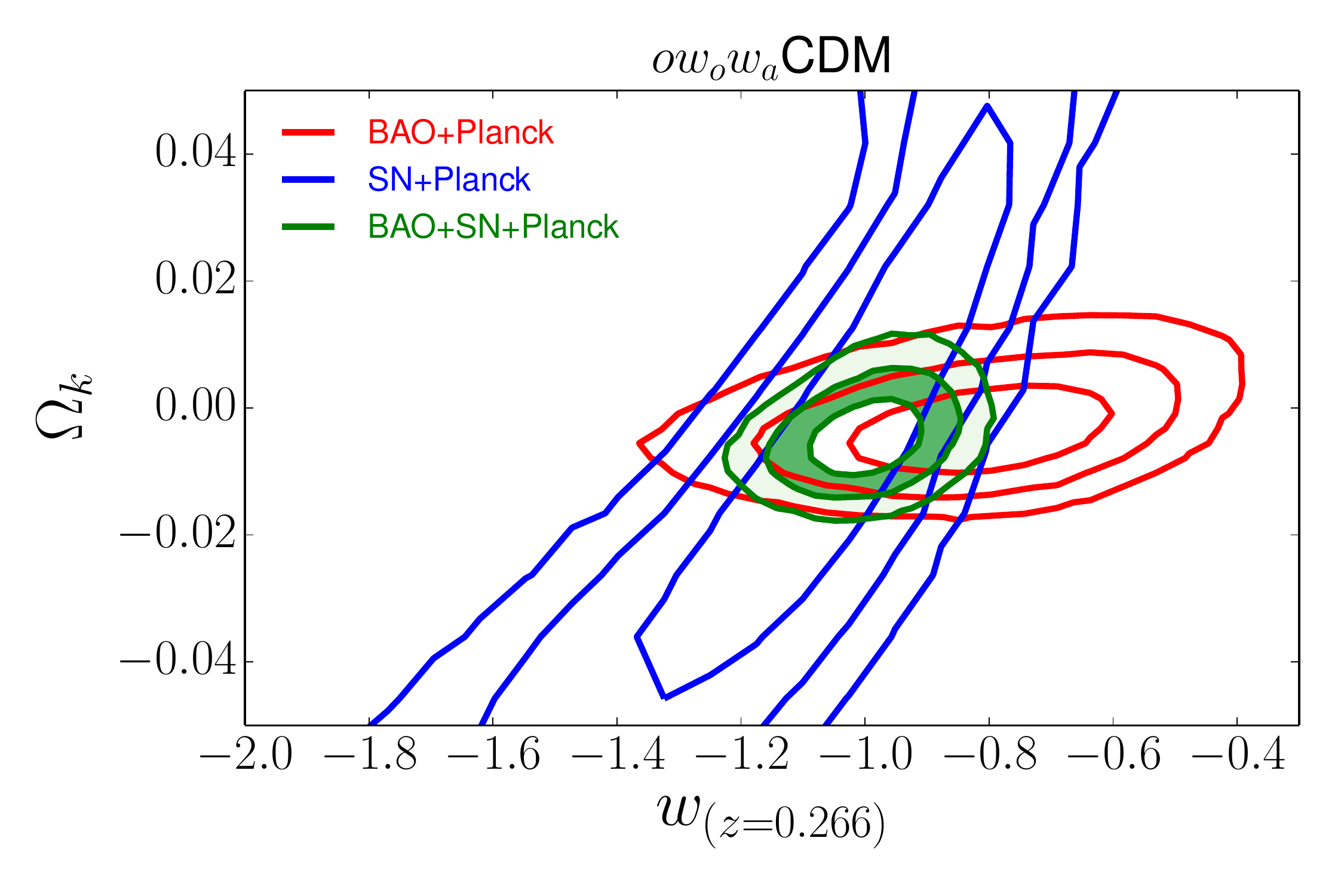} \\
  \end{tabular}
  \caption{Constraints on interesting parameter combinations in
  a variety of dark energy models: \lcdm\ (upper
    left), \olcdm\ (upper right), \wlcdm\ (middle left), \owcdm\ (middle
    right), \wwacdm\ (bottom left) and \owwacdm\ (bottom right). 
    Curves show 68\%, 95\%, and 99.7\% confidence contours for
    the data combinations indicated in the legend.
    In the top panels the red contours are almost
    fully obscured by the green contours because the 
    BAO+Planck combination is already as constraining as
    the BAO+SN+Planck combination, but for models with 
    freedom in dark energy the SN and BAO constraints 
    are complementary.
    The bottom panels, with evolving $w(z)$, display
    the value of $w$ at $z=0.266$, the ``pivot'' redshift
    where $w$ is best constrained by BAO+SN+Planck in the
    \wwacdm\ model.
    For our BAO+SN+Planck contours, the white zone
    interior to the dark green annulus marks the 68\% confidence
    region, and the outer edge of the dark annulus is 95\%.
    }
  \label{fig:models}
\end{figure*}

More quantitative constraints appear in Figure~\ref{fig:models}
and Table~\ref{tab:constraints}.  
We begin with \lcdm\ and continue to the progressively more
flexible models described in Table~\ref{tab:models}.
For CMB data, we now use the compression of Planck or WMAP9
constraints described in Section~\ref{sec:cmb}.

We include all of the BAO data listed in Table~\ref{tab:data}.
Omitting the \lyaf\ BAO data makes almost no difference to the
central values or error bars on model parameters, though it has
a significant impact on goodness-of-fit as we discuss later.

For \lcdm\ and \olcdm, the combination of Planck CMB constraints and
BAO is remarkably powerful, a point already emphasized by
\cite{PlanckXVI}.  Adding SN data makes negligible difference to the
parameter constraints of these models; SN+Planck constraints have
nearly identical central values to BAO+Planck, but larger errors.  In
\lcdm, substituting BAO+SN+WMAP9 for BAO+SN+Planck has a tiny effect,
shifting $\Omega_m = 0.302 \pm 0.008$ to $\Omega_m = 0.300 \pm 0.008$
with a small compensating shift in $h$.  Figure~\ref{fig:models}
illustrates the extremely tight curvature constraint that comes from
combining CMB and BAO data: for \olcdm\ we find $\Omega_k = -0.003 \pm
0.003$ using Planck CMB or $\Omega_k = -0.004 \pm 0.004$ using WMAP9.

Supernovae play a much more important role in models that allow
$w \neq -1$, as their high precision relative distance measurements provide
strong constraints on low-redshift acceleration.  For both 
\wcdm\ and \owcdm, SN+Planck and BAO+Planck constraints are 
perfectly consistent but complementary, and the combination of
all three data sets provides much tighter error bars than 
any pairwise combination.  For \wcdm\ we find $w = -0.97 \pm 0.05$.
For \owcdm\ the curvature constraint from BAO is particularly
important, lifting the degeneracy between $w$ and $\Omega_k$
that arises for SN+CMB alone; we find $w = -0.98 \pm 0.06$ and
$\Omega_k = -0.002 \pm 0.003$.
Substituting WMAP9 for Planck again produces only slight shifts
to central values and a minor increase of error bars.

Even with these powerful BAO, SN, and CMB data sets, constraining the
evolution of $w$ is difficult.  The constraint on the evolution
parameter from BAO+SN+Planck is $w_a = -0.2 \pm 0.4$ in \wwacdm\ and
weakens to $w_a = -0.6 \pm 0.6$ in \owwacdm.  Both results are
consistent with constant $w$, but they allow order unity changes of
$w$ at $z < 1$.  This data combination still provides a good
constraint on the value of $w$ at a ``pivot'' redshift $z_p=0.266$
where it is uncorrelated with $w_a$ (determined specifically for
\wowacdm for BAO+SN+Planck combiations): $w(0.266) = -0.97 \pm 0.05$ in
\wowacdm\ and $-0.99\pm 0.06$ in \owowacdm.  

We note that the degradation of our ability to constrain the evolution
of the equation of state is not accompanied with significant
degradation in our ability to measure the curvature of space: the
constraint on curvature remains tight even when allowing an evolving
equation of state, $\Omega_k = -0.005 \pm 0.004$.

By decoupling the time dependence of $w$ from its present-day value,
the $w_0-w_a$ model allows flexible evolution of the dark energy density.
Adopting a particular form for the dark energy potential reduces
this freedom, and one can construct physically motivated models 
that have evolving dark energy but do not require an additional
free parameter to describe it.
\cite{Gott2011} advocate an interesting example of this model class,
in which dark energy is a slowly rolling scalar field with a
$\frac{1}{2} m^2\phi^2$ potential, analogous to the inflaton
of chaotic inflation models.
\cite{Gott2011} show that this model yields
$\delta w(z) \equiv 1+w(z) \approx \delta w_0 \times H_0^2/H^2(z)$ 
and therefore \cite{Slepian2014}
\begin{eqnarray}
 \frac{H^2(z)}{H^2_0} &\approx&  \Omega_m(1+z)^3  \nonumber \\
        &+& \OmegaDE\left[ \frac{(1+z)^3}{\Omega_m(1+z)^3+\OmegaDE}  
      \right]^{\delta w_0/\OmegaDE},
\label{eqn:slowroll}
\end{eqnarray}
where 
the approximation is first-order in $\delta w_0 = 1+w_0$.

\begin{table*}
\small
\begin{tabular}[t]{llcccccc}
\hline
Model & Data & $\Omega_m$ & $\Omega_b h^2$ & $h$ & $\Omega_k$ & $w$ & $w_a$ \\
\hline

\lcdm &  BAO+Planck & 0.303 (8)  & 0.0223 (3)  & 0.682 (7)  & -- & -- & -- \\
\lcdm &  SN+Planck & 0.295 (16)  & 0.0224 (3)  & 0.688 (13)  & -- & -- & -- \\
\lcdm &  \textbf{BAO+SN+Planck} & 0.302 (8)  & 0.0223 (3)  & 0.682 (6)  & -- & -- & -- \\
\lcdm &  BAO+SN+WMAP & 0.300 (8)  & 0.0224 (5)  & 0.681 (7)  & -- & -- & -- \\

\olcdm &  BAO+Planck & 0.301 (8)  & 0.0225 (3)  & 0.679 (7)  & -0.003 (3)  & -- & -- \\
\olcdm &  SN+Planck & 0.30 (4)  & 0.0224 (4)  & 0.68 (4)  & -0.002 (10)  & -- & -- \\
\olcdm &  \textbf{BAO+SN+Planck} & 0.301 (8)  & 0.0225 (3)  & 0.679 (7)  & -0.003 (3)  & -- & -- \\
\olcdm &  BAO+SN+WMAP & 0.295 (9)  & 0.0226 (5)  & 0.677 (8)  & -0.004 (4)  & -- & -- \\

\wlcdm &  BAO+Planck & 0.311 (13)  & 0.0225 (3)  & 0.669 (17)  & -- & -0.94 (8)  & -- \\
\wlcdm &  SN+Planck & 0.298 (18)  & 0.0225 (4)  & 0.685 (17)  & -- & -0.99 (6)  & -- \\
\wlcdm &  \textbf{BAO+SN+Planck} & 0.305 (10)  & 0.0224 (3)  & 0.676 (11)  & -- & -0.97 (5)  & -- \\
\wlcdm &  BAO+SN+WMAP & 0.303 (10)  & 0.0225 (5)  & 0.674 (12)  & -- & -0.96 (6)  & -- \\

\owcdm &  BAO+Planck & 0.308 (17)  & 0.0225 (4)  & 0.671 (19)  & -0.001 (4)  & -0.95 (11)  & -- \\
\owcdm &  SN+Planck & 0.28 (8)  & 0.0225 (4)  & 0.73 (11)  & 0.01 (3)  & -0.97 (18)  & -- \\
\owcdm &  \textbf{BAO+SN+Planck} & 0.303 (10)  & 0.0225 (4)  & 0.676 (11)  & -0.002 (3)  & -0.98 (6)  & -- \\
\owcdm &  BAO+SN+WMAP & 0.299 (11)  & 0.0227 (5)  & 0.671 (12)  & -0.004 (4)  & -0.96 (6)  & -- \\

\wwacdm &  BAO+Planck & 0.34 (3)  & 0.0224 (3)  & 0.639 (25)  & -- & -0.58 (24)  & -1.0 (6)  \\
\wwacdm &  SN+Planck & 0.292 (23)  & 0.0224 (4)  & 0.693 (24)  & -- & -0.90 (16)  & -0.5 (8)  \\
\wwacdm &  \textbf{BAO+SN+Planck} & 0.307 (11)  & 0.0223 (3)  & 0.676 (11)  & -- & -0.93 (11)  & -0.2 (4)  \\
\wwacdm &  BAO+SN+WMAP & 0.305 (11)  & 0.0224 (5)  & 0.674 (12)  & -- & -0.93 (11)  & -0.2 (5)  \\

\owwacdm &  BAO+Planck & 0.34 (3)  & 0.0225 (4)  & 0.640 (25)  & -0.003 (4)  & -0.57 (23)  & -1.1 (6)  \\
\owwacdm &  SN+Planck & 0.29 (8)  & 0.0225 (4)  & 0.72 (11)  & 0.01 (3)  & -0.94 (21)  & -0.3 (9)  \\
\owwacdm &  \textbf{BAO+SN+Planck} & 0.307 (11)  & 0.0225 (4)  & 0.673 (11)  & -0.005 (4)  & -0.87 (12)  & -0.6 (6)  \\
\owwacdm &  BAO+SN+WMAP & 0.302 (11)  & 0.0227 (5)  & 0.670 (12)  & -0.006 (5)  & -0.88 (11)  & -0.5 (5)  \\

SlowRDE &  \textbf{BAO+SN+Planck} & 0.307 (10)  & 0.0224 (3)  & 0.676 (11)  & -- & -0.95 (7)  & -- \\

\hline
\end{tabular}
\caption{Cosmological parameter constraints from Galaxy+\lyaf\ BAO data 
combined with our compressed description of CMB constraints
from Planck+WP or WMAP9 and the JLA SN data.
Entries for which the parameter is fixed in the listed cosmological model 
are marked with a dash. 
For \wowacdm\ and \owowacdm, column 7 lists the value of 
$w$ at $z=0.266$, which is the ``pivot'' redshift for \wowacdm\ 
with the full data combination.
For SlowRDE, this column lists $w = \delta w_0 - 1$.
\label{tab:constraints}
}

\end{table*}

\begin{table*}
\small
\begin{tabular}{lcccccc}
  &  &  &\multicolumn{4}{c}{$\rho_{DE}/\rho_c$} \\
  & $\Omega_m$ & $H_0$ & $z < 0.5$ & $0.5< z < 1.0$ & $1.0< z < 1.6$  &
  $1.6<z$ \\
\hline
\hline
\\
Suzuki et al 2012 \cite{Suzuki2012}  & ... & ... & $0.731^{+0.015}_{-0.014}$
& $0.880^{+0.240}_{-0.210}$ & $0.330^{+1.900}_{-1.000}$ &
$0.700^{+2.400}_{-1.800}$ \\
\\
BAO+PLANCK         & $0.317_{-0.020}^{+0.009}$ &
$66.4_{-0.9}^{+1.5}$ & $0.667^{+0.026}_{-0.012}$ &
$0.844^{+0.194}_{-0.188}$ & $8.581^{+5.237}_{-6.278}$ &
$-0.921^{+0.897}_{-0.611}$ \\
\\
BAO+SN+PLANCK      & $0.307_{-0.014}^{+0.012}$ &
$67.3_{-1.0}^{+1.5}$ & $0.685^{+0.022}_{-0.016}$ &
$0.765^{+0.146}_{-0.165}$ & $5.154^{+4.761}_{-4.259}$ &
$-0.634^{+0.957}_{-0.601}$ \\
\\
\hline
\end{tabular}
\caption{Parameter constraints in the model in which $\rhoDE(z)$ is
held constant in four discrete bins of redshift.
Uncertainties on each parameter are marginalized over all others,
including $\Omega_b h^2$, which is not listed in the table. $H_0$ is in
km\,s$^{-1}$\,Mpc$^{-1}$.\label{tab:StepCDM}}
\end{table*}

\begin{figure}
  \centering
    \includegraphics[width=\linewidth]{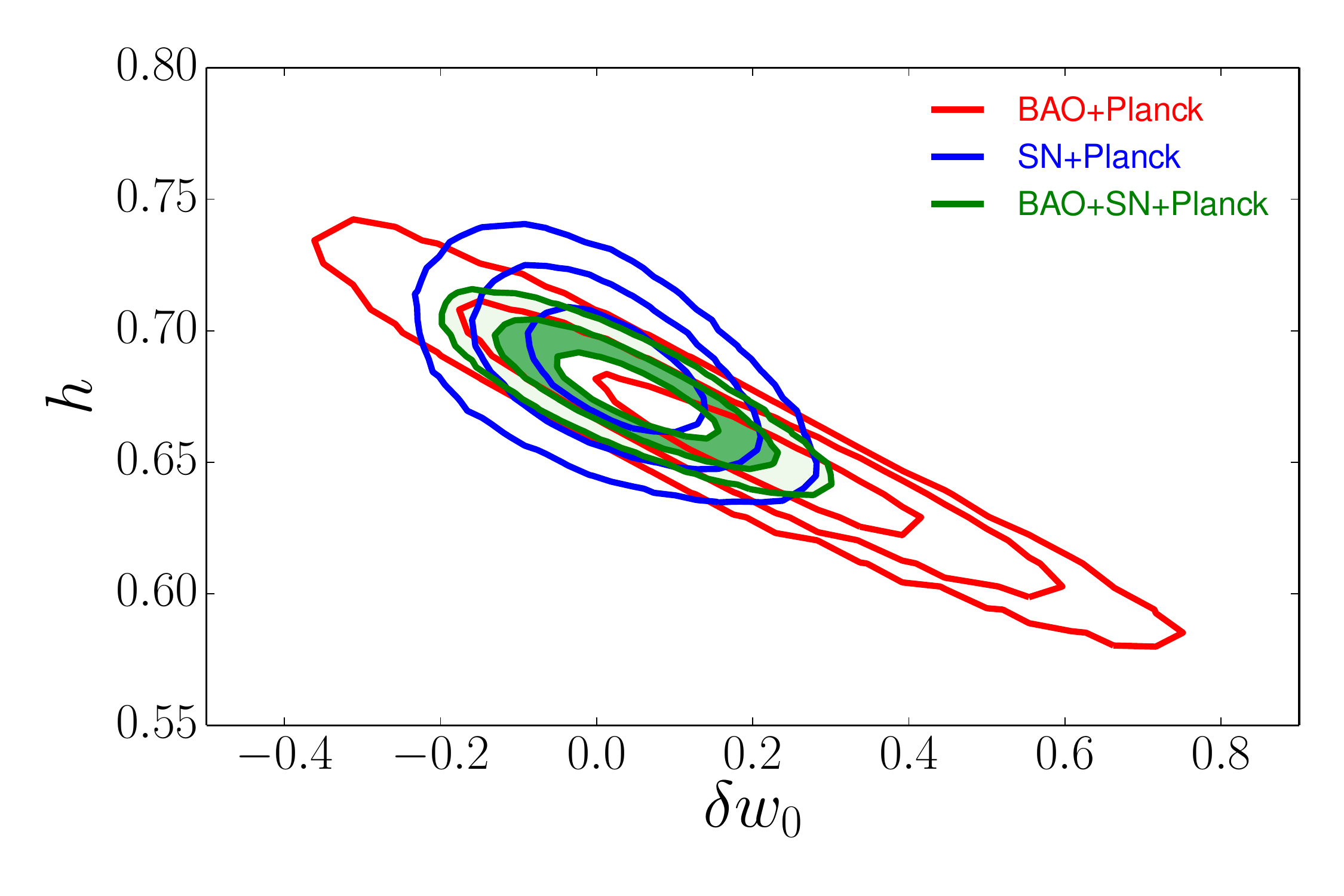}
  \caption{Constraints on $h$ and $\delta w_0 \equiv 1+w_0$ in
  the slow roll dark energy model (eq.~\ref{eqn:slowroll}),
  in the same format as Fig.~\ref{fig:models}.
  }
  \label{fig:SlowRDE}
\end{figure}

Figure~\ref{fig:SlowRDE} presents parameter constraints for the slow
roll dark energy scenario in a flat universe, a model that has the
same number of parameters as \wcdm.
BAO and SN both contribute to the constraints of the joint fit,
which yields $\delta w_0 = 0.05 \pm 0.07$,
$h = 0.675 \pm 0.011$, and $\Omega_m = 0.306 \pm 0.010$.
Results for this scenario are thus consistent with \lcdm\ but allow
small departures from $w_0 = -1$.

A striking feature of Table~\ref{tab:constraints} is that
the best-fit parameter values barely shift as additional freedom
is added to the models.  For the BAO+SN+Planck combination,
the best-fit $\Omega_m$ values range from 0.301 to 0.307
and the best-fit $h$ values from 0.676 to 0.682,
while combinations with WMAP9 favor just slightly lower values
of $\Omega_m$ and $h$.
More importantly, models that allow dark energy
evolution are all consistent with constant $w = -1.0$ at $\approx 1\sigma$,
and $\Omega_k$ is consistent with zero at $1\sigma$ 
in all cases that allow curvature.
The fact that models with additional freedom remain consistent
with \lcdm\ is a substantial argument in favor of this minimal model.

\begin{figure}
  \centering
    \includegraphics[width=\linewidth]{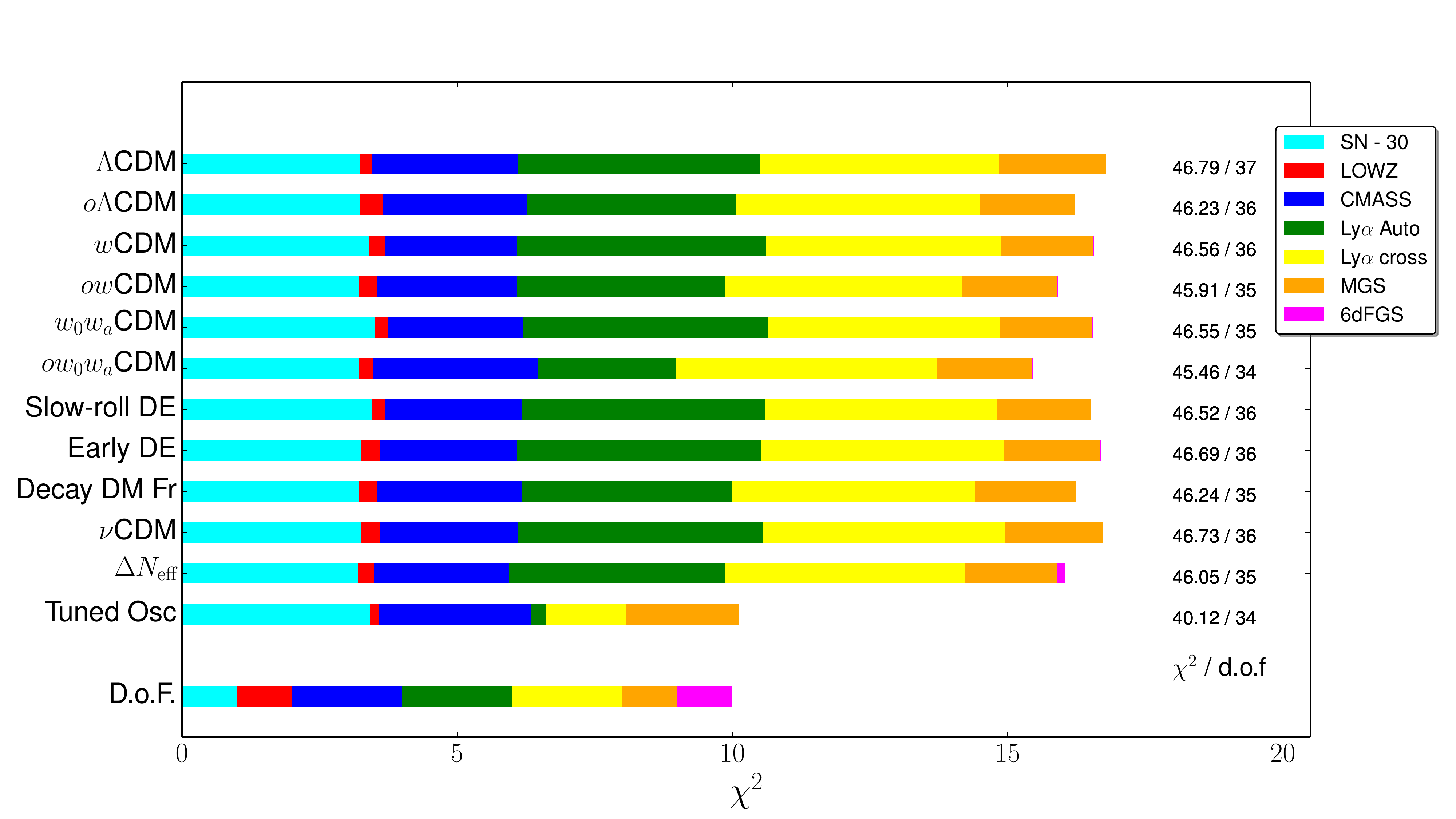}
    \caption{$\chi^2$ values for the best-fit versions of 
    cosmological models considered in the paper.
    Each bar represents the minimum $\chi^2$ for the model
    listed at the left axis, and colors show the $\chi^2$ 
    contributions of individual data sets.  
    For better visualization, we subtract 30 from the SN $\chi^2$.
    CMB contributions are not included but (with our 3-parameter
    compression) are always close to zero.
    The total $\chi^2$ and model degrees-of-freedom (d.o.f., 
    40 data points minus number of fit parameters, which includes
	the SNIa absolute magnitude normalization as well as cosmological
	quantities) are listed
    to the right of each bar.
    The bottom bar shows the number of d.o.f. associated with
    each data set.
	For the $\Delta\neff$ model we use {\tt cosmomc} rather than
	our compressed CMB description, but we again omit CMB
	contributions to $\chi^2$.
    }
  \label{fig:chisq}
\end{figure}

Figure~\ref{fig:chisq} illustrates the goodness-of-fit for
the models in Table~\ref{tab:constraints}, and for additional
models discussed below in Section~\ref{sec:alternatives}.
For the best-fit parameter values in each model, 
horizontal bars show the total $\chi^2$, with colors
indicating the separate contributions from the JLA SN data, the various
galaxy BAO data sets, 
and the \lyaf\ auto-correlation and cross-correlation measurements.
For visualization purposes, we have subtracted 30 from
the SN $\chi^2$, which would otherwise dominate the total
length of these bars because there are 31 SN data points and
many fewer in other data sets.  The constraints on 
$\omega_{cb}$, $\omega_b$, and $D_M(1090)/r_d$ from the
CMB are sufficiently tight that parameter variations within
the allowed range have minimal impact on other observables.
Our minimization yields $\chi^2 \approx 0$ for the CMB data
in essentially every case, since all the models have enough
parameters to fit the three (compressed) CMB constraints perfectly.
For this plot, we have chosen to omit the CMB constraints from
both the $\chi^2$ sum and the degrees-of-freedom (d.o.f.)
computation, though these constraints are still used
when determining model parameters.

The bottom bar in Figure~\ref{fig:chisq} indicates the number of
d.o.f. associated with each data set: 31 for SNe, one each
for the $D_V$ measurements from LOWZ, MGS, and 6dFGS, two
for the $D_M$ and $D_H$ measurements from CMASS, and two
each ($D_M$ and $D_H$) for \lyaf\ auto- and cross-correlation, totaling 40.
Numbers to the right of each model bar list the $\chi^2$ of the model
fit and the corresponding d.o.f. after subtracting the number of fit
parameters.  For \lcdm, for example, we count as free parameters
$\Omega_m$, $h$, and the SNIa absolute magnitude normalization $M_0$,
yielding d.o.f.$=40-3=37$.  We omit
$\omega_b$ because it is determined almost entirely by the CMB data, 
which we have excluded from the $\chi^2$ sum.  The total $\chi^2$ for this
model is 46.79, with a one-tailed $p$-value (probability of obtaining
$\chi^2 \geq 46.79$) of 0.13 for 37 d.o.f.
Thus, if we consider all of the data collectively, the fit
of the \lcdm\ model is acceptable, and for any of the more
complex models considered so far the reduction in $\chi^2$ is
smaller than the number of additional free parameters in the model.

As already emphasized in our discussion, the \lcdm\ model does not give a
good fit to the \lyaf\ BAO data.  This tension is evident in
Figure~\ref{fig:chisq} in the length of the yellow and green $\chi^2$
bars relative to the corresponding d.o.f.  Combining the
Lyman-$\alpha$ auto- and cross-correlation
measurements into a single likelihood
because they measure the same quantities,
the \lcdm\ $\chi^2_{{\rm Ly}\alpha} = 8.3$ for two d.o.f.
has a $p$-value of 0.016, consistent with
Figure~\ref{fig:chain_vs_contour1}.  It is unclear how much to make of
this mild tension in the context of a fit that yields
adequate-to-excellent agreement with multiple other data sets and an
acceptable $\chi^2$ overall.  It is evident that none of the more
complex models considered so far allows a significantly better fit to
the \lyaf\ BAO data.  The partial exception is \owowacdm, which has
the most freedom to adjust high-redshift behavior relative to
low-redshift behavior, but even here the reduction in $\chi^2$
relative to \lcdm\ is only 1.33 (coming almost entirely from \lyaf),
for three additional model parameters.  Omitting the \lyaf\ data makes
almost no difference to the best-fit parameter values or their error
bars in any of these models, which are driven mainly by the
high-precision CMB, CMASS, and SN constraints.

To conclude this section, we examine a model
in which dark energy is characterized by specifying its energy density
in four discrete bins of redshift:
$z<0.5$, $0.5<z<1.0$, $1.0 < z < 1.6$, and $z > 1.6$.
This step-wise model is a useful complement to models that
specify $\rhoDE(z)$ through parameterized descriptions of $w(z)$.
The bins are chosen to be the same as the ones considered in
a similar analysis by \cite{Suzuki2012} (their Table 8),
who combined Union 2.1 SN data,
WMAP7 CMB data \citep{2011ApJS..192...18K},
BAO from the combined analysis of SDSS DR7 and 2dFGRS
\citep{2002MNRAS.337.1068P},
and the distance-ladder $H_0$ measurement of \cite{Riess11}.
We fit simultaneously for $\rhoDE(z)$ in each of these bins and
for the values of $\Omega_m$, $\Omega_b$, and $H_0$, assuming
$\Omega_k=0$ to match \cite{Suzuki2012}.
Within an individual bin, $\rhoDE$ is held constant and $H(z)$
evolves according to the Friedmann equation, but there are
discontinuities in $H(z)$ at bin boundaries to accommodate the
discontinuous changes in $\rhoDE(z)$.  The matter density evolves
as $\rho_m(z) = \Omega_m\rhocrit (1+z)^3$, where $\Omega_m$ and
$\rhocrit$ denote $z=0$ values as usual.

\begin{figure}
\centering
\includegraphics[width=0.9\columnwidth]{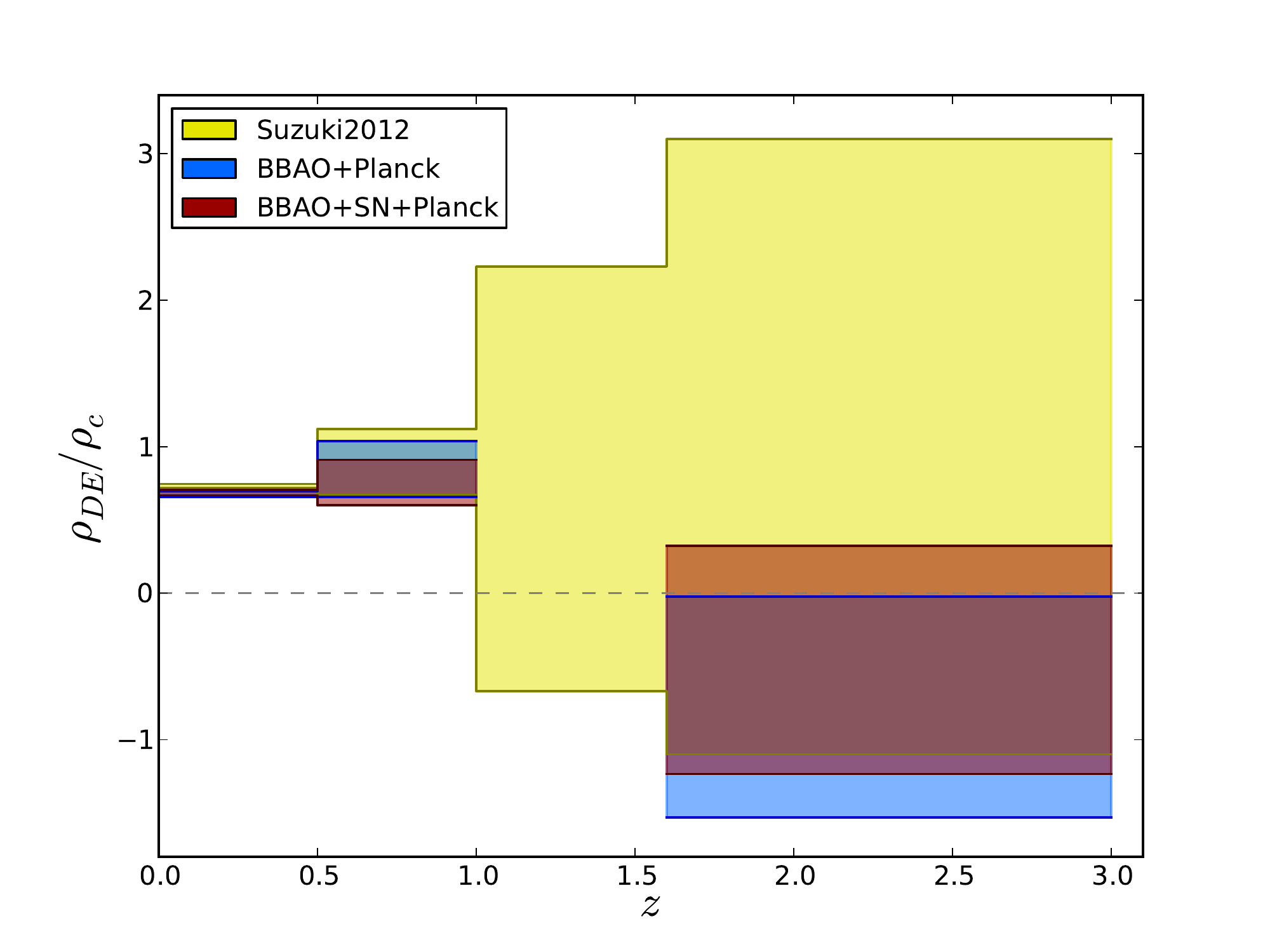}
\caption{Constraints on $\rho_{DE}(z)$ assumed to be constant 
within redshift bins, in units of the present-day critical density $\rho_c$. 
Shaded areas represent 68\% confidence levels.
Yellow bands show constraints in the same bins from \cite{Suzuki2012}.
Our constraints in the $z = 1.0-1.6$ bin are omitted.}
\label{fig:StepCDM}
\end{figure}

Constraints on this model
from our BAO+Planck and BAO+SN+Planck data combinations
appear in Figure~\ref{fig:StepCDM} and Table~\ref{tab:StepCDM}.
As in our other models that allow time-varying dark energy,
BAO and SN data both contribute significantly to the parameter
constraints.
Our results show a clear detection of non-zero dark energy density
in each of the first two redshift bins at $z<1$, and they are
consistent with a constant energy density across this redshift range.
Compared to \cite{Suzuki2012}, we obtain a significantly tighter
constraint in the $0.5 < z < 1.0$ bin, where the CMASS BAO
measurement makes an important difference, but a slightly looser
constraint in the $z < 0.5$ bin, where we do not incorporate
a direct $H_0$ measurement.
We obtain much poorer constraints in the $1<z<1.6$ bin
because the JLA sample contains only 8 SNe with $z>1$ compared to 29 
for the Union~2.1 sample.  
At $z > 1.6$ our constraint is stronger thanks to the \lyaf\ BAO
measurement, but the uncertainty is large nonetheless, and
the low \lyaf\ value of $H(z)$ leads to a preference for
negative dark energy density in this bin, although consistent
with zero at $1\sigma$.

\section{Alternative Models}
\label{sec:alternatives}

We now turn to models with more unusual histories of the 
dark energy, matter, or radiation components.
In part we want to know what constraints
our combined data can place on interesting physical quantities, such
as neutrino masses, extra relativistic species, dark energy that is
dynamically significant at early times, or dark matter that decays
into radiation over the history of the universe.  We also want to see
whether any of these alternative models can resolve the tension with
the \lyaf\ measurements at $z=2.34$, which persists in all of the
models considered in Section~\ref{sec:deconstraints}.
We begin with the early dark energy model, because understanding
the origin of the constraints on this model informs the discussion
of subsequent models.

\subsection{Early Dark Energy}
\label{sec:ede}

In typical dark energy models, including all of those
discussed in Section~\ref{sec:deconstraints}, dark energy is
dynamically negligible at high redshifts because its
energy density grows with redshift much more slowly
than $(1+z)^3$.
However, some scalar field potentials yield a dark
energy density that tracks the energy density of 
the dominant species during the radiation and matter
dominated eras, then asymptotes towards a cosmological
constant at late times \cite{Albrecht00,Hebecker01}.
These models ameliorate, to some degree, the
``coincidence problem'' of constant-$w$ models
because the ratio of dark energy density to total
energy density varies over a much smaller range.

As a generic parameterized form of such early dark energy models,
we adopt the formulation of Doran \& Robbers \cite{Doran06}, 
in which the density
parameter of the dark energy component evolves with $a=(1+z)^{-1}$ as
\begin{equation}
\OmegaDE(a) = {\OmegaDE - \OmegaDE^e\left(1-a^{-3w_0}\right)
               \over \OmegaDE + \Omega_m a^{3w_0}}
			   + \OmegaDE^e\left(1-a^{-3w_0}\right)~,
\label{eqn:ede}
\end{equation}
where $\OmegaDE$ and $\Omega_m$ denote $z=0$ values as usual
and $\OmegaDE^e$ is the dark energy density parameter at early times.
A flat universe is assumed, with $\OmegaDE+\Omega_m=1$.
The quantity $w_0$ is the effective value of the equation-of-state
parameter today.  At high redshift ($a \ll 1$),
the denominator of the first term is $\gg 1$, and $\Omega_d(a)$
approaches the constant value $\OmegaDE^e$.
This in turn requires a dark energy density that scales
as $a^{-3}$ in the matter-dominated era and as $a^{-4}$ 
in the radiation-dominated era, though it is $\OmegaDE(a)$
rather than $\rho_{\rm de}(a)$ that is specified explicitly.
For $w_0=-1$, the model approaches \lcdm\ as $\OmegaDE^e$ goes to zero.
There are other generic forms of models with early dark energy,
as well as non-parametric descriptions (see discussion by
\cite{Samsing12}).

\begin{figure}
  \centering
    \includegraphics[width=\linewidth]{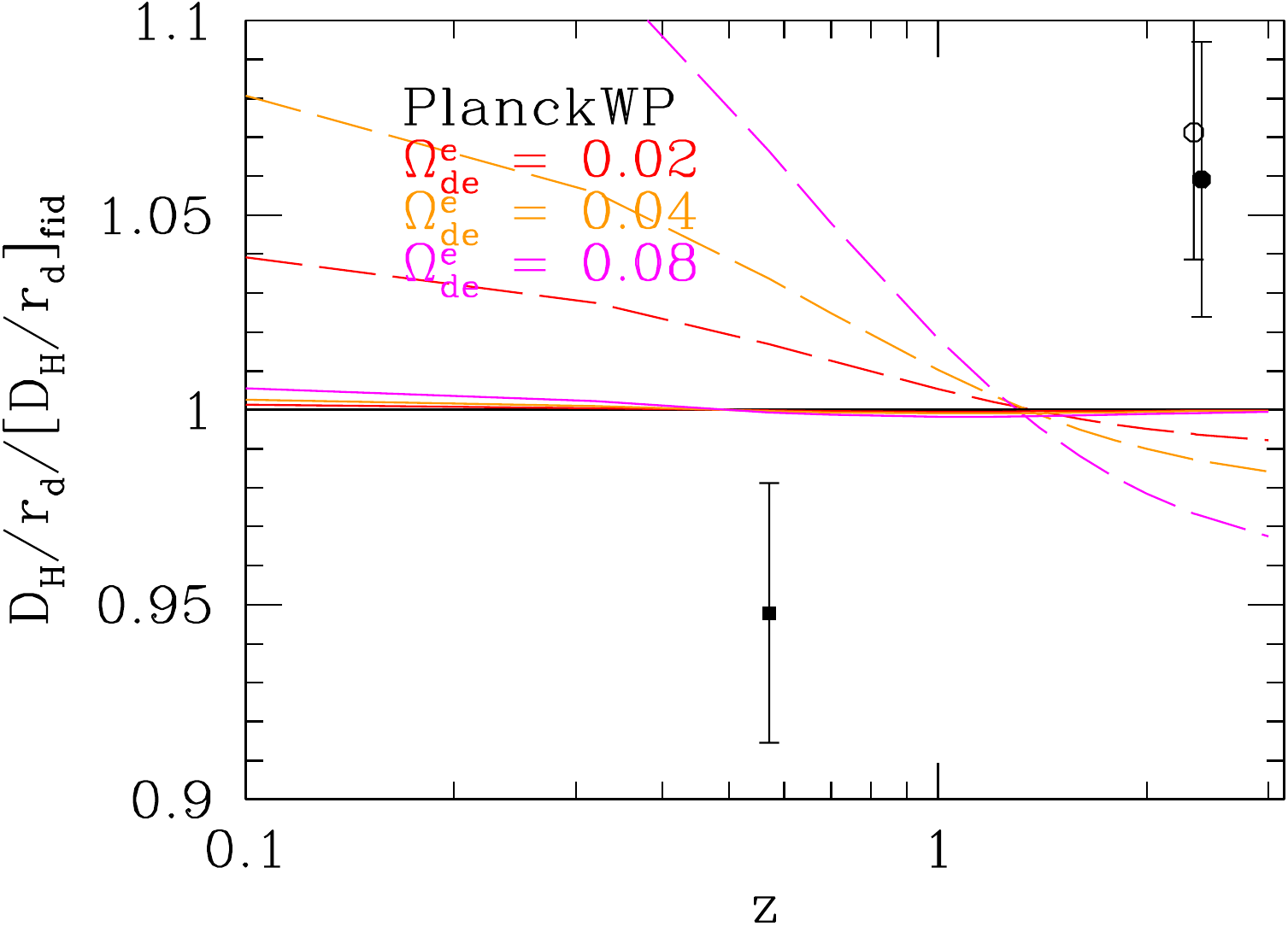} \\
    \includegraphics[width=\linewidth]{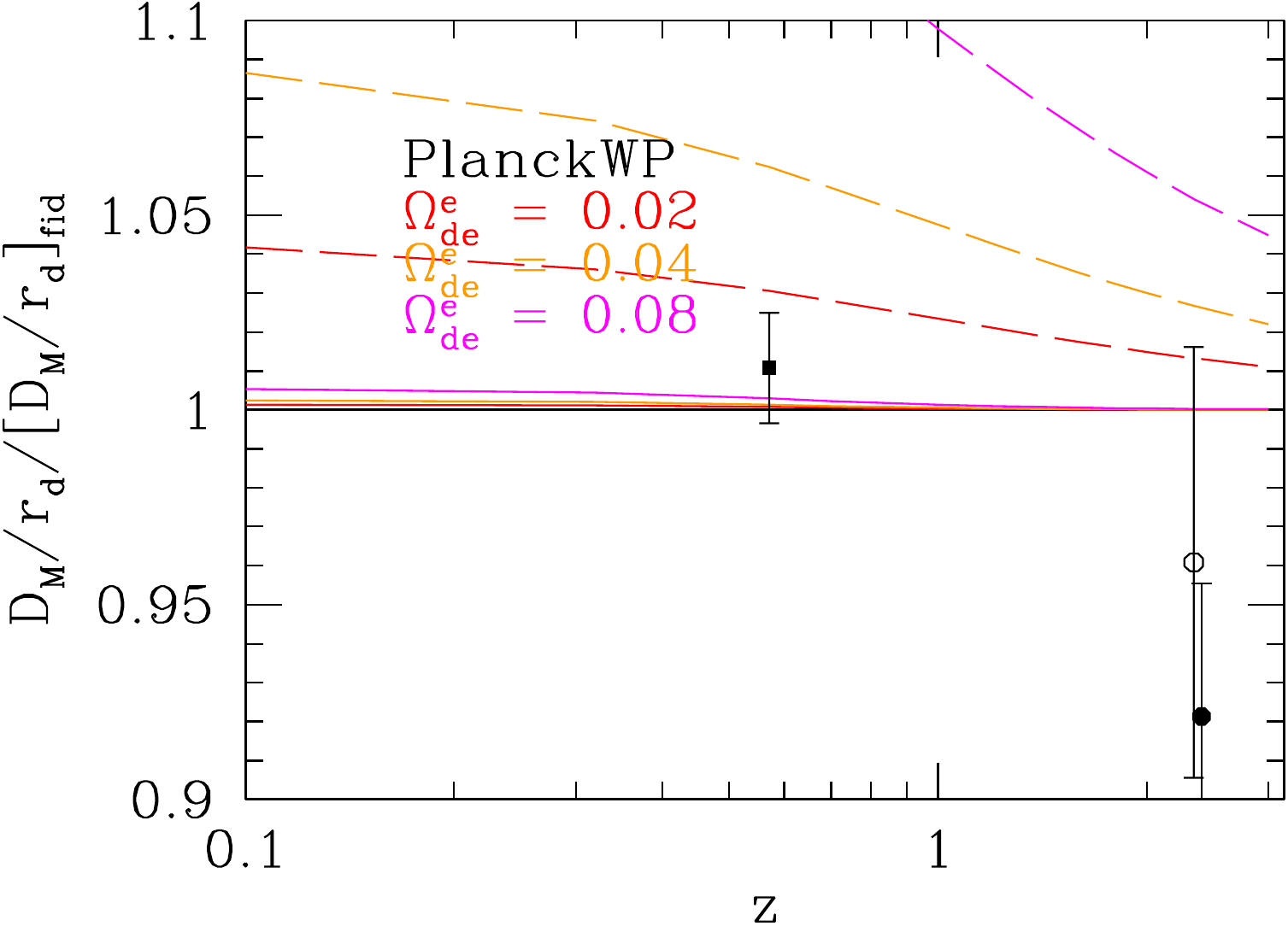}
  \caption{Predicted BAO scales for early dark energy models with
  CMB observables $\Omega_b h^2$, $\Omega_m h^2$, and $D_M(1090)/r_d$ 
  held fixed to the values of the best-fit Planck+WP \lcdm\ model. 
  We adopt equation~(\ref{eqn:ede}) with $w_0=-1$.
  Solid lines show the case in which $r_d$ is rescaled by
  $(1-\OmegaDE^e)^{1/2}$ to represent the effect of early dark
  energy in the pre-recombination era, while dashed lines show the
  case in which $r_d$ is held fixed at the fiducial model value
  of $r_d=147.49$ Mpc.  We show ratios of $D_H(z)/r_d$ (top) or
  $D_M(z)/r_d$ (bottom) relative to the fiducial ($\OmegaDE^e=0$) model.
  Points with error bars show the BAO measurements from
  CMASS galaxies at $z=0.57$ (filled square) and from \lyaf\ auto-correlation
  (open circle) and cross-correlation (filled circle) at $z=2.34$.  
  For visual clarity, the \lyaf\ cross-correlation points have been slightly 
  shifted in redshift.}
  \label{fig:ededistances}
\end{figure}

\begin{figure}
  \centering
    \includegraphics[width=\linewidth]{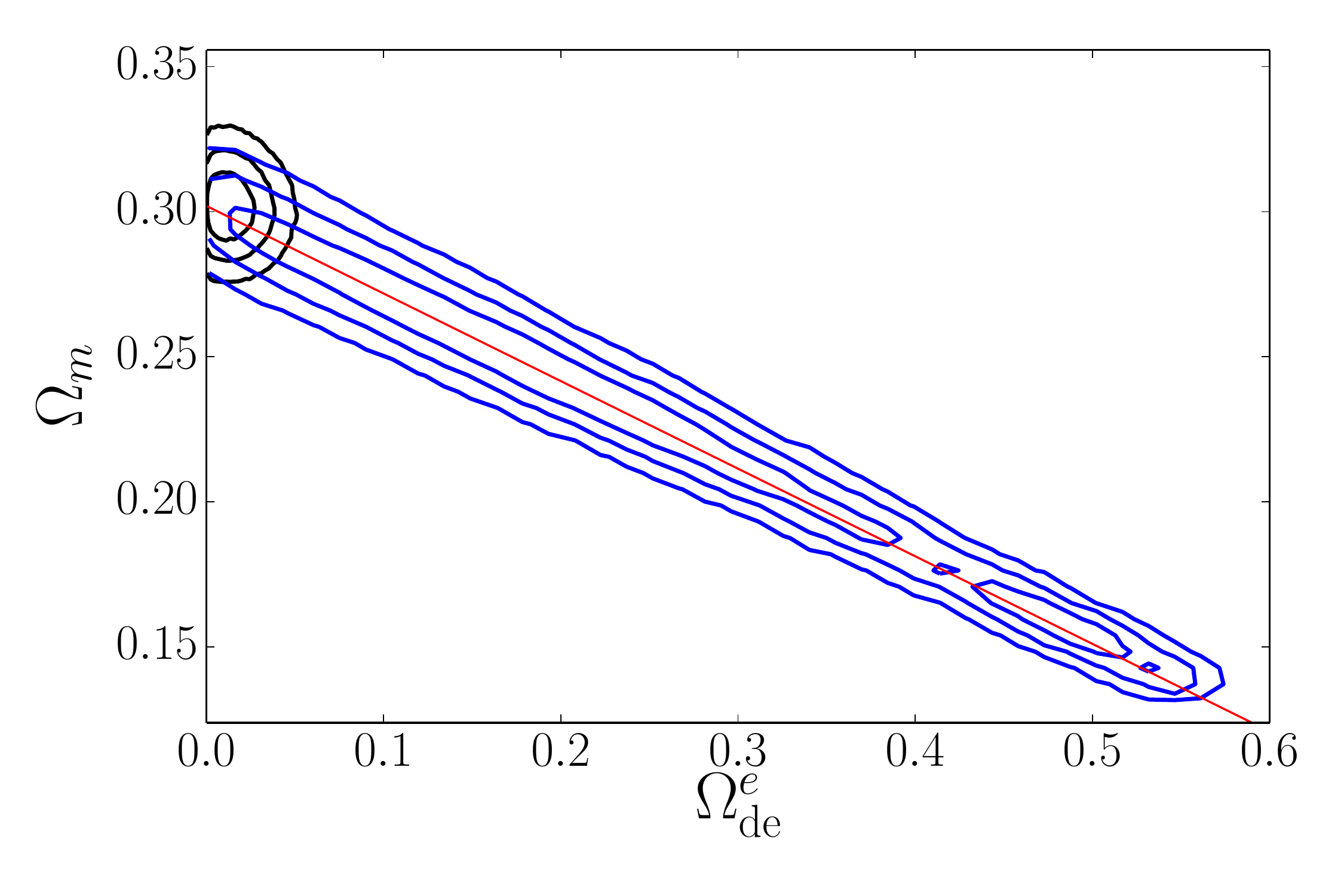} 
  \caption{Constraints in the $\OmegaDE^e-\Omega_m$ plane from
  the combination of our compressed CMB description with galaxy+\lyaf\
  BAO data.  Black contours (68\%, 95\%, and 99.7\%) show the tight
  constraints on early dark energy for models with fixed $r_d=147.49$ Mpc.
  Blue contours show the constraints for models with
  $r_d \propto (1-\OmegaDE^e)^{1/2}$ as expected if $\OmegaDE^e$ is
  constant into the radiation-dominated era.  
  The red solid line traces the parameter degeneracy 
  $\Omega_m = \Omega_{m,{\rm fid}}(1-\OmegaDE^e)$ predicted
  by the approximate scaling arguments described in the text.
  }
  \label{fig:edeconstraints}
\end{figure}

If dark energy is important in
the pre-recombination era, then the boosted energy density
in this era reduces the sound horizon 
by a factor $(1-\OmegaDE^e)^{1/2}$ relative
to a conventional model with the same parameters
\cite{Doran06,Samsing12}.
Pre-recombination dark energy also influences the detailed shape
of the CMB anisotropy spectrum by altering the early integrated
Sachs-Wolfe contribution and the CMB damping tail \cite{Hojjati13}.
Our analysis here incorporates the rescaling of the sound horizon,
but we continue to use the compressed CMB description 
of Section~\ref{sec:cmb} and therefore
ignore the more detailed changes to the power spectrum shape.
Because of the exquisite precision of CMB measurements, the
power spectrum shape may impose tighter constraints
on early dark energy than the expansion history measurements
employed here (see, e.g., \cite{Hojjati13}).
However, those constraints are more dependent on the specifics
of the models being examined, both the dark energy evolution
and other parameters that describe the inflationary spectrum,
tensor fluctuations, relativistic energy density, and reionization.

Figure~\ref{fig:ededistances}
plots the evolution of $D_H(z)/r_d$ and $D_M(z)/r_d$ for models
with $\OmegaDE^e = 0,$ 0.02, 0.04, and 0.08.
We always adopt $w_0=-1$, but we constrain 
$h$, $\Omega_b$, and $\Omega_m = 1-\OmegaDE-\Omega_{\nu+r}$
by fixing $\Omega_b h^2$, $\Omega_m h^2$, and
$D_M(1090)/r_d$ to the values in the best-fit
Planck+WP \lcdm\ model, in effect forcing the errors in our
compressed CMB description to zero.
Solid curves incorporate the expected
$(1-\OmegaDE^e)^{1/2}$ reduction of $r_d$.
Dotted curves show the case in which we instead keep $r_d$ fixed
at its fiducial value of 147.49 Mpc.
The latter case would be physically relevant in a model where dark energy is
dynamically negligible in the pre-recombination era
but approaches the evolution of equation~(\ref{eqn:ede})
later in the matter dominated era.
To highlight model differences, we scale all values to those
of the fiducial \lcdm\ model, which corresponds to $\OmegaDE^e=0$.

Remarkably, for the rescaled $r_d$ case, the predicted values 
of $D_H(z)/r_d$ and $D_M(z)/r_d$ change by less than 0.5\% at
all redshifts, even for $\OmegaDE^e = 0.08$.
We can understand this insensitivity by considering the low and
high-redshift limits for the simplified case of a flat cosmology
with only matter and dark energy.  The matter density at redshift $z$ is
\begin{equation}
\begin{split}
\rho_m(z) &= \rhocrit \times \Omega_m (1+z)^3 \\
          &= {3 \left(100\hubunits\right)^2 \over 8\pi G}
		      \times \left(\Omega_m h^2\right) (1+z)^3 ~,
\end{split}
\label{eqn:rhomz}
\end{equation}
where $\rhocrit$ denotes the $z=0$ value as usual and we have used
$H_0 = 100h\hubunits$ but relocated $h$ to the second factor.
Using $\rho_m(z)+\rhoDE(z) = \rho_{\rm crit}(z) = 3H^2(z)/8\pi G$ implies
\begin{equation}
\begin{split}
H(z) = &\left(100\hubunits\right) \times 
  \left[\left(\Omega_m h^2\right)(1+z)^3\right]^{1/2} \\
  &\times \left[1+\rhoDE(z)/\rho_m(z)\right]^{1/2} ~.
\end{split}
\label{eqn:edeHz}
\end{equation}
For a cosmological constant, $\rhoDE(z)/\rho_m(z) \propto (1+z)^{-3}$,
so the ratio tends rapidly to zero at high redshift, but for the early 
dark energy model this ratio asymptotes instead to 
$\OmegaDE^e/\Omega_m(z) \approx \OmegaDE^e/(1-\OmegaDE^e)$.  
Thus, at fixed
$\Omega_m h^2$, $H(z)$ is higher in the early dark energy model by a factor
$(1-\OmegaDE^e)^{-1/2}$, and $D_H(z)$ is smaller by the same factor.
This reduction in $D_H(z)$ exactly compensates the $(1-\OmegaDE)^{1/2}$
rescaling of $r_d$,
leaving $D_H(z)/r_d$ independent of $\OmegaDE^e$.

At low redshift, conversely, 
\begin{equation}
D_M(z) = {c \over H_0}\int_0^z {H_0 \over H(z')}dz'
\end{equation}
depends mainly on $H_0$, since the evolution of $H_0/H(z)$ is
insensitive to moderate changes in $\Omega_m$ and $\OmegaDE$
for $z \ll 1$.  Therefore, to keep the value of $D_M(1090)/r_d$
fixed to the CMB constraint, one must increase $H_0$ by approximately
$(1-\OmegaDE^e)^{-1/2}$ so that both the low and high-redshift
contributions to $D_M(1090)$ shrink by the factor required to compensate
the change in $r_d$.  This change again forces $D_H(z)/r_d$ to nearly
the same value as the fiducial model with $\OmegaDE^e = 0$.

These scaling arguments are not perfect because they break down at
intermediate redshifts and because a change in $H_0$ at fixed
$\Omega_m h^2$ implies a change in $\Omega_m$, which itself affects
the low-redshift evolution of $H_0/H(z)$.  Nonetheless, the full
calculation in Figure~\ref{fig:ededistances}
demonstrates that for $\OmegaDE^e$ as large as 0.08 there is minimal
change in $D_H(z)$ at any redshift, and minimal
change in $H(z)$ in turn implies minimal change in $D_M(z)$.
However, the values of $H_0$ are larger, and $\Omega_m$ correspondingly
smaller, for the successively higher $\OmegaDE^e$
curves in Figure~\ref{fig:ededistances}. 
The combination $H_0 r_d$ is nearly constant, decreasing by just 0.14\%,
0.24\%, and 0.49\% for $\OmegaDE^e=0.02$, 0.04, and 0.08, respectively.

Reversing these arguments explains why $D_H(z)$ and $D_M(z)$ change
rapidly with $\OmegaDE^e$ if $r_d$ stays fixed instead of rescaling
(dashed curves in Fig.~\ref{fig:ededistances}).
In this case, $D_M(1090)$ must stay fixed to keep the angular
scale of the acoustic peaks unchanged, so the decrease of 
high-redshift contributions to 
$D_M(z)$ by $(1-\OmegaDE^e)^{1/2}$ requires
a compensating increase of $D_M(z)$ at low redshift.  This requires
a large fractional reduction in $H_0$, since most of the contribution
to $D_M(1090)$ comes from high redshift (e.g., 75\% from $z>1$).
This in turn leads to large deviations in $D_H(z)/r_d$ and $D_M(z)/r_d$
at low redshift.  At high redshift, $D_H(z)$ is again smaller
by $(1-\OmegaDE^e)^{1/2}$, but this now leads to a deviation in
$D_H(z)/r_d$ because it is no longer compensated by a smaller $r_d$.
Even for $\OmegaDE^e = 0.02$, fixing CMB observables requires a 4.3\%
reduction in $H_0$.  Furthermore, adding early dark energy in this
case moves model predictions further from the CMASS and
\lyaf\ measurements of $D_H(z)/r_d$ {\it and} further from the
\lyaf\ $D_M(z)/r_d$.  We therefore expect tight constraints on
$\OmegaDE^e$ in the case of fixed $r_d$.

Figure~\ref{fig:edeconstraints} presents constraints on these
early dark energy models from our MCMC analysis, with $w_0$
fixed to $-1$.  We now account for uncertainties in the CMB
constraints, using the compressed description of Section~\ref{sec:cmb}.
Note that we assume that the CMB constraints on $\omega_m$ and $\omega_b$
are not altered by the introduction of early dark energy, which might not
hold in a complete analysis that uses the full CMB spectrum.
The non-rescaled case is tightly constrained as expected,
with a $2\sigma$ upper limit $\OmegaDE^e < 0.031$. 
SNe do not significantly improve these constraints,
although they would play a larger role if we allowed $w_0$ as a free
parameter.
To summarize, adding early dark energy with fixed $r_d$ worsens agreement
with our BAO measurements, and a dynamically significant value
of $\OmegaDE^e$ is ruled out.

For rescaled $r_d$, which is the physically expected case if 
$\OmegaDE^e$ remains constant back into the radiation-dominated
era, we instead find a valley of near-perfect degeneracy between
$\Omega_m$ and $\OmegaDE^e$.
For $H_0 r_d =\,{\rm const.}$ and fixed $\Omega_m h^2$,
the expected degeneracy line is 
$\Omega_m \propto r_d^{-2} \propto (1-\OmegaDE^e)$, marked
by the red solid line in Figure~\ref{fig:edeconstraints}.  
This prediction describes our numerical MCMC results extremely well.
Along this line, there are
models with $\Delta\chi^2 < 1$ relative to the best-fit
\lcdm\ ($\OmegaDE^e=0$) model, at least out to $\OmegaDE^e = 0.32$.
Including SNe again makes minimal difference to our constraints
because the models along the degeneracy line predict
nearly identical $D_M(z)$.

Although these models are degenerate with respect to our
geometrical constraints, they predict different values of $H_0$
and different measures of structure growth.
Intriguingly, non-zero $\OmegaDE^e$ with rescaled $r_d$
reduces tension
with distance-ladder measurements of $H_0$ and with the level of
matter clustering inferred from cluster masses, weak lensing,
and redshift-space distortions.  We discuss the impact on structure
growth measures in Section~\ref{sec:growth}.

As already emphasized, the detailed shape of the CMB power spectrum
may impose much tighter constraints on early dark energy;
e.g., for the specific case of the Doran-Robbers model, 
\cite{Hojjati13} infer $\OmegaDE^e < 0.012$ at 95\% confidence.
However, the degeneracies identified here in the expansion history
constraints are striking, and highlight the potential value of
early dark energy studies that fully explore 
degeneracies with other parameters that affect the CMB power
spectrum shape.

\subsection{Decaying Dark Matter}
\label{sec:ddm}

\begin{figure}
  \centering
    \includegraphics[width=\linewidth]{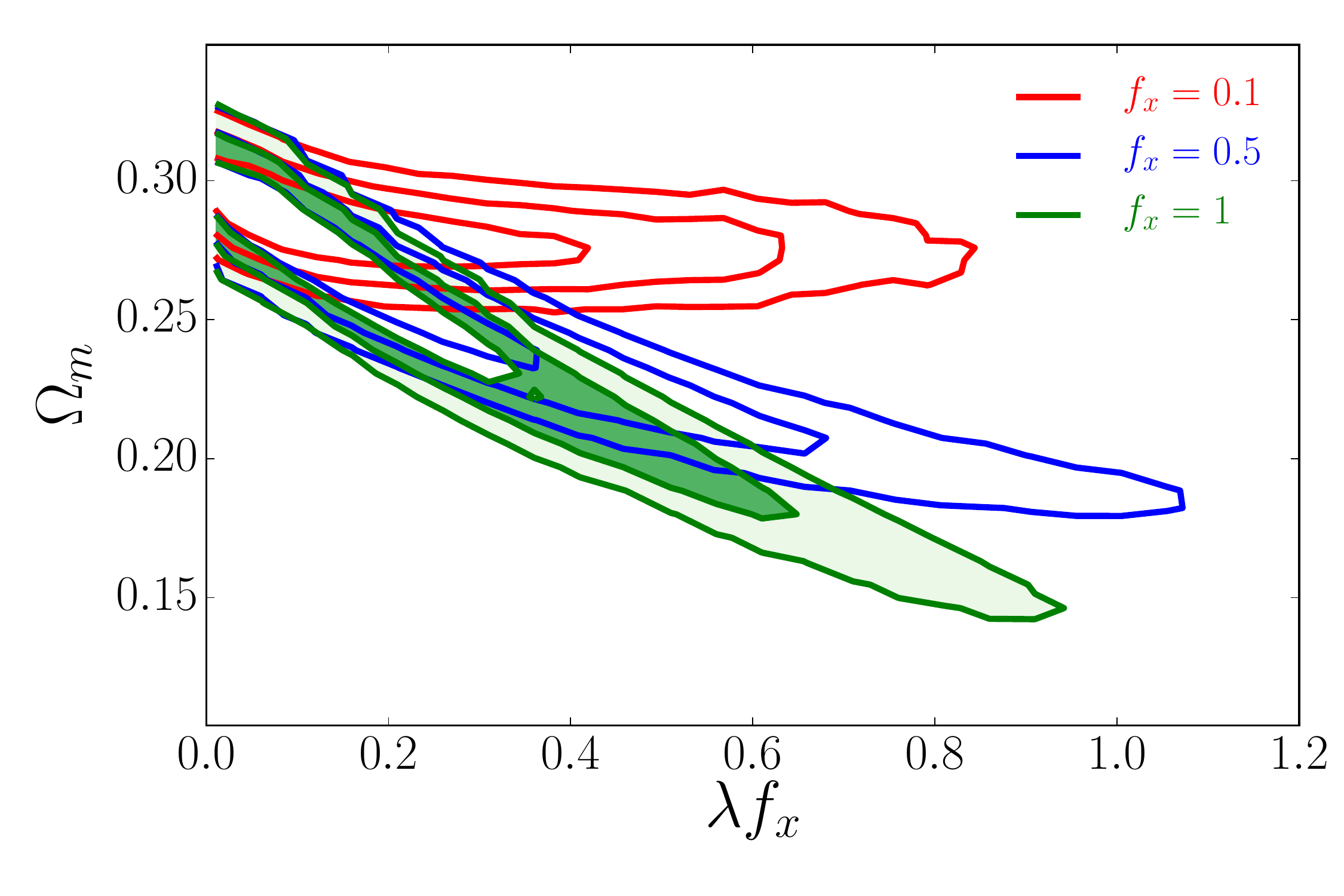} 
    \caption{Constraints on decaying dark matter from CMB, SN and
      galaxy+\lyaf\ BAO data.  For various choices of the fraction
      $f_x$ of dark matter in the decaying component, we plot
      posterior probability distributions for the product $\lambda
      f_x$, where $\lambda$ is the decay constant, assuming a flat
      prior on $\lambda$ vs value of $\Omega_m$ today (which, by
      definition, includes both the non-decaying component and the
      undecayed fraction of the decaying component).  Removing SN data
      does not significantly relax these constraints. }
  \label{fig:decay}
\end{figure}

If dark matter is a metastable particle that decays into undetected
radiation on a timescale comparable to $H_0^{-1}$, where the
undetected radiation could be neutrinos or some other low-mass
particle that interacts weakly enough to avoid detection (note that
decay of a significant fraction of the dark matter into photons would
need to have a very small branching ratio to be consistent with upper
limits on cosmic backgrounds), then the matter density will decrease
faster at low redshift than simple $(1+z)^3$ dilution (for an early
discussion, see \cite{Turner1985}).  While the radiation density is
boosted by dark matter decay, it subsequently decreases as $(1+z)^4$,
so the total energy density at low redshift is lower in a decaying
dark matter (DDM) model than it would be for stable dark matter with
the same high-redshift density.  We initially considered this model as
a potential explanation of the low $H(2.34)$ inferred from the \lyaf\
BAO.  The heights of the acoustic peaks constrain the value of
$\Omega_m h^2$ at the recombination epoch, but the reduced matter
density at low redshift implies a lower value of $H^2(z) = (8\pi G
/3)\rho_{\rm crit}(z).$ The sound horizon scale $r_d$ is unchanged
because the pre-recombination densities are unchanged.  However, the
full impact of introducing DDM is complex, because the values of $h$
and $\Omega_m$ must change to keep $D_M(1090)/r_d$ at its precisely
measured value, and because these changes and the dark matter decay
itself affect the galaxy BAO observables and the \lyaf\ value of
$D_M(2.34)$.

We assume exponential decay of the dark matter (i.e., a constant decay
rate), so that the governing equations for the decaying matter density
(marked with subscript $x$) and decay products' radiation density
(marked with subscript $g$) are:
\begin{eqnarray}
 \dot \rho_x &=& -3H\rho_x -\lambda H_0 \rho_x \label{eq:rhox}, \\
 \dot \rho_g &=& -4H\rho_g + \lambda H_0 \rho_x \label{eq:rhor}.
\end{eqnarray}
The decay rate $\lambda$ is dimensionless, and $\lambda^{-1}$ is the
decay time in units of $H_0^{-1}$.  However, we allow for the
possibility that there are two kinds of dark matter, only one of which
is susceptible to decay, so we introduce an additional parameter $f_x
= \lim_{z \rightarrow \infty} \Omega_x(z)/\Omega_{\rm dm}(z)$ that is the
ratio of this decaying component to the total dark matter density in
the infinite past.\footnote{The earlier \texttt{arXiv} version of this
  paper had $f$ defined as a fraction of total matter density
  today.}  The other components of the model remain the same as
those used for \lcdm.  Initial conditions are chosen so that there is
no energy density in the decay product radiation in the infinite past.
We discuss some details of our solution technique in
Appendix~\ref{app:decaying-dark-matter}.

An important subtlety in this analysis is that that the CMB
peaks constrain the dark-matter density at the time of recombination,
and hence the $\omega_b$ and $\omega_c$ densities that we feed into
the compressed CMB likelihood corresponds to the densities the system
would have had if the decay did not take place. Of course, the
distance to the last scattering surface is still affected by the
changes in the expansion history due to decaying dark matter. 

Figure~\ref{fig:decay} shows two-dimensional posterior probability
distributions in the $\lambda f_x - \Omega_m$ plane for several values of
$f_x$. Although the data prefer no decaying dark matter, we see strong
degeneracies that extend to surprisingly large values of
$\lambda$.  For $f_x=1$, decay of nearly 50\% of the primordial dark matter 
is allowed at 95\% confidence.
As expected, $\Omega_m$ is negatively correlated
with $\lambda f_x$: the CMB constrains $\omega_c$ in the early
universe, and if more dark matter decays then $\Omega_m$ today is lower.
There is also a weak correlation with the Hubble parameter (not shown),
with $h$ rising by $\sim 0.01$ for $\lambda f_x \sim 0.5$.

To gain some understanding of this degeneracy, 
one can calculate the effective $w$ of the
fluid composed of the combined decaying dark matter and the resulting
radiation. At $f_x=1$, $\Omega_m=0.23$ and $\lambda=0.4$ (the edge of
our 68\% contour), the effective $w$ takes values $0.07$ at
$z=0$, falling to $0.03$ at $z=2$ and $10^{-4}$ at $z=100$. 
With no decay, this component (which starts at the same energy
density, as fixed by the CMB) would evolve with $w=0$.
The surprisingly small corrections to the total $w$ make it hard to
constrain the decaying dark matter from expansion history data alone.

Using combinations of CMB, SN, and large-scale structure data sets,
\cite{2009JCAP...06..005D} obtained a limit $\Gamma^{-1} > 100\,$Gyr
for the dark matter decay constant, using methodology similar to that
described here but also including constraints from the full shape of
the CMB power spectrum and the amplitude of matter clustering.  A more
recent analysis by \cite{2014arXiv1407.2418A}, using the Planck+WP CMB
power spectrum and BAO measurements from BOSS and WiggleZ, obtained a
somewhat stronger limit of $\Gamma^{-1} > 160\,$Gyr.  For a Hubble
parameter $h=0.68$, our limit for $f_x=1$ corresponds $\Gamma^{-1} >
28\,$Gyr at 95\% confidence level.  A more detailed analysis by
\cite{Blackadder2014} calculates the velocity distributions of
daughter particles for varying assumptions about the decay products.
For a daughter relativistic fraction of 1\% and higher, they find
$\Gamma^{-1}> 10\,$Gyr based on analysis of Union 2.1 SNIa data in the
context of CMB determined cosmological model. From these results we
conclude that, somewhat surprisingly, the expansion history alone is
not sufficient to significantly constrain the decay of dark matter
into an unknown relativistic component.

We note that \cite{2008PhRvD..77j3511G} state a limit
$\Gamma^{-1} > 700\,$Gyr based on only the CMB acoustic scale and
SNIa data available in 2008.  We do not understand how these more
limited data could lead to a stronger bound on the decay time,
which suggests that the analysis of \cite{2008PhRvD..77j3511G} 
contains a hidden assumption.  

\subsection{Massive Neutrinos}
\label{sec:mnu}

\begin{figure*}
  \centering
    \includegraphics[width=\linewidth]{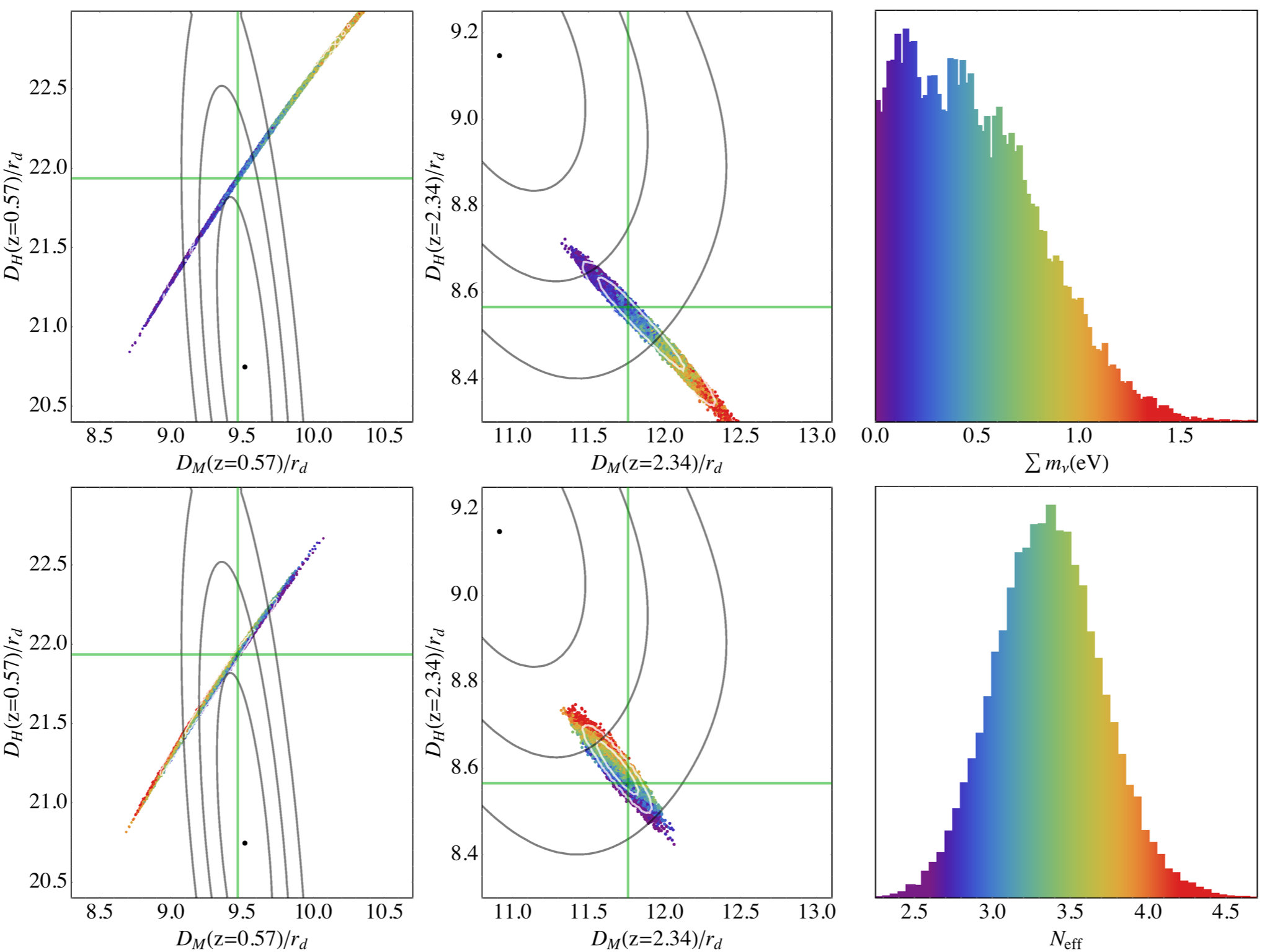}

    \caption{BAO constraints in the $D_M-D_H$ planes at $z=0.57$
      (left) and $z=2.34$ (middle) compared to predictions of
      CMB-constrained, flat \lcdm\ models in which the neutrino mass
      $\sum m_\nu$ or the number of relativistic species $\neff$ is a
      free parameter.  Black curves show 68\%, 95\%, and 99.7\%
      likelihood contours from the CMASS and \lyaf\ BAO measurements,
      relative to the best-fit values (black dots).  Colored points
      represent individual models from Planck+WP+ACT/SPT MCMC chains,
      which are color-coded by the value of $\sum m_\nu$ (top row) or
      $\neff$ (bottom row) as illustrated in the right panels.  Green
      cross-hairs mark the predictions of the flat \lcdm\ model with
      $\sum m_\nu = 0.06\,{\rm eV}$ and $\neff = 3.046$ that best fits
      the CMB data.  White curves show 68\% and 95\% likelihood
      contours for the CMB data alone.  CMB results in the top row are
      marginalized over the lensing parameter $A_L$.  }
  \label{fig:chain_vs_contour2}
\end{figure*}

\begin{figure}
  \centering
    \includegraphics[width=\linewidth]{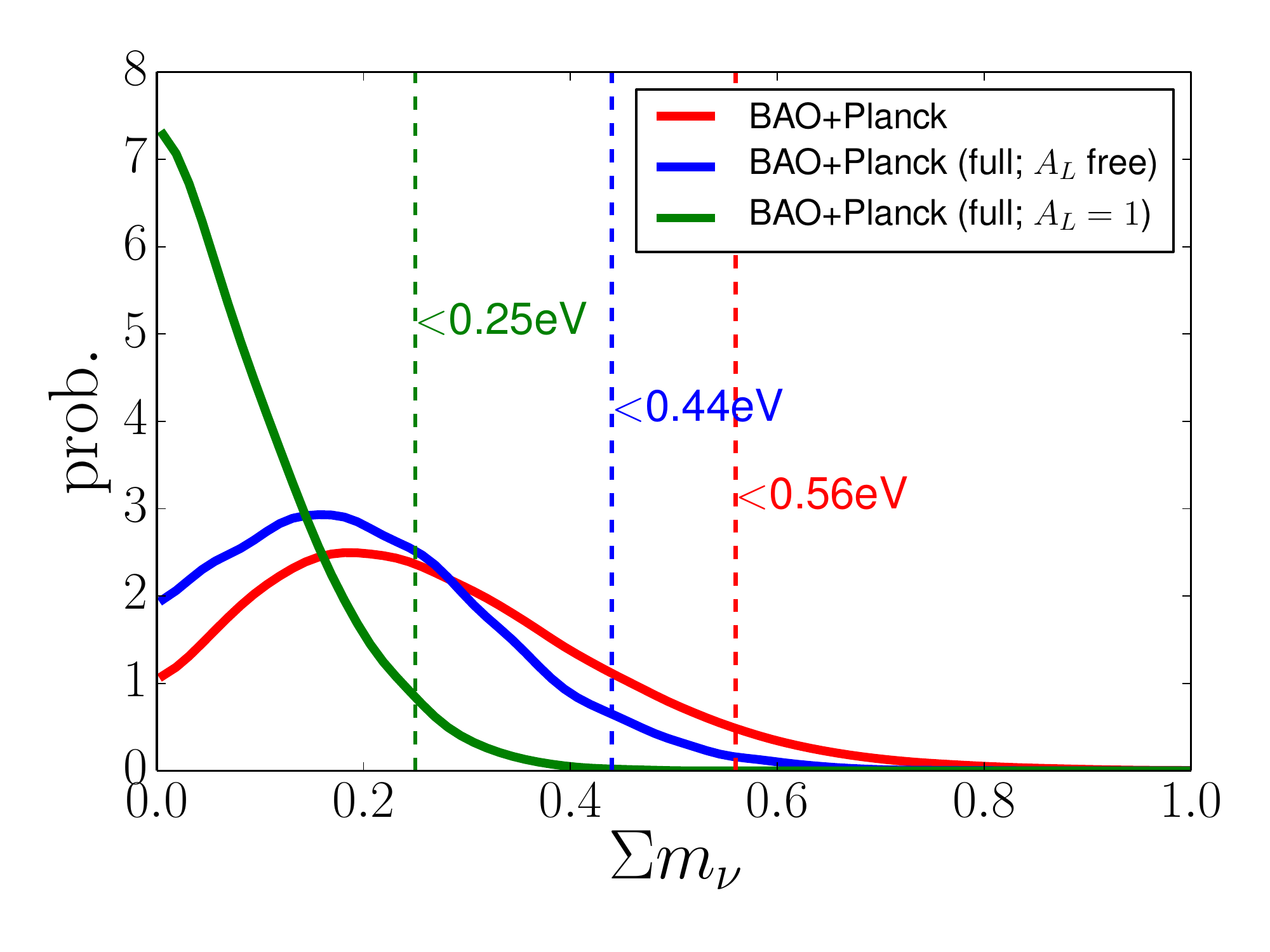} 
    \includegraphics[width=\linewidth]{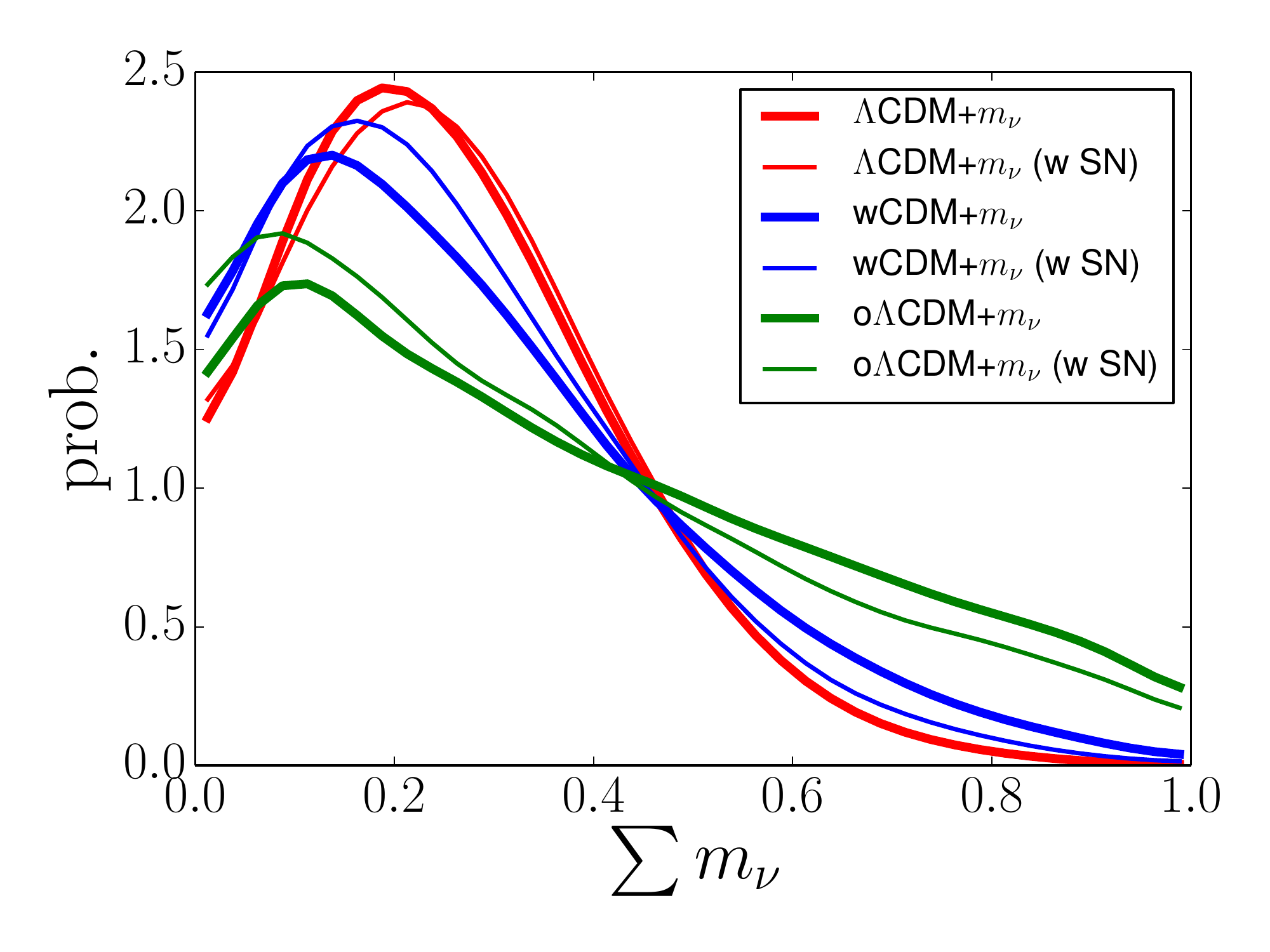} 

    \caption{Neutrino mass constraints for several combinations of
      data and model freedom.  In the top panel, the red curve shows
      the posterior pdf (with a flat prior) on $\sum m_\nu$ from the
      expansion history constraint on the neutrino mass, based on the
      combination of BAO with our compressed CMB description, which
      yields $\sum m_\nu < 0.56\eV$ at 95\% confidence.
      The green curve shows the result obtained by replacing our
      compressed CMB description with the full Planck+WP power spectrum
      using \texttt{cosmomc}, which strengthens the upper limit to
      $\sum m_\nu < 0.25\eV$.  The blue curve adopts the same data combination
      but additionally marginalizes
      over the parameter $A_L$, demonstrating that
      the difference between the green and red curves is driven mainly
      by the lensing amplitude information in the Planck data.  In the
      lower panel, curves show the posterior pdf of $\sum m_\nu$ from
      our usual BAO+Planck (thick) or BAO+SN+Planck (thin) geometrical
      constraints, assuming \lcdm, \wcdm, or \olcdm\ (red, blue,
      green, respectively). }
  \label{fig:mnu}
\end{figure}

In addition to constraining dark energy and space curvature, measuring
neutrino masses is a key objective of precision cosmology.  Given CMB
constraints that $\omega_m \approx 0.14$, the fractional contribution of
neutrinos to the low-redshift
matter density is $\omega_\nu/\omega_m \approx 0.07
(\sum m_\nu / 1\,{\rm eV})$, so neutrino masses have a
noticeable cosmological impact if they are a significant fraction of
an eV.  Atmospheric and laboratory measurements constrain the mass
splittings among the three standard-model neutrino species to be 
$m_2^2-m_1^2=7.54^{+0.26}_{-0.22} \times 10^{-5}$ eV$^2$ and
$m_3^2- (m_1^2+m_2^2)/2= \pm 2.43^{+0.06}_{-0.10} \times 10^{-3}$
eV$^2$ \cite{Fogli:2012ua}.  This sets the minimum neutrino mass in a
normal hierarchy, where $m_1 < m_2 \ll m_3$, to $\sum m_\nu =
58.4^{+1.2}_{-0.8}\rm\ meV$, which motivates our assumption that $\sum
m_\nu=0.06\eV$ in the standard \lcdm\ model.  In the case of an inverted
hierarchy, where $m_1\simeq m_2 \gg m_3$, the sum of the neutrino
masses must exceed $0.1\eV$, and for the degenerate neutrino
mass case where $m_1 \simeq m_2 \simeq m_3$, the minimum mass sum
is approximately $0.15\eV$.  These masses are well within
reach of the cosmological experiments in the coming decade.

Neutrinos affect the CMB and large-scale structure differently from
cold dark matter because they are still relativistic at the epoch of
matter-radiation equality, because their linear clustering is
suppressed on scales below $k_{\rm sup} = 2\pi/\lambda_{\rm sup} =
0.018 \sqrt{m_\nu/1{\rm eV}} \hmpc$ \cite{Lesgourgues2006}, and
because their high thermal velocities prevent them from clustering in
small potential wells even in the non-linear regime
\cite{LoVerde2014,LoVerde2014b}. The relative suppression in the
linear matter power spectrum is linear in the fraction density in
neutrinos $f_\nu$ and is about $\Delta P / P \sim -8 f_\nu \sim
0.063 \left(\sum m_\nu/0.1{\rm eV}\right)$ (at best fit values of
$\Omega_m$ and $h$).

Measurements of matter clustering can constrain $\sum m_\nu$
by detecting the suppression of small scale power.
Expansion history measurements, which we focus on here,
can constrain $\sum m_\nu$ because of their transition from a
relativistic species whose energy density scales as $(1+z)^4$
to a non-relativistic species whose energy density scales as $(1+z)^3$,
effectively the converse of decaying dark matter.
Specifically, the CMB acoustic peaks constrain $\omega_{cb}$
almost independently of $\omega_\nu$, but the matter density that
affects late-time expansion rates and distances is $\omega_{cb}+\omega_\nu$.
With other parameters held fixed, a $0.2\eV$ neutrino mass-sum increases the 
late-time matter density by 1.4\%, which is significant given
the extremely precise CMB measurement of $D_M(1090)/r_d$ and
precise distance scale measurements from BAO.  In practice, changing
$\sum m_\nu$ leads to adjustments in other parameters to seek
a global best fit.  Although the neutrino mass influences the
sound horizon (eq.~\ref{eqn:rd}), this impact is small for
the range of $\sum m_\nu$ allowed by our constraints
($-0.26\%$ for $\sum m_\nu = 0.5\eV$).

The top row of Figure~\ref{fig:chain_vs_contour2} compares the CMASS
and \lyaf\ BAO constraints to the predictions of CMB-constrained, flat
\lcdm\ models with $\sum m_\nu$ as a free parameter, in the same
format as Figure~\ref{fig:chain_vs_contour1}.  The Planck CMB chain used 
here is {\tt base\_mnu\_planck\_lowl\_lowLike\_highL\_Alens},
where we have selected a chain that
marginalizes over the lensing amplitude parameter $A_L$ for
reasons discussed below.
The base Planck \lcdm\ model
adopts $\sum m_\nu = 0.06\eV$, but this chain allows $\sum m_\nu$ down to
zero.  While $\sum m_\nu > 0.5\eV$ is allowed by the CMB data alone,
this mass significantly worsens agreement with the BAO data, both at $z=0.57$
and at $z=2.34$.  Higher $m_\nu$ increases $\Omega_m h^2$ and thus
decreases $D_H(2.34)$, moving further from the \lyaf\
measurement.  Additionally, because of the tight CMB constraint on
$D_M(1090)/r_d$, the reduction in $c/H(z)$ at high redshift must be
compensated by changes in $h$ and $\Omega_m$ that raise $D_M$ at low
redshift, so $D_M(2.34)$, $D_M(0.57)$, and $D_H(0.57)$ all
increase, again moving away from the BAO measurements.  Conversely,
moving towards $\sum m_\nu = 0$ slightly improves agreement with the
\lyaf\ BAO because $D_H(2.34)$ increases while $D_M(2.34)$, which
has a large contribution from lower redshifts, decreases.  However,
the same change worsens agreement with the CMASS BAO, which are
already well fit by the base model with $\sum m_\nu = 0.06\eV$.

The red curve in the upper panel of Figure~\ref{fig:mnu} shows
the purely geometric constraint that arises from combining just
the compressed CMB description with galaxy and \lyaf\ BAO data.
This constraint is surprisingly tight at $\sum m_\nu
<0.56\eV$ (at 95\% c.l.), and it is independent of mass constraints 
based on the suppression of structure
growth by neutrino free-streaming.  Adding
SN data does not significantly improve this constraint.  The
geometrical constraint on neutrino mass weakens if we allow either
curvature or $w \neq -1$, as shown by the blue and green curves in the
lower panel of Figure~\ref{fig:mnu}.  Because neutrinos influence
the observables only via the effect of $\omega_m$ on
distances and expansion rates, adding another degree of freedom
introduces degeneracy.  In these cases, including SN data 
does improve the neutrino mass constraint,
as shown by the thin curves.  The improvement
is not dramatic, indicating that our multiple BAO measurements can
break the degeneracy themselves to a significant degree.

The 95\% upper limits on $\sum m_\nu$ for the various models and data
combinations we have considered are listed in Table~\ref{tab:mnu}.
Direct measurements of matter clustering at low redshift can be a
powerful diagnostic of neutrino masses because they are sensitive to
the distinctive effect of suppressing small scale power, which is not
easily mimicked by other parameter variations.  We return to this
point in Section~\ref{sec:growth}.

In the top panel of Figure~\ref{fig:mnu}, 
the green curve shows the result of full
fitting using the \texttt{cosmomc} machinery.  The constraint
tightens significantly to $\sum m_\nu<0.25$eV (at 95\% c.l.).  This
number is the same as the Planck collaboration constraint
for Planck+BAO \cite{PlanckXVI}, even though we
use more and better constraining BAO data.  This exercise demonstrates
that the compression into $\omega_{cb}$, $\omega_b$, and
$D_M(1090)/r_d$ is missing important information that the CMB
provides on neutrino mass. The extra information is in the constraint
on the amplitude of low-redshift matter clustering that comes from the
smoothing of the acoustic peaks in the CMB power spectrum
by weak gravitational lensing.
The BAO information in this case fixes the
matter density of the universe, thus allowing inference on the
amplitude of matter fluctuations and hence neutrino mass to be
determined from the smoothness of the peaks (the CMB only constraint
is $\sum m_\nu <0.93$eV \citep{PlanckXVI}). The role of CMB lensing is further
demonstrated by the blue curve in the top panel.
Here we have run a \texttt{cosmomc} chain that, in addition to having
$\sum m_\nu$ as a free parameter, marginalizes over a parameter $A_L$
that multiplies the predicted lensing signal, effectively removing the
lensing information. (The base model fixes $A_L=1$.)  The
result here is very similar to that found by using the compressed
CMB description.

For \lcdm, fitting the CMB temperature power spectrum from Planck
(together with WMAP polarization and high-$l$ data from ground-based
experiments) yields $A_L = 1.23 \pm 0.11$ \citep{PlanckXVI}, showing
that the lensing signal measured in the Planck power spectrum is
significantly stronger than the predicted signal based on
extrapolating the observed CMB fluctuations forward in time.  A larger
neutrino mass suppresses the low-redshift clustering, exacerbating
this tension, which is why the $\sum m_\nu$ limit is considerably
tighter when $A_L$ is fixed to unity.  However, the Planck measurement
of lensing through the CMB 4-point function does agree with $A_L=1$
\citep{PlanckXVI}.  These internal tensions on the lensing signal
within the CMB data alone suggest that one should be cautious in using
them to constrain $\sum m_\nu$.  The red curve is thus a more
conservative inference, using only geometric constraints plus the CMB
constraints on $\omega_{cb}$ and $\omega_b$.  It is impressive that
even a 3.5\% contribution of neutrinos to $\omega_m$ ($\sum m_\nu =
0.5\eV$) is enough to be substantially disfavored by these
expansion history measurements.

\begin{table}
  \centering
  \begin{tabular}{l|c|c}
    Combination & Model & 95\% limit on $\sum m_\nu$ \\
    \hline
BAO+Planck(full) & \lcdm &  0.25eV \\ 
BAO+Planck(full) & \lcdm\ + free $A_L$ &  0.43eV \\ 
\smallskip
BAO+Planck & \lcdm &  0.56eV \\ 
BAO+Planck & \wcdm &  0.68eV \\ 
BAO+Planck & \olcdm &  0.87eV \\ 
BAO+SN+Planck & \lcdm &  0.56eV \\ 
BAO+SN+Planck & \wcdm &  0.61eV \\ 
BAO+SN+Planck & \olcdm &  0.84eV \\ 
  \end{tabular}
  \caption{95\% confidence limits for the sum of the mass of neutrino species.
  The first two cases use full Planck CMB chains, while other cases adopt
  our compressed CMB description.}
  \label{tab:mnu}
\end{table}

\subsection{Extra Relativistic Species}
\label{sec:neff}

If the universe contains extra relativistic degrees of freedom
beyond those in the standard model, these increase the expansion
rate during the radiation dominated era and shift the epoch of
matter-radiation equality, thereby altering the sound horizon,
the shape of the matter power spectrum, and the history of
recombination.  Extra radiation is usually parameterized by
the quantity $\Delta\neff$, where $\Delta \neff=1$ corresponds to 
the amount of radiation that an extra massless
thermalized neutrino species (i.e., a fermion that thermally decouples
before electron-positrion annihilation) would produce.
In general, however, there is no requirement that 
$\Delta \neff$ be an integer.
The standard model has $\neff=3.046$ and $\Delta\neff=0$.

\begin{figure}
 \centering
    \includegraphics[width=\linewidth]{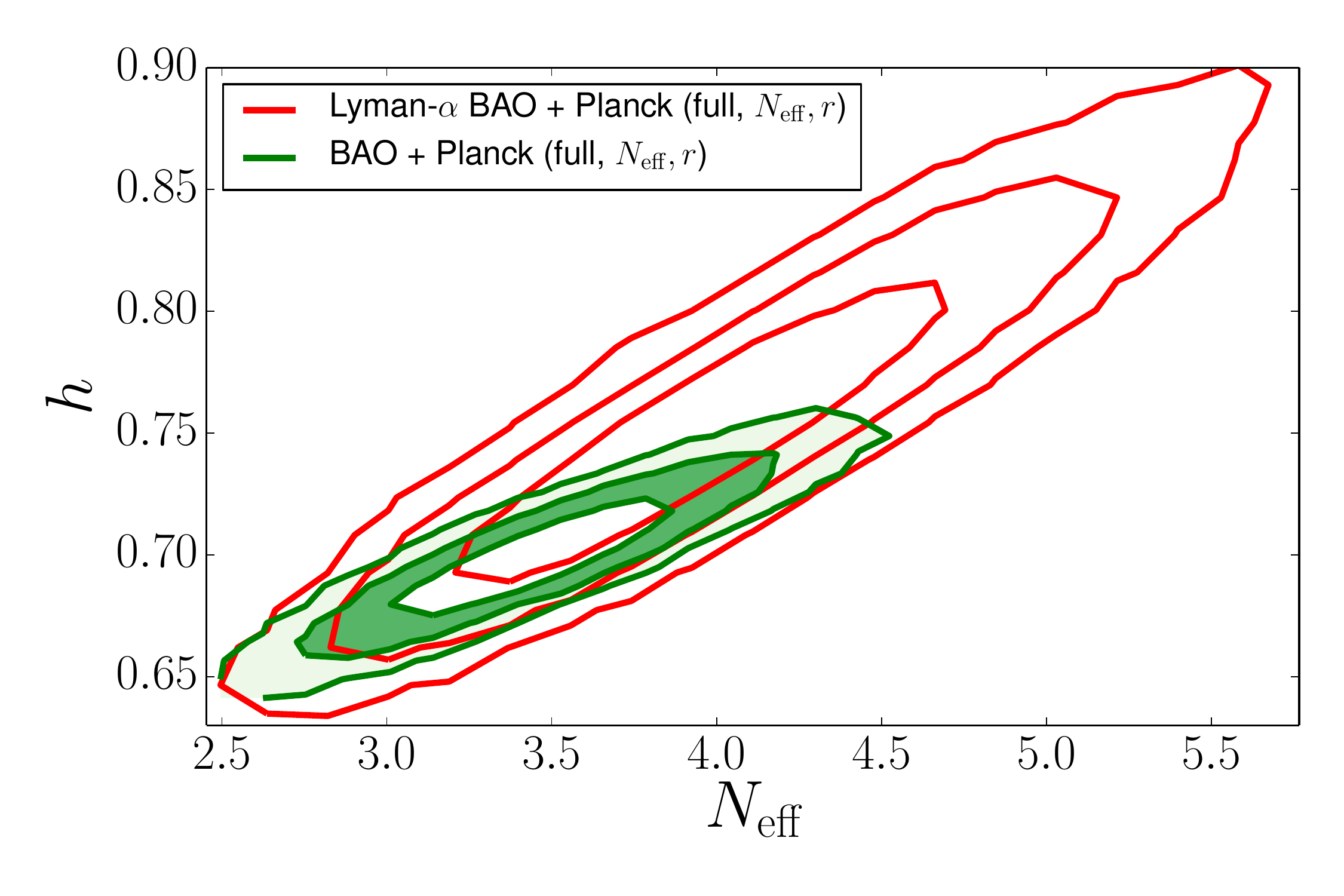} 
  \caption{Constraints on the effective number of relativistic
  species $\neff$, assuming $w=-1$ and $\Omega_k=0$.
  These contours use full Planck+WP CMB constraints
  computed with {\tt cosmomc}, combined with \lyaf\ BAO
  only (red) or with our full set of BAO measurements (green).
  We marginalize over the tensor-to-scalar ratio $r$.
  }
  \label{fig:neff}
\end{figure}

The bottom row of Figure~\ref{fig:chain_vs_contour2} shows the
probability distribution for $\neff$ from CMB data alone
(right panel, from the chain 
\texttt{base\_nnu\_planck\_lowl\_lowLike\_highL}).
which peaks at $\neff \approx 3.4$, with a 95\% confidence 
range $2.7 \leq \neff \leq 4.04$.  
The middle panel shows that values of $\neff$ at the upper
end of this range can noticeably improve consistency with
the \lyaf\ BAO measurement, pushing the predicted values of
$D_H(2.34)/r_d$ up and $D_M(2.34)/r_d$ down so that
they lie within the 95\% likelihood contour of the BAO data.
However, high values of $\neff$ reduce the predicted
values of $D_H/r_d$ and $D_M/r_d$ at $z=0.57$, worsening
agreement with the galaxy BAO measurements.

We can understand these trends by arguments similar to those given for
early dark energy in Section~\ref{sec:ede}.  An increase of $\Delta\neff=1$
reduces the sound horizon $r_d$ by 3.2\% (eq.~\ref{eqn:rdneff})
because of the higher expansion rate in the early universe.
Maintaining the precisely measured value of $D_M(1090)/r_d$ requires
changes in $\Omega_m$ and $h$ to reduce $D_M(1090)$ by the same
factor.  Some of this reduction can be accomplished by raising
$\Omega_m h^2$, and thus raising $H(z)$ at high redshift, but
$\Omega_m h^2$ is already tightly constrained by the heights of the
acoustic peaks.  Therefore, the fractional change to $D_H(2.34)$ is
much lower than the fractional change to $r_d$, and the value of
$D_H(2.34)/r_d$ rises.  To maintain $D_M(1090)/r_d$, the value
of $H_0$ (which controls the low-redshift contribution to the $D_M$
integral) must increase by {\it more} than the drop in $r_d$.  Because
$D_M(2.34)$ is an integral over all $z < 2.34$, the ratio
$D_M(2.34)/r_d$ drops even as $D_H(2.34)/r_d$ rises.  At $z=0.57$,
where the value of $H(z)$ retains sensitivity to $H_0$, {\it both}
$D_H/r_d$ and $D_M/r_d$ drop as $\neff$ increases. 

A change in $\neff$ has multiple effects on the CMB, which renders
our compression into a 3-variable matrix questionable.
For this section of the paper, therefore, we have run \texttt{cosmomc}
chains using the full Planck+WP CMB information, 
while still treating the BAO
data as measurements of $D_M/r_d$ and $D_H/r_d$. 
We have checked that using our \texttt{simplemc} chains,
which adopt the compressed CMB description, yields qualitatively
similar but quantitatively different results.
We assume a flat universe with a cosmological constant,
but we treat the tensor-to-scalar ratio $r$ as a free parameter
and marginalize over it, since there is no theoretical reason
to expect $r=0$.  While allowing free $r$ would not alter the
compressed CMB constraints used elsewhere in the paper, it
has an impact here because $r$ and $\neff$ have partially
degenerate effects on the shape of the CMB power spectrum.

Red contours in Figure~\ref{fig:neff} show confidence intervals
in the $\neff-h$ plane obtained by 
combining the CMB data with \lyaf\ BAO alone.
The allowed range of $\neff$
is larger here than in Figure~\ref{fig:chain_vs_contour2} because
we do not fix $r=0$.  
As one would expect from Figure~\ref{fig:chain_vs_contour2},
the addition of \lyaf\ BAO pulls the preferred value of $\neff$
upward, with a best-fit value of $\neff \approx 4$ and $\neff=3$
significantly disfavored.  
However, the galaxy BAO measurements prefer $\neff \approx 3$ and
have higher precision, so when they are added (green contours)
the allowed range shifts downward to $\neff = 3.43 \pm 0.26$ (68\%)
or $\neff = 3.43 \pm 0.53$ (95\%).
Higher $\neff$ correlates with higher $h$ for the reasons discussed above,
and with $\neff$ and $r$ 
as an additional degree of freedom our BAO+CMB constraint
is $h = 0.71 \pm 0.017$.
Adding SN data does not significantly shift these contours
once galaxy BAO are included.
As shown in Figure~\ref{fig:chisq}, introducing $\Delta\neff$ as a
parameter reduces $\chi^2$ by 0.75 relative to \lcdm, with nearly
all of the change coming from a slightly better fit to the \lyaf\
BAO data.

One complication in constraining $\Delta\neff$ models is that
changing the radiation density can alter the broadband power
spectrum shape enough to affect the BAO fitting procedure itself.
\cite{Anderson14} find that the compression of BAO data into
the $\apar-\aperp$ plane (in other words, the fact that the
inferred values of $D_M/r_d$ and $D_H/r_d$ are independent
of the adopted fiducial model) breaks down in the presence of extra
radiation at a $\sim 0.4$\% level.  This effect is negligible
compared to the statistical errors in the \lyaf\ BAO data, but
it is not completely negligible relative to the galaxy BAO errors.
A more exact treatment of the $\neff$ constraints therefore
requires refitting the BAO data themselves, but we would expect
only small shifts relative to the constraints reported here.
We plan to revisit this question when the final BOSS measurements
are available.

\subsection{A Tuned Oscillation}
\label{sec:tun}

\begin{figure}
  \centering
    \includegraphics[width=\linewidth]{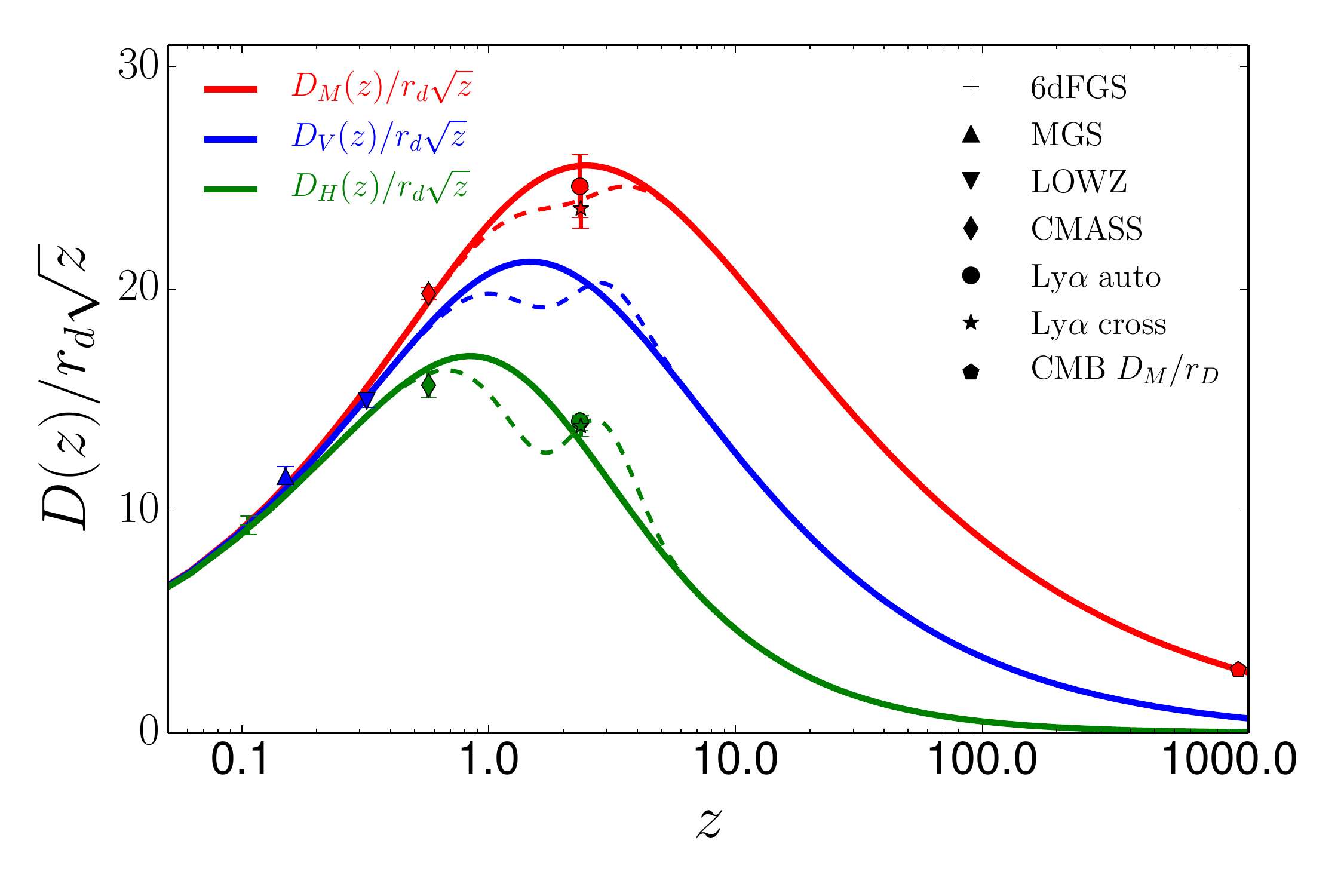} 
  \caption{A ``tuned oscillation'' model in which a Gaussian perturbation of
  the \lcdm\ $D_M(z)$ is introduced to allow a good simultaneous fit 
  to the galaxy and \lyaf\ BAO data.
  Solid lines show the same \lcdm\ model plotted in 
  Figure~\ref{fig:hubblediagram}, while the dashed line
  shows the perturbed model.
  }
  \label{fig:weird}
\end{figure}

\begin{figure}
  \centering
    \includegraphics[width=\linewidth]{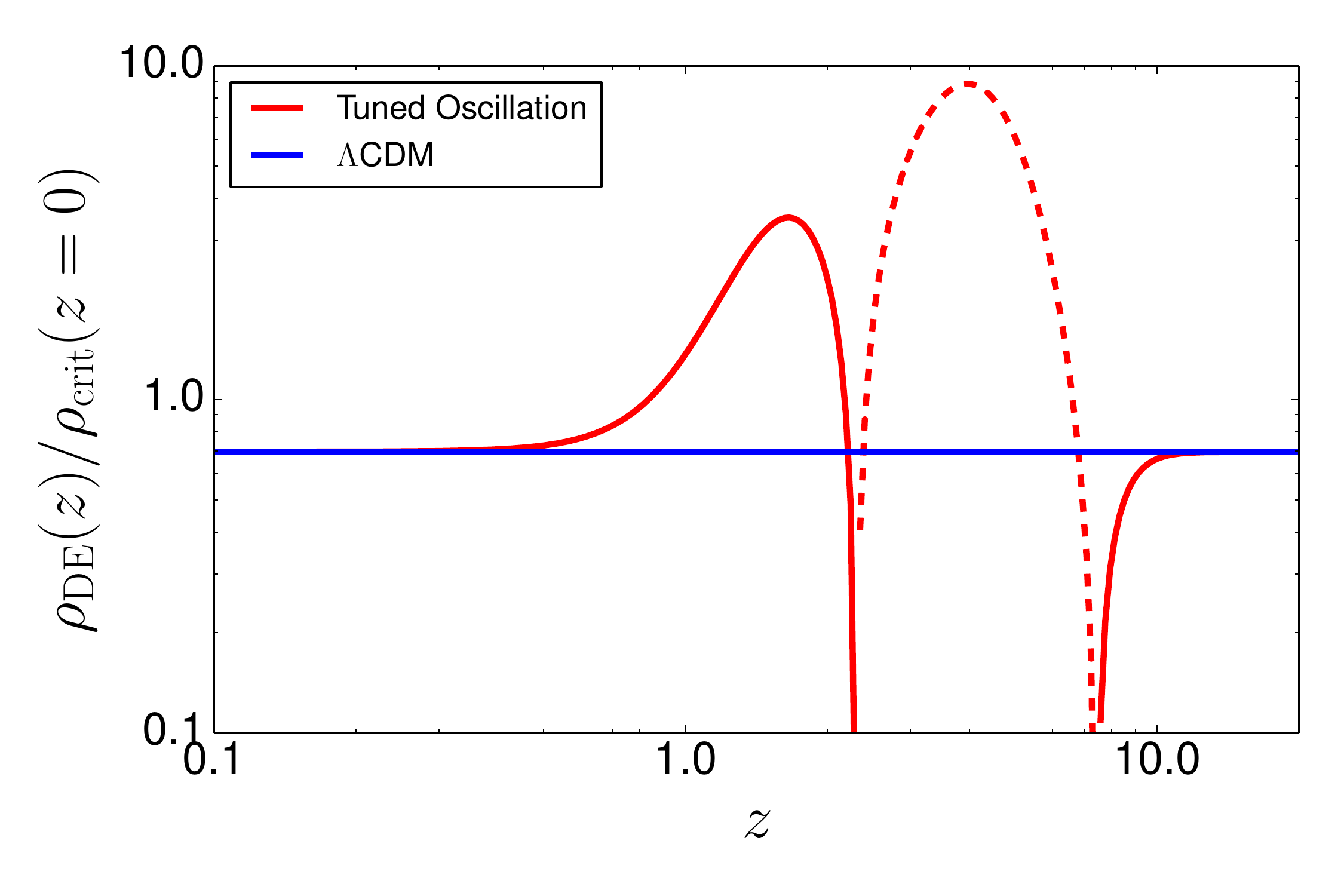} 
    \caption{Implied variation in the energy density of the dark
      energy component for the model shown in Figure
      \ref{fig:weird}. The dotted line corresponds to the density becoming
      negative. These plots illustrate the difficulty of concurrently
      fitting the \lyaf\ and galaxy BAO constraints on $D_M$ and $D_H$ 
	  while satisfying the CMB constraint on $D_M$. 
  }
  \label{fig:weird2}
\end{figure}

While the $\Delta\neff$ model moderately reduces tension with the
\lyaf\ BAO data, {\it none} of the models we have considered
produces a truly good fit to all the measurements.  To understand
what is required to achieve a good fit, we have constructed an artificial
model that maintains the mathematical link between $D_M(z)$ and $D_H(z)$
but has the freedom needed to fit all the BAO and CMB data
at the $\approx 1\sigma$ level.  
We consider a flat universe whose angular diameter
distance is given by 
\begin{equation}
  D_M^{(w)} (z) = D_M^{(\Lambda{\rm CDM})} (z) \left[1+A \times G(\log z;
    \log z_o, \sigma_o) \right],
\end{equation}
where $G(\log z; \log z_m,\sigma_o)$ denotes a Gaussian in $\log z$
with mean $z_o$ and variance $\sigma_o$ and $A$ is an amplitude
parameter. It is clear that such model can, for a sufficiently
localized Gaussian perturbation, fit the low-redshift and CMB data
with sufficient goodness of fit. Given three parameters it has
enough flexibility to also match our $z=2.34$ points.  
Figure~\ref{fig:weird} plots the best fitting model of this form.
The Hubble parameter undergoes an oscillation between 
$z \approx 4$ and $z \approx 0.8$, which allows it to
match the \lyaf\ and CMASS values of $D_H$ and to 
change $D_M$ at $z=2.34$ without altering the low and high
redshift values.

This model reduces the overall $\chi^2$ by 6.6 with three extra degrees of
freedom, a considerable improvement over any other model we have
investigated (see Figure~\ref{fig:chisq}).
Generically, any small perturbation to the Friedmann equation 
that is able to improve fits to our \lyaf\ data runs afoul of
CMB and/or galaxy BAO constraints. 
This model works because it is fine-tuned to change distances
near $z=2$ but not upset the distance to the last-scattering surface.
However, the model is physically extreme,
as demonstrated in Figure~\ref{fig:weird2}, where we have
converted $H(z)$ into an implied dark energy density
via the Friedmann equation.  Producing the desired oscillation
in $H(z)$ requires a {\it negative} $\rhoDE$ between $z=6$ and $z=2$
(see eq.~\ref{eqn:edeHz}).
\cite{Sahni2014} argue that the BOSS \lyaf\ measurements may be
explained in a modified gravity model that alters the 
Friedmann equation itself in a physically motivated way, but
more work is needed to determine whether any such model can
provide a good fit to all of the BAO measurements while
satisfying CMB constraints.

The difficulty in finding a well motivated model that matches
the BOSS \lyaf\ measurements suggests that the tension with 
these measurements may be a statistical fluke, or a consequence
of an unrecognized systematic that either biases the central values
of $D_M(z)$ and $D_H(z)$ or causes their error bars to be
underestimated.  Analyses of the final BOSS data set will
address both of the latter points, as they will allow more
exhaustive investigation of analysis procedures and tests 
against larger suites of mock catalogs.
Addressing the first point will require high-redshift
BAO measurements from new data sets, such as the
\lya\ emission-line galaxy survey of HETDEX \cite{Hill2004}
or a much larger \lyaf\ sample from DESI \cite{Levi2013}.

\section{Comparison to Structure Growth Constraints}
\label{sec:growth}

The Planck cosmology papers highlighted a tension
between the predictions of the CMB-normalized \lcdm\ 
model and observational constraints on matter clustering
at low redshifts, from cluster abundances, weak gravitational lensing,
or redshift-space distortions.  We now revisit this issue
with our updated BAO and SN constraints and our broader
set of models, to see whether these tensions persist and
whether they are significantly reduced in some classes of models.

\begin{figure*}
  \centering
  \includegraphics[width=\linewidth]{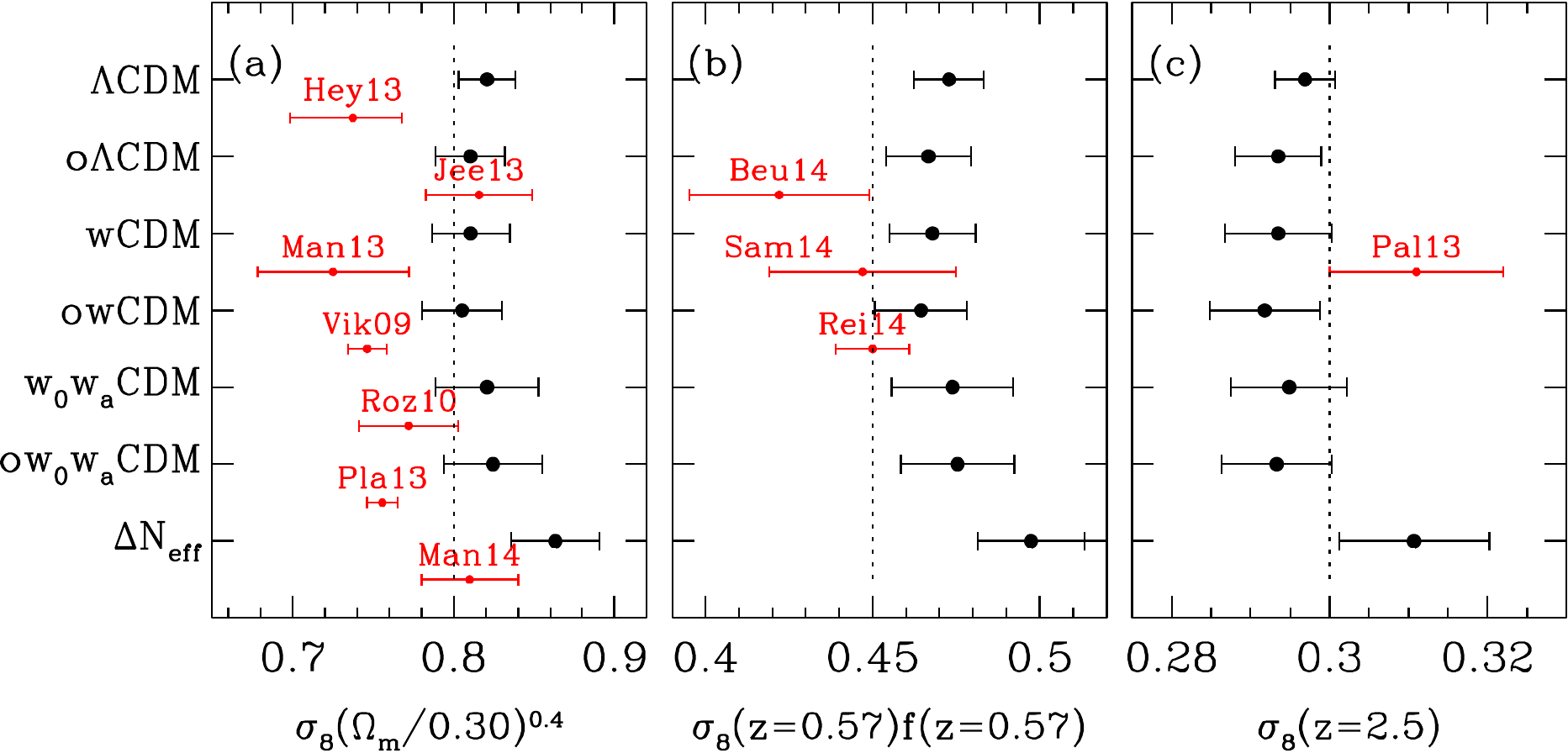}
    \caption{Predictions of matter clustering from our BAO+SN+CMB
    constrained models compared to observational estimates.
	The vertical location of the observational estimates (red points)
	is arbitrary.
	Labels for the model points (black) in all panels
    are indicated along the left vertical axis.
    Panel (a) shows the $z=0$ parameter combination 
    $\sigma_8(\Omega_m/0.3)^{0.4}$, which approximately describes
    the quantity best constrained by low-redshift measurements
    of the cluster mass function or weak lensing.
    Black points show the mean and $1\sigma$ range computed from
    our model chains, and red points show observational estimates
    discussed in the text.  Panel (b) presents a similar
    comparison for $\sigma_8(z=0.57)f(z=0.57)$, constrained by
    redshift-space distortions in CMASS galaxy clustering.
    Panel (c) compares $\sigma_8(z=2.5)$ to an estimate
    from the BOSS \lyaf\ 1-d power spectrum.
    Observational sources are the cosmic shear measurements of
    Hey13 \cite{Heymans13} and Jee13 \cite{Jee13}, 
	the galaxy-galaxy lensing measurement of
    Man13 \cite{Mandelbaum13}, the cluster mass function measurements of
    Vik09 \cite{Vikhlinin09}, Roz10 \cite{Rozo10}, Pla13 \cite{PlanckXX}, 
    and Man14 \cite{2014arXiv1407.4516M}, the RSD measurements of
    Beu14 \cite{Beutler14}, Sam14 \cite{2014MNRAS.439.3504S},
    and Rei14 \cite{Reid14}, and the \lyaf\ 
    power spectrum measurement of Pal13 \cite{Palanque13}.
    Dotted vertical lines are provided for visual reference.
  }
  \label{fig:growth}
\end{figure*}

Low-redshift measurements of cluster abundances and weak lensing 
most tightly constrain the parameter combination $\sigma_8\Omega_m^\alpha$
with $\alpha \approx 0.4-0.6$ (see discussions in \cite{Weinberg13}
and references therein).
As representative but not exhaustive
examples of constraints at $z \approx 0$ we adopt:
$\sigma_8(\Omega_m/0.27)^{0.46} = 0.774_{-0.041}^{+0.032}$
from tomographic cosmic shear measurements in the CFHTLens 
survey \cite{Heymans13};
$\sigma_8(\Omega_m/0.27)^{0.5} = 0.86 \pm 0.035$ from 
cosmic shear measurements in the 
Deep Lens Survey\footnote{The authors do not quote their results
in this form, so this constraint has been eyeballed from
their Fig.~25} \cite{Jee13};
$\sigma_8(\Omega_m/0.27)^{0.57} = 0.77 \pm 0.05$ from
the combination of galaxy-galaxy lensing and galaxy clustering
in the SDSS \cite{Mandelbaum13};
$\sigma_8(\Omega_m/0.25)^{0.47} = 0.813 \pm 0.013$ from the
mass function of X-ray clusters observed with {\it Chandra}
and {\it ROSAT} \cite{Vikhlinin09};
$\sigma_8(\Omega_m/0.25)^{0.41} = 0.832 \pm 0.033$
from stacked weak lensing of clusters in the SDSS \cite{Rozo10};
and 
$\sigma_8(\Omega_m/0.27)^{0.3}=0.78 \pm 0.01$ from Sunyaev-Zeldovich
clusters in Planck, where we have taken the value quoted for a 20\%
X-ray mass bias \cite{PlanckXX}.
These estimates are shown as red points with $1\sigma$ error bars in
Figure~\ref{fig:growth}a, where we have scaled the amplitudes 
to $\Omega_m=0.30$ using the formulas listed above and retained
the original fractional errors.
We compare to model predictions of $\sigma_8(\Omega_m/0.3)^{0.4}$,
treating 0.4 as a representative slope for these constraints.

Recently \cite{2014arXiv1408.4742M} have completed an independent analysis of
tomographic cosmic shear in the CFHTLens data and confirmed the
findings of \cite{Heymans13}.  
Conversely, \cite{2014arXiv1407.4516M} have
performed a cluster mass function analysis using extensive
weak lensing calibration of X-ray cluster masses and found
$\sigma_8(\Omega_m/0.3)^{0.17} = 0.81 \pm 0.03$, which 
corresponds to higher $\sigma_8$ for $\Omega_m \approx 0.3$ than the 
other cluster studies listed above.
We also show this point in Figure~\ref{fig:growth}a, with the
caution that the scaling with $\Omega_m$ reported by 
\cite{2014arXiv1407.4516M} is quite different from that
of the other analyses.

For redshift-space distortion (RSD), 
the point labeled Beu14 in Figure~\ref{fig:growth}b shows
the recent BOSS CMASS measurement
by \cite{Beutler14}, which yields 
$f(z)\sigma_8(z) = 0.422 \pm 0.027$ at $z = 0.57$, where
$f(z) \approx [\Omega_m(z)]^{0.55}$ is the linear fluctuation growth rate.
This analysis fits simultaneously for redshift-space distortion and
the Alcock-Paczynski (AP) effect \cite{Alcock79}.
Here we have used the error for fixed value of the AP parameter
$D_M(z) H(z)$ because the geometry is well constrained by our
BAO+SN+CMB data, so that the fractional error in the AP parameter
is much smaller than the 6.4\% error on $f(z)\sigma_8(z)$.
The point labeled Sam14 shows the estimate $f(z)\sigma_8(z) = 0.447 \pm 0.028$
from the same data set using a power spectrum analysis instead of a 
correlation function analysis.  Since the data are the same, the
difference from Beu14 provides an indication of the uncertainties
associated with modeling systematics.
Other analyses of redshift-space clustering in BOSS
\cite{2013MNRAS.433.3559C,2013MNRAS.433.1202S} and the WiggleZ survey \cite{2011MNRAS.415.2876B}
yield compatible results.
We also plot an estimate of $f(z)\sigma_8(z) = 0.450 \pm 0.011$ 
from an analysis of smaller scale redshift-space distortions in
the CMASS sample by \cite{Reid14}, which adopts more aggressive
modeling assumptions and achieves a substantially smaller statistical
error.

At higher redshift, the 1-dimensional flux power spectrum
of the Lyman-$\alpha$ forest probes the underlying matter clustering,
with the tightest constraints on comoving scales of 
a few Mpc \cite{Croft02,McDonald05,Palanque13}.
Here we take the result from the BOSS analysis of 
\cite{Palanque13}, who find $\sigma_8 = 0.83\pm 0.03$
when fitting a \lcdm\ model to the 1-d $P(k)$ at redshifts
$z = 2.2-4.0$.  We translate this result to a constraint on
$\sigma_8(z=2.5) = 0.311 \pm 0.011$ by using the growth factor
at $z=3$ for their central value of $\Omega_m = 0.26$.
This measurement is indicated by
a point with $1\sigma$ error bar in
Figure~\ref{fig:growth}c.

Given the wide range of models that we wish to consider and
the several-percent errors on the observational data, we
have opted for an approximate method of computing clustering
amplitude predictions that is accurate at the 0.5-percent
level or better.
Following the strategy of \cite{Hu04} and \cite{Weinberg13},
we first use CAMB calculations to calibrate a Taylor expansion
for the value of $\sigma_8(z=9)$ about a fiducial Planck \lcdm\ model,
finding
\begin{equation}
\begin{split}
\sigma_8(&z=9) = 0.1058 \times 
  \left({A_s \over 2.196\times 10^{-9}}\right)^{1/2} 
  \left({\Omega_m h^2 \over 0.1426}\right)^{0.520} \\
  &\times
  \left({\Omega_b h^2 \over 0.02205}\right)^{-0.294} 
  \left({h\over 0.673}\right)^{0.683} 
  \left({\neff \over 3.046}\right)^{-0.24} \\
  &\times
  e^{0.3727(n_s-0.96)}
  \left(1-\Omega_k\right)^{0.175}.
\end{split}
\label{eqn:sig8fit}
\end{equation}
Here $A_s$ is the amplitude of primordial curvature perturbations
at the scale $k_0 = 0.05\,$Mpc$^{-1}$.
This formula updates equations~(46)-(47) of \cite{Weinberg13},
which were expanded about a WMAP7 fiducial model.
The fairly strong $h$-dependence arises because of the conversion
from a power spectrum predicted in Mpc units based on cosmological
parameter values to an amplitude defined on a scale of $8\hmpc$.
We have made numerous checks
of this formula against full CAMB calculations
for models in the parameter ranges allowed by Planck + WP data, 
finding accuracy of better than 0.1\% for \lcdm, 
for \olcdm\ with $-0.2<\Omega_k < 0.2$,
and for \wcdm\ with $-1.2<w<-0.8$, and 
accuracy better than 0.5\% for $\Delta\neff$ models
with $2.5 < \neff < 4.5$.
While equation~(\ref{eqn:sig8fit}) correctly reflects the response
of $\sigma_8(z=9)$ to an isolated change in $\neff$, in practice
the CMB-preferred values of $A_s$, $\omega_m$, $\omega_n$, $h$, and
$n_s$ all increase when $\neff$ increases, with the result that
higher $\neff$ models end up predicting higher clustering amplitudes.
We have not attempted to incorporate the effects of non-zero 
neutrino mass in this formula because the suppression of clustering
by neutrino free streaming is redshift and scale dependent
(see \cite{Rossi14} for useful representations).

Except in early dark energy models, the value of $\sigma_8(z=9)$
is essentially independent of dark energy parameters because
dark energy is dynamically insignificant at $z > 9$
(e.g., $\rho_\Lambda/\rho_m < 0.003$ for a cosmological constant).
To evolve $\sigma_8$ forward to $z=3$, 0.57, or 0, we use
the approximate integral formulation of the growth factor
from equation (16) of \cite{Weinberg13}, which simply integrates
the growth rate approximation of \cite{Linder05},
$f(z) \approx [\Omega_m(z)]^\gamma$ with 
$\gamma = 0.55 + 0.05[1+w(z=1)]$.
Spot checks against exact calculations with \texttt{cosmomc}
indicate that this approach is accurate to 0.3\% or better
for models with $m_\nu = 0$ and other 
parameters in the range allowed by our CMB+BAO+SN data,
although it becomes less accurate for more extreme parameter
values (especially of $\Omega_k$).
For $\sum m_\nu = 0.06\eV$, CAMB yields a ratio 
$\sigma_8(z=0)/\sigma_8(z=9)$ that is 0.5\% lower than 
for $m_\nu = 0$, with little dependence on other parameters,
so we also multiply all of our low-redshift $\sigma_8$ values
by 0.995 to account for this effect.

We determine the mean values and error bars on the predicted 
growth observables for our models by computing the posterior-weighted
mean and $1\sigma$ dispersion of $\sigma_8\Omega_m^{0.4}$,
$\sigma_8(z=0.57)[\Omega_m(z=0.57)]^{0.55}$, or
$\sigma_8(z=2.5)$ for the parameter values in our MC chains,
using the above approximations for $\sigma_8$.  
Because our chains do not actually use or include values of $A_s$,
we compute $\sigma_8$ for the fiducial value in 
equation~(\ref{eqn:sig8fit}) and add a fractional error 
(based on the Planck+WP column in Table~5 of \cite{PlanckXVI})
of 1.25\% in quadrature to the MCMC error to account for the 2.5\%
error in $A_s$, which is proportional to $\sigma_8^2$.
Inspection of Planck chains indicates only weak correlations
between $A_s$ and other cosmological parameters, so the approximation
of an independent error contribution added in quadrature
should be adequate.
We also add in quadrature a fractional error of 0.3\% to 
represent potential errors of our approximate growth calculations,
though our spot checks indicate higher accuracy than this.

Figure~\ref{fig:growth}a shows a persistent offset between
the predicted amplitude of matter clustering and the 
majority of observational
estimates from weak lensing and cluster masses.
For \lcdm, where the model predictions are best constrained,
the statistical significance of the tension with any given
data set is usually only $\approx 2\sigma$ or smaller.
However, the sign of discrepancy is usually the same, so the
overall significance is high unless multiple analyses are
affected by a common systematic.
The important exceptions are the cosmic shear measurement
from the Deep Lens Survey \cite{Jee13} and 
the recent cluster analysis of
\cite{2014arXiv1407.4516M}, which both agree well 
with the \lcdm\ prediction. The authors of \cite{2014arXiv1407.4516M}
emphasize that theirs was
a ``blind'' analysis in which technical choices about 
data cuts and procedures were made without knowing their eventual
impact on inferred cosmological parameter values.
Our predicted value of $\sigma_8\Omega_m^{0.4}$ is somewhat
lower than the value inferred by \cite{PlanckXVI}
from CMB data alone (Planck+WP+highL),
in part because the BAO data pull
towards lower $\Omega_m$, and in part because our compressed
CMB description does not include the lensing information in the
Planck power spectrum, which pulls towards higher $\sigma_8$.
For more flexible dark energy models, central values of
$\sigma_8\Omega_m^{0.4}$ remain within the $1\sigma$ range
found for \lcdm, and the error bars are moderately larger.
The tension with the data is moderately reduced in these models,
but not eliminated.  A formal assessment of this reduced disagreement
is difficult because the true level of systematic uncertainty
in the low-redshift measurements is itself uncertain, but 
at a qualitative level this reduction appears too small to favor
adopting one of these more complex models.

Figure~\ref{fig:growth}b shows an offset of similar magnitude between the
predictions of our standard dark energy models and the redshift-space
distortion measurement of \cite{Beutler14} at $z=0.57$.
However, the statistical significance of this tension is low because
of the statistical error on the measurement, and the power spectrum
analysis of \cite{2014MNRAS.439.3504S} yields a higher central value.
The more precise
determination from \cite{Reid14}, which draws on simulation-based
modeling of non-linear scales, overlaps the \lcdm\ prediction
at $\approx 1\sigma$.
A recent analysis that combines galaxy clustering and galaxy-galaxy
lensing of CMASS galaxies \cite{More14}
yields constraints in the $\sigma_8-\Omega_m$
plane that are consistent with the \lcdm\ model predictions,
but the current statistical errors are large enough to be
compatible with the central values from any of the redshift-space
distortion analyses shown here.
In contrast to the lower redshift results, 
Figure~\ref{fig:growth}c shows that 
the predicted clustering amplitude at $z=2.5$
in our standard models
is {\it lower} than that inferred from the \lyaf\ power spectrum,
though consistent at $\approx 1\sigma$.

In the $\Delta\neff$ model, where we assume a \lcdm\ cosmology
but allow extra relativistic species, the preferred value of
$\neff$ is higher than the standard value of 3.046, as shown
previously in Figure~\ref{fig:neff}.  Because of the
correlation of $\neff$ with other cosmological parameters
in CMB fits, the central value of the clustering amplitude
predictions shifts upwards, while the freedom in $\neff$
broadens the error bar relative to standard \lcdm.
These changes noticeably increase the tension with
the clustering measurements at $z=0$ and the RSD measurements
at $z=0.57$, though they improve agreement with the \lyaf\ 
power spectrum at $z=2.5$.  Overall, current clustering
measurements provide moderate evidence against extra
relativistic species, though a firmer understanding of
systematic uncertainties in these measurements will be needed
to draw solid conclusions.  

Massive neutrinos have a redshift- and scale-dependent impact on
matter clustering, which changes between the linear and non-linear regimes.
We have not attempted a full examination of free-$\sum m_\nu$ 
models in this section because the summary of the observational
results in terms of $\sigma_8-\Omega_m$ constraints may not
adequately capture the effect of massive neutrinos on the
clustering observables.  A value of $\sum m_\nu = 0.5\eV$,
near the 95\% upper bound inferred from our compressed CMB description
and BAO constraints, would lower the predicted value of $\sigma_8$ in a
Planck+WP-normalized \lcdm\ model by about 12\% relative to 
$\sum m_\nu = 0.06\eV$.  A value $\sum m_\nu = 0.25\eV$, near the
upper bound that we find when combining the full Planck likelihood
with BAO data, would produce a 6\% suppression of $\sigma_8$. These
numbers are somewhat different from what a naive expectation based on
linear suppression would indicate because CMB degeneracies are important
at these relatively large neutrino mass fractions.
Even the lower value is enough to remove the tension 
seen in Figure~\ref{fig:growth}a.  However, the corresponding
decrease in $\sigma_8(z=2.5)$ produces a significant discrepancy with the
\lyaf\ measurement in Figure~\ref{fig:growth}c, and a full analysis
that models the \lyaf\ power spectrum based on 
hydrodynamic simulations with a massive neutrino component
leads to a stringent upper limit on neutrino mass
\cite{Palanque14}.

As discussed in Section~\ref{sec:ede}, our geometric constraints
are nearly degenerate with respect to the presence of an early
dark energy component, provided this early dark energy is present
in the radiation-dominated epoch as well as the matter-dominated
epoch and therefore shrinks the scale of the sound horizon.
Increasing the early dark energy fraction reduces the value 
of $\Omega_m$ (see Fig.~\ref{fig:edeconstraints})
and will also suppress growth of structure relative
to \lcdm.  Predictions of structure for early dark energy are
subtle because of the combined impacts of CMB normalization,
the imprint of early dark energy fluctuations on the CMB itself,
and the post-recombination growth rate.  We therefore defer
detailed investigation of early dark energy models to future
work and make the qualitative observation that an early dark energy 
component will go in
the direction of reducing tensions with low redshift clustering 
measurements.

\section{Conclusions}
\label{sec:conclusions}

In the decade since the first observational detection of baryon
acoustic oscillations, BAO analysis has emerged as one of the sharpest
tools of precision cosmology. Its power arises from 
the grounding of its absolute distance scale in straightforward
underlying physics, from the distinctiveness of a feature that is localized
in scale and thus not easily mimicked by observational systematics,
and from its insensitivity to non-linear gravitational evolution and galaxy
formation physics (a consequence of the large scale of BAO). The
principal challenge of the method is that one must map enormous cosmic
volumes to obtain good statistical precision.  Building on the legacy
of 2dFGRS, SDSS-I/II, 6dFGS, and WiggleZ, BOSS has made major progress
on this challenge, with distance scale measurements of $1-2\%$
precision at $z=0.32$, 0.57, and 2.34.  The combination of BAO
measurements with Planck+WP CMB data and the JLA SNIa compilation
leads to numerous significant
constraints on dark energy, space curvature, and the
cosmic matter and radiation density.

If we treat BAO as an {\it uncalibrated} standard ruler,
assuming only that it is constant in time, then the combination
of galaxy and \lyaf\ BAO measurements yields a strong 
($>3\sigma$) detection of dark energy, independent of 
any other cosmological data.
If we assume that the angular acoustic scale of the CMB
represents the same standard ruler, then the resulting
constraints in an \olcdm\ model collapse around
a flat universe dominated by dark energy,
with $\Omega_m=0.292\pm0.18$, $\Omega_k=-0.010\pm 0.016$.
Thus, high-precision measurements of a common
standard ruler at $z < 0.7$, $z = 2.34$, and $z=1090$
already lead to strong constraints on the cosmological model.

BAO become much more powerful when we incorporate the absolute
calibration of the sound horizon $r_d$ using CMB measurements
of the matter, baryon, and radiation energy density
(eq.~\ref{eqn:rd}).
With Planck+WP CMB data, residual uncertainties in
$\omega_m$ and $\omega_b$ leave only 0.4\% uncertainty in
the acoustic scale $r_d = 147.49 \pm 0.59$ Mpc
assuming a standard radiation background with three
neutrino species.
One particularly interesting application of this calibration
is to combine galaxy BAO measurements with the high-precision
measurements of {\it relative} distances from Type Ia SNe
to infer $H_0$.  The addition of the SN data makes the inferred
value of $H_0$ insensitive to uncertainties in the dark 
energy model, which would otherwise affect the extrapolation
of the distance scale from the moderate redshifts of the
BAO measurements down to $z=0$.  With our standard BAO and
SN data sets, this inverse distance ladder measurement yields
$H_0 = 67.3 \pm 1.1 \hubunits$, where the 1.7\% uncertainty
includes the Planck+WP uncertainty in $r_d$.
This value agrees perfectly with the value inferred from current
CMB data under the much stronger assumption of a flat \lcdm\
model, an important consistency test of the standard cosmology.
It is lower than most recent estimates using a Cepheid-based distance ladder.
Our measurement of $H_0$ does rely on the
assumption of a standard cosmic radiation background, and
the directly constrained parameter combination is $H_0 r_d$.
A convincing discrepancy with conventional distance-ladder
determinations of $H_0$ could not be resolved by appealing
to the late-time behavior of dark energy.
It would point instead to 
non-standard physics in the pre-recombination universe,
such as extra relativistic degrees of freedom or early dark energy,
which can shrink $r_d$ and thus raise the inferred value of $H_0$.  

The full combination of CMB, BAO, and SN data places strong constraints
on dark energy and space curvature, as summarized in 
Figure~\ref{fig:models} and Table~\ref{tab:constraints}.
In models that allow both $w \neq -1$ and non-zero curvature,
the BAO and SN data are highly complementary.
For the \owcdm\ model, we find $w = -0.98 \pm 0.06$ and
$\Omega_k = -0.002 \pm 0.003$.  For models with $w(a) = w_0 + w_a(1-a)$,
the constraint on evolution remains poor,
with $w_a = -0.6 \pm 0.6$ in \owwacdm, but the value
of $w$ at the pivot redshift where it is best constrained
remains close to $-1$.
A striking feature of Table~\ref{tab:constraints} is that
as degrees of freedom are added to the cosmological model
the best-fit values of parameters barely change, always remaining
close to those of flat \lcdm.

These models are fit to a total of 43 observables:
three in our compressed description of the CMB,
31 for the compressed SN data, five for the galaxy BAO data
($D_V$ from 6dFGS, MGS and BOSS LOWZ, $D_M$ and $D_H$ from 
BOSS CMASS), and four for the \lyaf\ BAO
($D_H$ and $D_M$ from forest auto-correlation
and from quasar-forest cross-correlation).
The \lcdm\ model, with three free parameters ($\Omega_\Lambda$, $h$, and
the absolute magnitude normalization for SNIa),
has $\chi^2 = 46.79$ for 40 d.o.f., which is statistically 
acceptable.\footnote{Note that Figure~\ref{fig:chisq} omits the
CMB data from the $\chi^2$ accounting, so it has three fewer d.o.f.}
The decrease in $\chi^2$ for the alternative models is not enough to 
justify the addition of parameters; for example, the addition of
three free parameters in \owwacdm\ reduces $\chi^2$ by only 1.33.
However, the best-fit models in all of these cases are in significant 
tension with the \lyaf\ measurements on their own, typically at the 
$2-2.5\sigma$ level.  The \lyaf\ data have little impact
on the best-fit parameter values in any of these models,
not because they agree well with the model predictions
but because parameter changes that would significantly
improve agreement with the \lyaf\ run afoul of the
higher precision galaxy BAO measurements.  Moreover,
because the \lyaf\ measurements have lower $D_H$ but
higher $D_M$ than expected in the best-fit \lcdm\ model,
many parameter changes that would improve the fit to $D_H$
worsen the fit to $D_M$, and vice versa.

We have examined several models with non-standard dark energy or
dark matter histories or non-standard radiation backgrounds.
Early dark energy that has constant $\OmegaDE^e$ in the 
radiation and matter dominated eras (before evolving
towards a cosmological constant at low redshift)
alters the sound horizon $r_d$ and evolution of 
$H(z)$ and $D_M(z)$.  Remarkably, the
cancellation of these effects leaves the BAO observables
$D_H(z)/r_d$ and $D_M(z)/r_d$ nearly unchanged (including
at $z=1090$), even for $\OmegaDE^e$ as large as 0.3,
so that the observations incorporated in our fits still
allow a substantial early dark energy component.
Because of the smaller $r_d$, such model fits yield a
higher $H_0$ and lower $\Omega_m$, and the suppression of
growth by early dark energy is likely to reduce the amplitude
of low-redshift matter clustering.  
Full CMB power spectrum analyses yield stronger
but more model-dependent constraints on early dark energy
through its influence on the shape of the acoustic peaks
and the structure of the damping tail \cite{Hojjati13}.
Nonetheless, the ability of these models to match expansion
history constraints while improving agreement with local $H_0$
and structure growth measurements suggests that they merit
further investigation.

If ``early'' dark energy becomes important only
after recombination, so that it does {\it not} alter the acoustic
scale, then CMB+BAO data impose strong constraints, with 
$\OmegaDE^e < 0.03$ at 95\% confidence.
A similar conclusion applies to other physical effects that
distort the low-redshift distance scale relative to $r_d$
and the distance to last scattering.
In particular, we considered models in which a component of dark
matter decays into radiation on cosmological timescales,
boosting $\Omega_r$ and decreasing $\Omega_m$ at low
redshifts.  The BAO+CMB data limit the 
fraction of dark matter that can decay by $z=0$ to below 
3\% (95\% confidence).

With respect to expansion history, massive neutrinos are in some
sense the converse of decaying dark matter: they are relativistic
at the epoch of recombination, but at low redshift they increase
the matter density $\Omega_m$ relative to the value $\Omega_{cb}$
inferred from the CMB acoustic peaks.
The purely geometric constraints that come from BAO and our
compressed CMB description yield a 95\% confidence upper limit of 
$\sum m_\nu < 0.56\eV$ assuming \lcdm, with moderately weaker
limits for models that allow $w \neq -1$ or non-zero $\Omega_k$
(see Table~\ref{tab:mnu}).  If we use full Planck CMB chains
in place of our compressed description we obtain the significantly
stronger limit $\sum m_\nu < 0.25\eV$, a difference driven by the
relatively high amplitude lensing signal detected in the Planck
power spectrum.  Measurements of low-redshift matter clustering
can yield more sensitive limits on neutrino masses, and potentially
a measurement of $\sum m_\nu$ through its impact on structure growth,
but the expansion history constraints are robust and impressively
stringent on their own.

Adding relativistic degrees of freedom can noticeably
improve the agreement with the \lyaf\ BAO, and if we combine
only CMB and \lyaf\ data the preferred $\neff$ is $\approx 4$.
However, increasing $\neff$ worsens agreement with the
galaxy BAO data, and when we consider our full data
combination we find $\neff = 3.43 \pm 0.26$.
Increasing $\neff$ reduces the value of $r_d$ and thereby leads
to a higher inferred $H_0$; for a model with free $\neff$ and
free tensor-to-scalar ratio $r$ we find a marginalized constraint
$H_0 = 70.1 \pm 1.7 \hubunits$. 
We caution that modifying the radiation background alters
the shape of the acoustic peaks, an effect not accounted
for in the BAO measurements used here; we expect this effect
to be smaller than our statistical errors, but perhaps
not negligible.

Among these alternative models, only the model with
free $\neff$ can reduce the tension with the \lyaf\ data,
and even there the reduction is small once the galaxy BAO
constraints are also imposed.
We did construct a model with a tuned
oscillation in $D_M(z)$ that reproduces both the
$D_M/r_d$ and $D_H/r_d$ measurements from the \lyaf\ BAO
while continuing to satisfy all other constraints.
However, this model requires non-monotonic evolution of $H(z)$
and thus of $\rho_{\rm tot}(z)$, which is difficult to
achieve in any model with non-negative dark energy density.
The artificiality and physical implausibility of this
model, and the failure of our more physically motivated models,
illustrate how difficult it is to obtain a good fit
to the BOSS \lyaf\ BAO measurements.
This difficulty suggests that the tension of simpler models
with the \lyaf\ data is a statistical fluke,
or perhaps reflects an unrecognized systematic
in the BAO measurement, but it highlights the importance
of further measurements of $D_M$ and $H(z)$ at high redshifts.

The cosmological constraints considered here are essentially
geometrical, tied to the expansion history of the homogeneous
universe.  As a further test, we have computed the predictions
of our models for low redshift measurements of matter clustering.
Confirming previous findings, but now with tighter
cosmological parameter constraints, we find that a \lcdm\ model
normalized to the observed amplitude of CMB anisotropies
predicts cluster masses, weak lensing signals, and redshift-space
distortions that are higher than most observational estimates.
The tension with individual data sets is only $\approx 2\sigma$,
and the measurements themselves may be affected by systematics.
However, the direction of the discrepancy is consistent across
many analyses (though not all of them).
The additional freedom in standard
dark energy models does not reduce this tension because the
parameter values allowed by our data are always close to those of \lcdm.
Massive neutrinos can reduce the tension by suppressing
structure growth on small scales (lowering $\sigma_8$),
an effect that is small but not negligible for neutrino masses
in the range allowed by our fits.  
Conversely, increasing $\neff$ above the standard value of 3.046 
leads to higher predicted values of $\sigma_8$ because of correlation 
with other cosmological parameters, thus amplifying the tension.
As previously noted,
early dark energy may reduce the tension
with the clustering data, both because it suppresses growth
of structure during the matter dominated era and because
the reduced $r_d$ value leads to higher $h$ and lower $\Omega_m$
when combined with CMB constraints.
Our standard \lcdm\ fits produce good agreement with the matter
clustering amplitude inferred from the \lyaf\ power spectrum
at $z \approx 2.5$; this agreement is itself an important
constraint on neutrino masses or other physical mechanisms
that reduce small scale clustering \cite{Palanque14}.

The application of the BAO technique to large cosmological surveys has
enabled the first percent-level measurements of absolute distances
beyond the Milky Way.  In combination with CMB and SN data, these
measurements yield impressively tight constraints on the cosmic 
expansion history and correspondingly stringent tests of dark energy
theories.  Over the next year, the strength
of these tests will advance significantly with the final results
from BOSS and with CMB polarization and improved temperature maps from Planck.
In the longer term, BAO measurements will gain in precision and
redshift range through a multitude of ongoing or planned spectroscopic surveys,
including SDSS-IV eBOSS, HETDEX, SuMIRE, DESI, WEAVE, Euclid, and 
WFIRST.\footnote{See \cite{Weinberg13b} for a brief summary
of these projects and references to more detailed descriptions.}
These data sets also enable precise measurements of matter
clustering through redshift-space distortion analyses, the shape
of the 3-dimensional power spectrum, and other clustering statistics.
In combination with the expansion history constraints, these measurements can
test modified gravity explanations of cosmic acceleration and
probe the physics of inflation, the masses of neutrinos, and 
the properties of dark matter.  
In parallel with these large spectroscopic surveys,
supernova measurements of expansion history are gaining in precision,
data quality, and redshift range, and weak lensing constraints
on matter clustering are advancing to the percent and sub-percent
level as imaging surveys grow from millions of galaxy shape measurements
to hundreds of millions, and eventually to billions.
From the mid-1990s through the early 21st century, improving cosmological
data sets transformed our picture of the universe.
The next decade --- of time and of precision ---
could bring equally surprising changes to our understanding of the cosmos.

\section*{Acknowledgements}

We thank Eric Linder for useful discussions of early dark energy and
structure growth. We also thank Savvas Koushiappas and Gordon
Blackadder for alerting us to an error in the decaying dark matter
section of the preprint version of this paper and answering
our questions as we corrected it.

Funding for SDSS-III has been provided by the Alfred P. Sloan
Foundation, the Participating Institutions, the National Science
Foundation, and the U.S. Department of Energy Office of Science. The
SDSS-III web site is \texttt{http://www.sdss3.org/}.

SDSS-III is managed by the Astrophysical Research Consortium for the
Participating Institutions of the SDSS-III Collaboration including the
University of Arizona, the Brazilian Participation Group, Brookhaven
National Laboratory, Carnegie Mellon University, University of
Florida, the French Participation Group, the German Participation
Group, Harvard University, the Instituto de Astrofisica de Canarias,
the Michigan State/Notre Dame/JINA Participation Group, Johns Hopkins
University, Lawrence Berkeley National Laboratory, Max Planck
Institute for Astrophysics, Max Planck Institute for Extraterrestrial
Physics, New Mexico State University, New York University, Ohio State
University, Pennsylvania State University, University of Portsmouth,
Princeton University, the Spanish Participation Group, University of
Tokyo, University of Utah, Vanderbilt University, University of
Virginia, University of Washington, and Yale University.

\appendix
\section{List of institutions} 

The following is the list of institutions corresponding to the
list of authors on the front page.

\vspace*{0.3cm}

\noindent
$^{1}$ APC, Astroparticule et Cosmologie, Universit\'e Paris Diderot, CNRS/IN2P3, CEA/Irfu, Observatoire de Paris, Sorbonne Paris Cit\'e, 10, rue Alice Domon \& L\'eonie Duquet, 75205 Paris Cedex 13, France \\
$^{2}$ Lawrence Berkeley National Laboratory, 1 Cyclotron Road, Berkeley, CA 94720, USA \\
$^{3}$ Department of Astronomy, University of Washington, Box 351580, Seattle, WA 98195, USA \\
$^{4}$ Apache Point Observatory, P.O. Box 59, Sunspot, NM 88349-0059, USA \\
$^{5}$ Center for Cosmology and Particle Physics, New York University, New York, NY 10003, USA \\
$^{6}$ Department of Physics and Astronomy, UC Irvine, 4129 Frederick Reines Hall, Irvine, CA 92697, USA \\
$^{7}$ Department Physics and Astronomy, University of Utah, UT 84112, USA \\
$^{8}$ Institute for Advanced Study, Einstein Drive, Princeton, NJ 08540, USA \\
$^{9}$ Institute of Cosmology \& Gravitation, Dennis Sciama Building, University of Portsmouth, Portsmouth, PO1 3FX, UK \\
$^{10}$ Observat\'orio Nacional, Rua Gal.~Jos\'e Cristino 77, Rio de Janeiro, RJ - 20921-400, Brazil \\
$^{11}$ Laborat\'orio Interinstitucional de e-Astronomia, - LIneA, Rua Gal.Jos\'e Cristino 77, Rio de Janeiro, RJ - 20921-400, Brazil \\
$^{12}$ Instituto de Fisica Teorica (UAM/CSIC), Universidad Autonoma de Madrid, Cantoblanco, E-28049 Madrid, Spain \\
$^{13}$ Department of Physics, Carnegie Mellon University, 5000 Forbes Avenue, Pittsburgh, PA 15213, USA \\
$^{14}$ Astrophysics, University of Oxford, Keble Road, Oxford OX13RH, UK \\
$^{15}$ Department of Physics, Yale University, 260 Whitney Ave, New Haven, CT 06520, USA \\
$^{16}$ Institut de Ci\`encies del Cosmos, Universitat de Barcelona, IEEC-UB, Mart\'\i~i Franqu\`es 1, E08028 Barcelona, Spain \\
$^{17}$ Laboratoire d’astrophysique, Ecole Polytechnique Fédérale de Lausanne (EPFL), Observatoire de Sauverny,CH-1290 Versoix, Switzerland \\
$^{18}$ Harvard-Smithsonian Center for Astrophysics, 60 Garden St., Cambridge, MA 02138, USA \\
$^{19}$ Department of Astronomy, University of Florida, Gainesville, FL 32611, USA \\
$^{20}$ CEA, Centre de Saclay, IRFU, 91191 Gif-sur-Yvette, France \\
$^{21}$ Department of Astrophysical Sciences, Princeton University, Ivy Lane, Princeton, NJ 08544, USA \\
$^{22}$ Key Laboratory for Research in Galaxies and Cosmology of Chinese Academy of Sciences, Shanghai Astronomical Observatory, Shanghai 200030, China \\
$^{23}$ LPNHE, CNRS/IN2P3, Universit\'e Pierre et Marie Curie Paris 6, Universit\'e Denis Diderot Paris 7, 4 place Jussieu, 75252 Paris CEDEX, France \\
$^{24}$ Department of Physics, Ohio State University, Columbus, Ohio 43210, USA \\
$^{25}$ Center for Cosmology and Astro-Particle Physics, Ohio State University, Columbus, Ohio, USA \\
$^{26}$ Leibniz-Institut f\"{u}r Astrophysik Potsdam (AIP), An der Sternwarte 16, 14482 Potsdam, Germany \\
$^{27}$ CPPM, Aix-Marseille Universit\'e, CNRS/IN2P3, Marseille, France \\
$^{28}$ Max-Planck-Institut f\"ur Astronomie, K\"onigstuhl 17, D69117 Heidelberg, Germany \\
$^{29}$ University College London, Gower Street, London WC1E 6BT, UK \\
$^{30}$ Instituci\'{o} Catalana de Recerca i Estudis Avan\c{c}ats, Barcelona, Spain \\
$^{31}$ Department of Physics and Astronomy, University of Wyoming, Laramie, WY 82071, USA \\
$^{32}$ Institut d'Astrophysique de Paris, UPMC-CNRS, UMR7095, 98bis boulevard Arago, 75014 Paris, France \\
$^{33}$ INAF, Osservatorio Astronomico di Trieste, Via G. B. Tiepolo 11, 34131 Trieste, Italy \\
$^{34}$ Instituto de Astrof{\'\i}sica de Canarias (IAC), C/V{\'\i}a L\'actea, s/n, E-38200, La Laguna, Tenerife, Spain \\
$^{35}$ Departamento Astrof\'{i}sica, Universidad de La Laguna (ULL), E-38206 La Laguna, Tenerife, Spain \\
$^{36}$ A*MIDEX, Aix Marseille Université, CNRS, LAM (Laboratoire d'Astrophysique de Marseille) UMR 7326, Marseille, France \\
$^{37}$ Campus of International Excellence UAM+CSIC, Cantoblanco, E-28049 Madrid, Spain \\
$^{38}$ Instituto de Astrof\'isica de Andaluc\'ia (CSIC), E-18080 Granada, Spain \\
$^{39}$ Department of Physics, University of California, 366 LeConte Hall, Berkeley, CA 94720, USA \\
$^{40}$ Department of Physics, Drexel University, 3141 Chestnut Street, Philadelphia, PA 19104, USA \\
$^{41}$ Department of Astronomy and Space Science, Sejong University, Seoul, 143-747, Korea \\
$^{42}$ Max-Planck-Institut f\"ur extraterrestrische Physik, Postfach 1312, Giessenbachstr., 85748 Garching, Germany \\
$^{43}$ Department of Physics, Kansas State University, 116 Cardwell Hall, Manhattan, KS 666506, USA \\
$^{44}$ National Abastumani Astrophysical Observatory, Ilia State University, 2A Kazbegi Ave. GE-1060 Tbilisi, Georgia \\
$^{45}$ Departamento de F\'isica Te\'orica, Universidad Aut\'onoma de Madrid, E-28049 Cantoblanco, Madrid, Spain \\
$^{46}$ Department of Astronomy and Astrophysics, The Pennsylvania State University, University Park, PA 16802, USA \\
$^{47}$ Institute for Gravitation and the Cosmos, The Pennsylvania State University, University Park, PA 16802, USA \\
$^{48}$ Department of Physics and Astronomy, Ohio University, 251B Clippinger Labs, Athens, OH 45701 \\
$^{49}$ Brookhaven National Laboratory, 2 Center Road,  Upton, NY 11973, USA \\
$^{50}$ Center for Astrophysics and Space Sciences, Department of Physics, University of California, 9500 Gilman Dr., San Diego, CA 92093 USA \\
$^{51}$ INFN/National Institute for Nuclear Physics, Via Valerio 2, 34127 Trieste, Italy \\
$^{52}$ Department of Astronomy, University of Wisconsin-Madison, 475 N. Charter Street, Madison, WI, 53706, USA \\
$^{53}$ Department of Physical Sciences, The Open University, Milton Keynes, MK7 6AA, UK \\
$^{54}$ Department of Astronomy, Ohio State University, Columbus, Ohio, USA \\
$^{55}$ PITT PACC, Department of Physics and Astronomy, University of Pittsburgh, Pittsburgh, PA 15260, USA \\
$^{56}$ Department of Astronomy, Case Western Reserve University, Cleveland, Ohio 44106, USA \\
$^{57}$ National Astronomy Observatories, Chinese Academy of Science, Beijing, 100012, P.R. China \\

\section{Decaying dark matter model}
\label{app:decaying-dark-matter}
We consider a model of dark matter decaying into radiation as 
\begin{eqnarray}
  \dot{\rho_x} &=& -3 H(t) \rho_x - \lambda H_0 \rho_x,\\
  \dot{\rho_g} &=& -4 H(t) \rho_g + \lambda H_0 \rho_x,
\end{eqnarray}
where $\rho_x$ and $\rho_g$ are the new decaying dark matter and
radiation components and the decay time constant
$\lambda$ is made dimensionless by expressing it in units of $H_0$.  The
Hubble parameter is given by the usual expression for \lcdm\ with two
extra components
\begin{equation}
  \left(\frac{H}{H_0}\right)^2 = \Omega_{cb} a^{-3} + \Omega_\Lambda
  +{\rho_{\nu+r}(z)}/{\rho_{\rm crit}} +
  \frac{\rho_x(a)+\rho_g(a)}{\rho_{\rm crit}}
\end{equation}

Writing $\rho_x=\rho_{\rm crit} r_x a^{-3}$ and $\rho_g=\rho_{\rm crit}
r_r a^{-4}$, the system of equations can be rewritten as
\begin{eqnarray}
    \frac{d\,r_x}{d\, \ln a}  &=& - \lambda r_x \left(\frac{H}{H_0}\right)^{-1},\\
    \frac{d\,r_r}{d\, \ln a}  &=& + a \lambda r_x \left(\frac{H}{H_0}\right)^{-1},\\
  \left(\frac{H}{H_0}\right)^2 &=& \Omega_{cb} a^{-3} + \Omega_\Lambda
  +{\rho_{\nu+r}(z)}/{\rho_{\rm crit}}  \nonumber \\ && + r_x(a) a^{-3} +r_r(a) a^{-4},
\end{eqnarray}
with initial conditions $r_x(a=1) =\Omega_x$ and $r_r(a=1)=\Omega_r$.
We can solve this system of differential equations starting at $a=1$
and going backwards in time for a given choice of $\Omega_x$,
$\Omega_r$ and $\lambda$.  

However, in our parametrization, boundary conditions are specified in
the infinite past.  We therefore use a minimizing routine that
determines the values of $\Omega_x$ and $\Omega_r$ today that are
required to obtain the right fraction of decaying dark matter fraction
and zero initial density in the decay product in the infinite past (assumed
to be $a\sim 10^{-4}$ in the code). At each step in minimization, the
evolution equations are solved numerically and a suitable penalty
function is evaluated.

\bibliographystyle{revtex}
\bibliography{cosmo,cosmo_preprints}

\end{document}